\documentclass[superscriptaddress, aps, preprintnumbers,
amsmath, amssymb, sort&compress, nofootinbib, 10pt, paper, floatfix]{revtex4-2}

\usepackage{amsfonts,amsmath,amssymb,ascmac,bm,tensor, microtype}
\usepackage{fnpct} 
\usepackage{comment}
\usepackage{ifpdf}
\usepackage{slashed}
\usepackage{color}
\usepackage[mathscr]{eucal}
\usepackage[utf8]{inputenc}
\usepackage{cancel}
 
\ifpdf
  \usepackage{graphicx}     
  \usepackage[bookmarksopen,colorlinks=true,linkcolor=bblue,citecolor=bblue,urlcolor=ppink]{hyperref}
\else     
\fi

\definecolor{red}{rgb}{1,0,0}
\definecolor{darkred}{rgb}{0.6,0,0}
\definecolor{darkgreen}{rgb}{0.992447,0.623778,0.034597}
\definecolor{ppink}{rgb}{1,0.4,0.4}
\definecolor{bblue}{rgb}{0.284602,0.317763,0.963947}
\definecolor{purple}{rgb}{0.5 ,0, 0.7}

\definecolor{dgreen}{rgb}{0 ,0.5, 0.5}

\newcommand{\vev}[1]{ \left< {#1} \right> }
\newcommand{\dd}{\mathrm{d}}

\newcommand{\GW}{\text{GW}}
\newcommand{\GeV}{\text{GeV}}

\newcommand{\tot}{\text{tot}}
\newcommand{\ee}{\text{e}}
\newcommand{\hh}{\text{h}}

\newcommand{\eq}{{\text{eq}}}

\newcommand{\NL}{{\text{NL}}}
\newcommand{\Pl}{{\text{Pl}}}
\newcommand{\pbh}{{\text{PBH}}}
\newcommand{\dm}{{\text{DM}}}

\newcommand{\rr}{\text{r}}
\newcommand{\thre}{\text{th}}

\newcommand{\tils}{{\tilde \sigma}}

\newcommand{\rd}{\text{roll}}
\newcommand{\dec}{{\text{dec}}}
\newcommand{\decm}{{\text{dec},-}}
\newcommand{\decp}{{\text{dec},+}}

\newcommand{\en}{{\text{end}}}
\newcommand{\Rf}{{\text{R}}}
\newcommand{\Rs}{{\text{dec}}}
\newcommand{\Mpc}{\text{Mpc}}

\makeatletter
\newcommand\footnoteref[1]{\protected@xdef\@thefnmark{\ref{#1}}\@footnotemark}
\makeatother

\allowdisplaybreaks[1]

\begin{document}

\preprint{KEK-TH-2554, KEK-Cosmo-0324}

\title{
Axion Curvaton Model for the Gravitational Waves Observed by Pulsar Timing Arrays
}

\author{Keisuke Inomata}
\affiliation{William H. Miller III Department of Physics and Astronomy, Johns Hopkins University, 3400 N. Charles Street, Baltimore, Maryland, 21218, USA}
\affiliation{Kavli Institute for Cosmological Physics and Enrico Fermi Institute, University of Chicago, Chicago, IL 60637, USA}
\author{Masahiro Kawasaki}
\affiliation{ICRR, University of Tokyo, Kashiwa, 277-8582, Japan}
\affiliation{Kavli IPMU (WPI), UTIAS, University of Tokyo, Kashiwa, 277-8583, Japan}
\author{Kyohei Mukaida}
\affiliation{KEK Theory Center, Tsukuba 305-0801, Japan}
\affiliation{Graduate Institute for Advanced Studies (Sokendai), Tsukuba 305-0801, Japan}
\author{Tsutomu T.~Yanagida}
\affiliation{T. D. Lee Institute and School of Physics and Astronomy, Shanghai Jiao Tong University, 800 Dongchuan Rd, Shanghai 200240, China}
\affiliation{Kavli IPMU (WPI), UTIAS, University of Tokyo, Kashiwa, 277-8583, Japan}

\begin{abstract}
\noindent
The stochastic gravitational wave background (SGWB) recently detected by the PTA collaborations could be the gravitational waves (GWs) induced by curvature perturbations.
However, primordial black holes (PBHs) might be overproduced if the SGWB is explained by the GWs induced by the curvature perturbations that follow the Gaussian distribution.
This motivates models associated with the non-Gaussianity of the curvature perturbations that suppress the PBH production rate. 
In this work, we show that the axion curvaton model can produce the curvature perturbations that induce GWs for the detected SGWB while preventing the PBH overproduction with the non-Gaussianity.
\end{abstract}

\maketitle

\section{Introduction}

The stochastic gravitational wave background (SGWB) at nanohertz frequencies has been recently detected by the worldwide pulsar timing array (PTA) experiments, in particular by NANOGrav~\cite{NANOGrav:2023gor, NANOGrav:2023hde, NANOGrav:2023hvm} and by EPTA and IPTA~\cite{Antoniadis:2023rey, Antoniadis:2023utw, Antoniadis:2023zhi}, with the evidence of the Hellings-Downs pattern (see also the results of PPTA~\cite{Reardon:2023gzh, Zic:2023gta, Reardon:2023zen} and CPTA~\cite{Xu:2023wog} for the detection of the common signals).
This has given rise to a crucial query: what is the origin(s) of the detected SGWB?
GWs can be generally produced in many different situations because all matter in the universe couples to gravity.
To elucidate the origin(s) of the SGWB, it is essential to examine the characteristics of the SGWB, including its amplitude, spectral frequency dependence, polarization, and anisotropies.
It is worth mentioning that any evidence of anisotropies and statistical polarization in the SGWB has not been reported.
Because of this, the origins of the detected SGWB have been discussed in terms of its amplitude and spectral frequency dependence so far.
One of the prominent origins is supermassive black hole binaries (SMBHBs), which emit GWs through their evolution. 
However, the PTA signals have a tension to the simplest scenario where SMBHBs are circular and their energy loss is only due to GW emission~\cite{NANOGrav:2023hvm,Antoniadis:2023zhi,Ellis:2023dgf,Figueroa:2023zhu}.
Although the GW spectrum prediction can change if the eccentricity of the binaries and the environmental effects to the energy loss are taken into account, this tension motivates us to study other sources from cosmological events that can produce the SGWB in the early universe.
(See Refs.~\cite{NANOGrav:2023hvm,Ellis:2023oxs} for a comprehensive comparison between the SMBHB and the cosmological interpretations for the recent PTA results.)

In this work, we focus on one of the cosmological GWs, called scalar-induced gravitational waves (SIGWs).
These GWs arise from the nonlinear interaction between curvature (or scalar) and tensor perturbations~\cite{Ananda:2006af,Baumann:2007zm}. 
Since the nonlinear interaction appears at the second order in perturbations, SIGWs remain negligible if the power spectrum of curvature perturbations retains nearly scale-invariance from the Cosmic Microwave Background (CMB) scales to smaller scales.
However, they can be significant and detectable by the PTA experiments if the amplitude of the power spectrum on small scales is enhanced.
The SIGW interpretation of the recently detected SGWB has been discussed in Refs.~\cite{Madge:2023cak,Franciolini:2023pbf,Inomata:2023zup,Cai:2023dls,Zhu:2023faa,Wang:2023ost,Liu:2023ymk,Ebadi:2023xhq,Abe:2023yrw,Firouzjahi:2023lzg,You:2023rmn,Bari:2023rcw,HosseiniMansoori:2023mqh,Balaji:2023ehk,Basilakos:2023xof,Jin:2023wri,Das:2023nmm,Zhao:2023joc,Liu:2023pau,Yi:2023tdk,Frosina:2023nxu,Yuan:2023ofl,Bhaumik:2023wmw,Choudhury:2023wrm,Kawasaki:2023rfx,Yi:2023npi,Bhattacharya:2023ysp,Gangopadhyay:2023qjr,Chang:2023ist} (see also Refs.~\cite{Vaskonen:2020lbd,DeLuca:2020agl,Kohri:2020qqd,Domenech:2020ers,Papanikolaou:2020qtd,Inomata:2020xad,Kawasaki:2021ycf,Dandoy:2023jot} for the SIGW interpretation of the 12.5-year NANOGrav results~\cite{NANOGrav:2020bcs} and the IPTA Data Release 2~\cite{Antoniadis:2022pcn}).

Remarkably, the enhanced power spectrum produces not only GWs but also primordial black holes (PBHs).
Because of this, many works have been dedicated to exploring the relationship between PBHs and SIGWs~\cite{Saito:2008jc,Saito:2009jt,Inomata:2016rbd,Orlofsky:2016vbd,Nakama:2016gzw,Garcia-Bellido:2017aan,Di:2017ndc,Ando:2017veq,Cheng:2018yyr,Kohri:2018qtx,Chen:2019xse,Espinosa:2018eve,Kohri:2018awv,Cai:2018dig,Bartolo:2018evs,Bartolo:2018rku,Unal:2018yaa,Byrnes:2018txb,Inomata:2018epa,Clesse:2018ogk,Cai:2019amo,Cai:2019jah, Chen:2019xse,Domenech:2021ztg}.
Importantly, some studies, Refs.~\cite{Franciolini:2023pbf,Liu:2023ymk}, claim that, if the detected SGWB is from SIGWs during the radiation-dominated (RD) era and the curvature perturbations follow the Gaussian distribution, PBHs are overproduced and their energy density exceeds the current dark matter (DM) energy density.
While there remain uncertainties in the calculation of PBH abundance~\cite{Yoo:2020dkz,DeLuca:2023tun}, the results in Refs.~\cite{Franciolini:2023pbf,Liu:2023ymk} inspired investigations into PBH models with mechanisms that suppress the PBH abundance.
For instance, as demonstrated in Ref.~\cite{Harigaya:2023pmw}, if SIGWs are produced during the kination era, the PBH abundance can be significantly reduced while the SIGWs are still consistent with the PTA signals.
In Ref.~\cite{Balaji:2023ehk}, it has also been shown that the PBH overproduction can be avoided if the SIGWs are produced during the era dominated by a canonical scalar field that leads to the same equation of state as the RD era and the same sound speed of the fluid as the kination fluid.
In this work, we concentrate on an alternative model that predicts the non-Gaussianity of the curvature perturbations, which in turn suppresses PBH production.

The notion of non-Gaussianity as a solution for the PBH overproduction problem has been raised in Refs.~\cite{Franciolini:2023pbf,Liu:2023ymk}, though analyses for specific models have not been done yet.
It is worthwhile demonstrating a concrete UV model that simultaneously predicts the SIGWs for the PTA signals and the appropriate non-Gaussianity.
To this end, we focus on the axion curvaton model, where the axion works as a curvaton~\cite{Enqvist:2001zp,Lyth:2001nq,Moroi:2001ct}\footnote{See also Ref.~\cite{Pi:2021dft} for the curvaton model with a non-trivial kinetic term that causes a large enhancement of the curvature power spectrum.}.
The enhancement of the curvature power spectrum on the small scales is realized by the evolution of a complex U(1) scalar field during the inflation.
Depending on the dynamics of the complex field, the axion curvaton model can be classified into two types: type I~\cite{Kasuya:2009up,Kawasaki:2012wr,Kawasaki:2013xsa,Ando:2017veq,Ferrante:2023bgz} and II~\cite{Ando:2018nge,Kawasaki:2021ycf}.
In type I, the radial direction of the complex field (saxion) rolls down the potential with the decrease of its field value during the inflation. 
On the other hand, in type II, the saxion is first fixed at a small field value and, after a while, rolls down the potential with the increase of its field value.
Whether it is type I or type II, the potential for the phase direction field (axion) finally arises due to some nonperturbative effect after the inflation. Then, the axion starts to oscillate around the minimum and behaves as a curvaton.

In general, the curvaton model (not limited to the axion curvaton model) predicts the local-type non-Gaussianity, parameterized with the curvature perturbation $\zeta$ in real space~\cite{Sasaki:2006kq,Kawasaki:2011pd,Ando:2017veq}:\footnote{In this paper, we neglect the higher contributions to the non-Gaussianity such as $g_\NL \zeta_g^3$ or higher order terms to reduce the computational cost.
The full non-linear expression between $\zeta$ and $\zeta_g$ in the curvaton models can be seen in Refs.~\cite{Sasaki:2006kq,Ferrante:2022mui,Franciolini:2023pbf}.
Since the curvature perturbations that we focus on in this work are not in the non-perturbative regime, we can expect that the modification due to the higher-order terms is at most an $\mathcal O(1)$ change in the curvature power spectrum and the corresponding induced GW spectrum for the same PBH abundance.
This effect could be important when we discuss the PBH overproduction issue more precisely with the settlement of other uncertainties, such as those pointed out in Refs.~\cite{Yoo:2020dkz,DeLuca:2023tun}. 
}
\begin{align}
  \zeta(\bm x) \simeq \zeta_g(\bm x) + \frac{3}{5}f_\NL (\zeta_g^2(\bm x)- \vev{\zeta_g^2(\bm x)}),
\end{align}
where $\zeta_g$ follows the Gaussian distribution and $\vev{\cdots}$ denotes the ensemble average.
In the curvaton model, the non-Gaussianity parameter $f_\NL$ is related to the curvaton-radiation ratio at the curvaton decay $r$:
\begin{align}
  f_\NL &= \frac{5}{12}\left(-3 + \frac{4}{r} + \frac{8}{4+3r} \right), \\
  r &\equiv \frac{\rho^\sigma(\eta_\decm)}{\rho^\rr(\eta_\decm)},
   \label{eq:r}
\end{align}
where $\rho^\sigma$ and $\rho^\rr$ are the energy densities of the curvaton and radiation and $\eta_\decm$ is the conformal time just before the curvaton decay.
If $r > 1$, the universe is dominated by the curvaton before its decay. 
The suppression of the PBH production can be realized by a negative value of $f_\NL$~\cite{Byrnes:2012yx,Young:2013oia}, which requires the curvaton domination phase before its decay ($r>1.85$).\footnote{Recently, it has been shown that a negative $f_\NL$ cannot be realized in non-attractor single-field models of inflation without any spectator fields~\cite{Firouzjahi:2023xke}.}
This case has never been focused on in the previous works on the axion curvaton model.
In this work, we take the type II axion curvaton model as a fiducial example and show that it can really produce the SIGWs for the PTA signals while preventing PBH overproduction.

This paper is organized as follows. 
In Sec.~\ref{sec:setup}, we explain the type II axion curvaton model.
Then, in Sec.~\ref{sec:scales}, we summarize the scales of the perturbations, the inflation, and the reheating. 
In Sec.~\ref{sec:results}, we show the numerical results of the power spectrum of the curvature perturbations and the SIGWs with fiducial parameter sets.
In Sec.~\ref{sec:pbh}, we calculate the PBH mass spectrum and see that the non-Gaussianity in the axion curvaton model successfully prevents the PBH overproduction.
We conclude this paper in Sec.~\ref{sec:conclusion}.

\section{Type II axion curvaton model}
\label{sec:setup}

In this section, we briefly introduce our concrete setup, the type II axion curvaton model. 
We skip the derivation of some equations and refer readers to Refs.~\cite{Ando:2018nge,Kawasaki:2021ycf} for the details.\footnote{The difference between our setup and those in Refs.~\cite{Ando:2018nge,Kawasaki:2021ycf} is only in the form of the inflaton potential.
Unlike in this work, the quadratic potential $V_I = m^2_I I^2/2$ is used in Refs.~\cite{Ando:2018nge,Kawasaki:2021ycf}.}

The potential of the type II axion curvaton model is given by
\begin{align}
  V = \frac{\lambda}{4} \left( |\Phi|^2 - \frac{v^2}{2} \right)^2 + g I^2 |\Phi|^2 - v^3 \epsilon(\Phi + \Phi^*) + V_I(I),
  \label{eq:pot}
\end{align}
where $I$ is the inflaton, $\Phi$ is a complex scalar field, $\lambda$ and $g$ are coupling constants, $v$ is the parameter that determines the vacuum expectation value of $|\Phi|$ after the inflation, and $V_I$ is the inflaton potential. 
We take a sufficiently small coupling $g$ so that the quantum correction to the inflaton potential $\sim g^2 I^4$ is negligible.
The bias term, $\epsilon v^3 (\Phi + \Phi^*)$, avoids the cosmic string problem and the backreaction problem from the stochastic effects of $\Phi$.
In this work, we take the Starobinsky-type potential for $V_I$~\cite{Starobinsky:1980te} as a representative of concave potential consistent with the CMB observations~\cite{Planck:2018jri}:
\begin{align}
   V_I(I) = \frac{3}{4} M_\Pl^2 M^2 \left( 1 - \ee^{-\sqrt{2/3}I/M_\Pl}\right)^2.
\end{align}
Note that the specific form of the inflaton potential is not so important in the type-II axion curvaton model because the peak shape of the curvature power spectrum can be realized only if the inflaton evolves from a large value to a small value.
To be consistent with $\mathcal P_\zeta$ on the Planck results on CMB scales~\cite{Aghanim:2018eyx}, we take $M = 1.3 \times 10^{-5}M_\Pl$.

For convenience, we here decompose $\Phi$ into the radial ($\varphi$), called saxion, and the phase direction ($\theta$) as $\Phi = \varphi\, \ee^{i\theta}/\sqrt{2}$.
The phase direction works as the axion curvaton and we introduce the canonical field for the axion as $\tilde \sigma \equiv \theta \varphi$.
In the potential Eq.~(\ref{eq:pot}), the saxion $\varphi$ is fixed around the origin at first because of the large effective mass of $\varphi$ from the coupling term $g I^2 |\Phi|^2$.
Specifically, the effective mass of $\varphi$ can be expressed as 
\begin{align}
  m_\varphi^2 \equiv \frac{\partial^2 V}{\partial \varphi^2} = g I^2 - \frac{\lambda}{4} \left( v^2 - 3 \varphi^2 \right).
  \label{eq:m_varphi}
\end{align}
As the field value of the inflaton decreases, the effective mass of $\varphi$ also decreases and finally becomes tachyonic, which allows $\varphi$ to roll down toward the potential minimum, $\sim v/\sqrt{2}$. 
From Eq.~(\ref{eq:m_varphi}), we can see that $\varphi$ goes into the tachyonic region when the inflaton value becomes 
\begin{align}
  I \sim \frac{v}{2} \sqrt{\frac{\lambda}{g}},
  \label{eq:i_roll_down}
\end{align}
where we have used the fact that $\varphi \ll v$ before the roll down. 

During the inflation, the mass of the axion is raised only by the (U(1) breaking) bias term:
\begin{align}
  &-v^3 \epsilon(\Phi + \Phi^*) = -2 v^3 \epsilon \frac{\varphi}{\sqrt{2}} \cos\left(\frac{\tils}{\varphi} \right) \simeq -\sqrt{2}v^3 \epsilon \varphi  + \frac{1}{2} m_\tils \tils^2, \\
  &m_{\tilde \sigma}^2 = \sqrt{2}\epsilon \frac{v^3}{\varphi}.
\end{align}
We consider the situation where this mass is larger than the Hubble parameter ($m_{\tils} > H$) when $\varphi$ is fixed around the origin and it finally becomes $m_\tils < H$ after $\varphi$ rolls down the potential and has the large value.
Note that the background value of $\tilde \sigma$ (and $\theta$) is negligibly small because of $m_{\tilde \sigma} > H$ at the initial stage of the inflation.

We now discuss the perturbations of the phase direction. 
Due to the quantum fluctuations, the axion $\tils$ fluctuates by $H/(2\pi)$ at the horizon exit during the inflation. 
Using this fact, we can express the power spectrum of the phase direction $\theta$ at the horizon exit of each mode as 
\begin{align}
  \mathcal P_\theta(k,\eta_k) = \left(\frac{H(\eta_k)}{2\pi \varphi(\eta_k)} \right)^2,
\end{align}
where $\eta_k$ is the conformal time when the mode with the wavenumber $k$ exits the horizon.
This power spectrum evolves due to the mass of the phase direction even after the horizon exit of each mode. 
In particular, before the roll down of $\varphi$, the mass of the axion $(m_\tils > H)$ suppresses the fluctuations of $\tilde \sigma$.
To take into account this, we introduce the following factor:~\cite{Ando:2018nge}
\begin{align}
  \ln R_k \equiv \ln \left( \frac{\tilde \sigma_k(\eta_\en)}{\tilde \sigma_k(\eta_k)} \right) = - \frac{3}{2} \int^{\eta_\en}_{\eta_k} \dd \eta \, \mathcal H\, \text{Re} \left[ 1- \sqrt{1 - \left( \frac{2m_{\tilde \sigma}}{3H}\right)^2} \right],
    \label{eq:rk}
\end{align}
where $\mathcal H \equiv a H$ is the conformal Hubble parameter and $\eta_\en$ is the end of the inflation.
Specifically, we regard $\eta_\en$ as the time when $-\dot H/H^2$ (with the dot the time derivative) becomes unity.
Since $m_{\tils}$ becomes smaller than $H$ after $\varphi$ rolls down, the modes that enter the horizon after the roll down of $\varphi$ are not suppressed and $R_k \simeq  1$ for such modes. 
Then, we can approximate $R_k$ as
\begin{align}
  R_k \simeq \begin{cases}
  1\ \ &(\text{for}\  k \gtrsim k_\rd), \vspace{2pt}\\  
  \left(\cfrac{k}{k_\rd} \right)^{3/2} \ \ &(\text{for}\  k \lesssim k_\rd), \\
  \end{cases}
  \label{eq:R_k}
\end{align}
where $k_\rd$ corresponds to the inverse of the horizon scale at the time when $\varphi$ starts to roll down the potential.
Taking into account this suppression factor, we can express the power spectrum of the phase direction at the end of the inflation as
\begin{align}
    \mathcal P_\theta(k,\eta_\en) = R_k^2 \mathcal P_\theta(k,t_k) =  R_k^2 \left(\frac{H(\eta_k)}{2\pi \varphi(\eta_k)} \right)^2.
    \label{eq:p_theta_t_en}
\end{align}

We assume that some nonperturbative effect induces the mass of the axion curvaton after the inflation through the following potential:
\begin{align}
   V_\sigma = \Lambda^4 \left( 1 - \cos\Theta \right) \simeq \frac{1}{2}m_\sigma^2 \sigma^2,
\end{align}
where $\Theta = \theta - \theta_i$ with $\theta_i$ the value of $\theta$ at the minimum of $V_\sigma$, $\sigma$ is the axion after the inflation, defined by 
$\sigma \equiv v \Theta$, and $m_\sigma = \Lambda^2/v$.
Once $m_\sigma$ becomes larger than $H$, the curvaton starts to oscillate around the minimum and behaves as a matter (dust-like) fluid. 
The energy density of the oscillating curvaton decreases as $\propto a^{-3}$ while the radiation energy density decreases as $\propto a^{-4}$.
Because of this, if the axion oscillation lasts long enough before its decay to radiation, the axion may dominate the universe.

The final curvature power spectrum after the decay of the curvaton is given by~\cite{Ando:2018nge}
\begin{align}
  \mathcal P_\zeta(k) = \left( \frac{r}{4+3r}\right)^2 \left( \frac{2}{\theta_i}\right)^2 \mathcal P_\theta(k,\eta_k) R_k^2,
  \label{eq:cal_p_zeta}
\end{align}
where we have neglected the contributions from the inflaton fluctuations by assuming that the curvaton fluctuations dominate the curvature perturbations on small scales at the curvaton decay.
We use this expression of the power spectrum for the calculation of the SIGWs.
We use Eqs.~(\ref{eq:rk}) and (\ref{eq:p_theta_t_en}) to obtain $\mathcal P_\theta(k,\eta_\en)$ and $R_k$ in this equation.
This power spectrum has a peak around $k_\rd$.
The suppression of the large-scale side of the peak is due to $R_k$, which originates from the mass of the axion curvaton during the inflation. 
Meanwhile, the suppression on the small-scale side of the peak is due to $\mathcal P_\theta(k,\eta_\en)$, which originates from the roll down of $\varphi$ during the inflation.

Note that Eq.~(\ref{eq:cal_p_zeta}) is for the perturbations on the scales that enter the horizon after the curvaton decay. 
For the perturbations that enter the horizon \emph{before} the curvaton domination and its decay, we need to be careful about the evolution of the curvature perturbations even on superhorizon scales. 
On such scales, the perturbations of the curvaton during that time can be regarded as the isocurvature perturbations, and the formulas for the SIGWs, which we will introduce in Sec.~\ref{sec:results}, should be modified~\cite{Domenech:2021and}.
On the other hand, for the perturbations that enter the horizon \emph{after} the curvaton domination but \emph{before} its decay, we need to be careful about the suppression of the induced tensor during the decay of the curvaton~\cite{Inomata:2019zqy}. 
In this work, we focus on the scales that enter the horizon after the curvaton decay for simplicity and leave the analysis on the other scales for future work.

Before closing this section, we note that, although the curvature power spectrum is sharply suppressed on the small-scale side of the peak scale ($k > k_\rd$), the power spectrum becomes almost scale-invariant on the scales much smaller than the peak scale (but outside the horizon at the curvaton decay) because the significant change of $\varphi$ occurs only in the beginning of the roll down and $\varphi$ shortly becomes of $\mathcal O(v)$.
Combining Eqs.~(\ref{eq:R_k}), (\ref{eq:p_theta_t_en}), and (\ref{eq:cal_p_zeta}), we can roughly estimate the order of the power spectrum on such scales as 
\begin{align}
  \mathcal P_\zeta(k) \sim \mathcal O\left(\left( \frac{r}{4+3r}\right)^2 \left( \frac{2}{\theta_i}\right)^2 \left( \frac{H}{2\pi v} \right)^2\right) \  \ (\text{for} \ k \gg k_\rd).
  \label{eq:small_scale_pzeta}
  \end{align}
As we will see below, the perturbations on these scales can induce GWs that could be measured by future GW projects.

\section{Perturbation scale, inflation scale, and temperatures}
\label{sec:scales}

We here summarize the relation between the perturbation scale, the inflation scale, and the radiation temperatures. 
Throughout this work, we consider the following scenario.
First, the inflation occurs and ends at $\eta_\en$. Then, the universe behaves as a MD era (the first MD era) due to the inflaton oscillation until the decay of the inflaton.
Once the inflaton decays to radiation at $\eta_\Rf$ (the first reheating), the universe behaves as a RD era (the first RD era).
As mentioned in the introduction, we consider $r>1$ to avoid the PBH overproduction with a negative $f_\NL$, which leads to the domination of the curvaton in the energy density of the universe before its decay to radiation. 
The universe dominated by the curvaton behaves as a MD era (the second MD era).
After the curvaton decays to radiation (the second reheating) at $\eta_\Rs$, the universe behaves as a RD era (the second RD era) until the dark-matter domination at the equality time $\eta_\eq$ ($z_\eq \simeq 3.4\times10^3$ with $z_\eq$ the redshift at $\eta_\eq$). 
In the following, we assume that the first and the second reheating at $\eta_\Rf$ and $\eta_\dec$ occur instantaneously for simplicity.\footnote{
The effects of the non-instantaneous transition on the axion curvaton model are discussed in Ref.~\cite{Ferrante:2023bgz}.
Also, strictly speaking, an instantaneous transition from a MD era to a RD era at the perturbation level may enhance the SIGWs, through the Poltergeist mechanism~\cite{Inomata:2019ivs,Inomata:2020lmk}. In this work, we assume the instantaneous transition only at the background level just for simplicity and do not discuss the enhancement by the Poltergeist mechanism.
}

After some calculation, we obtain the following relation between the perturbation scale, the inflation energy scale, and the reheating temperature: 
(see Appendix~\ref{app:scales} for the derivation):
\begin{align}
  \frac{k}{k_\eq} = \, &1.3\times 10^{24}(1+r)^{-1/4} \left( \frac{\rho^\tot_\en}{3 \times 10^{-11}\, M_\Pl^4}\right)^{1/6} \left(\frac{T_\Rf}{10^{14}\,\GeV} \right)^{1/3}  \left(\frac{g^\rho_{\Rf}}{g^s_{\Rf}} \right)^{1/3} \left( \frac{g^\rho_{\decp}}{g^\rho_{\decm}} \right)^{1/4} \left( \frac{g^s_{\decp}}{g^s_{\decm}} \right)^{-1/3} \frac{k}{\mathcal H_\en},
  \label{eq:k_o_keq}
\end{align}
where $k_\eq (= 0.0103\,\Mpc^{-1}$~\cite{Aghanim:2018eyx}) is the inverse of the horizon scale at $\eta_\eq$, $\rho^\tot$ is the total energy density, $T_\Rf$ is the reheating temperature due to the inflaton decay, and $g^\rho$ and $g^s$ are the effective degrees of freedom for the energy and entropy density, respectively. 
We use the results in Ref.~\cite{Saikawa:2018rcs} for the temperature dependence of $g^\rho(T)$ and $g^s(T)$.
The subscript `$\bullet$' denotes the value at $\eta_\bullet$ with $\bullet$ being arbitrary characters (\textit{e.g.}, $\rho^\tot_\en = \rho^\tot(\eta_\en)$).
In particular, the subscripts '${\decm}$' and `${\decp}$' denote the value just before and after the curvaton decay, respectively.
Note that we cannot derive the concrete value of $\mathcal H_\en$ without fixing the thermal history after the inflation.
This is why the relation between $k_\eq$ and $\mathcal H_\en$ in Eq.~(\ref{eq:k_o_keq}) depends on the model parameters, which determine the thermal history.
In our concrete analysis of the axion curvaton model, we first numerically calculate the evolution of the perturbations with a fixed $k/\mathcal H_\en$ until the end of the inflation. Then, we relate the $k/\mathcal H_\en$ to the perturbation scale at the present time using Eq.~(\ref{eq:k_o_keq}).

We here obtain the expression of the temperature right after the curvaton decay. 
For simplicity, we assume that the curvaton starts to oscillate during the first MD (the inflaton oscillation) era. 
In this case, the ratio between the curvaton and the radiation at $\eta_\Rf$ can be expressed as\footnote{If the curvaton starts to oscillate after the first reheating, the quantities at $\eta_\Rf$ in Eqs.~(\ref{eq:ini_ratio})-(\ref{eq:horizon_tdec}) should be replaced by those at the time when the curvaton starts to oscillate.
In that case, the horizon scale at $\eta_\dec$ becomes larger.}
\begin{align}
  \frac{\rho^\sigma_\Rf}{\rho^\rr_\Rf} = \frac{v^2\theta_i^2}{6 M_\Pl^2}.
  \label{eq:ini_ratio}
\end{align}
Using this, we can express $r$, defined in Eq.~(\ref{eq:r}), as 
\begin{align}
  r &\equiv \frac{\rho^\sigma_\decm}{\rho^\rr_\decm} = \left( \frac{g^\rho_\Rf}{g^\rho_{\decm}} \right)\left( \frac{g^s_\Rf}{g^s_{\decm}} \right)^{-1} \frac{T_\Rf}{T_{\decm}} \frac{v^2\theta_i^2}{6 M_\Pl^2},
  \label{eq:r_exp}
\end{align}
where the subscript `$\decm$' denotes the value just before the curvaton decay. 
The temperature right after the curvaton decay can be expressed as 
\begin{align}
     T_\decp &= \left( \frac{a_\Rf}{a_\dec} \right)^{3/4} \left(\frac{g^\rho_\Rf}{g^\rho_\decp}\right)^{1/4} T_\Rf  \left(\frac{\rho^\sigma_\Rf}{\rho^\rr_\Rf}\right)^{1/4} \left(1+ r^{-1}\right)^{1/4}\nonumber \\
    &= \left( \frac{T_\Rf}{T_\decm} \right)^{-3/4} \left(\frac{g^s_{\Rf}}{g^s_{\decm}}\right)^{-1/4} \left(\frac{g^\rho_\Rf}{g^\rho_\decp}\right)^{1/4} T_\Rf \left(\frac{v^2\theta_i^2}{6 M_\Pl^2}\right)^{1/4} \left(1+ r^{-1}\right)^{1/4}\nonumber \\
    &= \left( \frac{g^\rho_\Rf}{g^\rho_{\decm}} \right)^{3/4} \left(\frac{g^s_{\Rf}}{g^s_{\decm}}\right)^{-1} \left(\frac{g^\rho_\Rf}{g^\rho_\decp}\right)^{1/4} T_\Rf \frac{v^2\theta_i^2}{6 M_\Pl^2}\frac{\left(1+ r\right)^{1/4}}{r},
\end{align}
where we have used Eq.~(\ref{eq:r_exp}) in the final equality.
The horizon scale at $T_\dec$ can be expressed as 
\begin{align}
  aH|_\dec &=  \left( \frac{g^s_\decp}{g^s_0} \right)^{-1/3} \left(\frac{T_\decp}{T_0}\right)^{-1} \left( \frac{\pi^2 g^\rho_\decp T_\decp^4}{90 M_\Pl^2} \right)^{1/2} \nonumber \\
  &= 1.71\times 10^{13}\, \Mpc^{-1} \, \left( \frac{g^s_\decp}{106.75} \right)^{-1/3} \left( \frac{g^\rho_\decp}{106.75} \right)^{1/2} \left(\frac{T_\decp}{10^6\, \GeV}\right),
  \label{eq:horizon_tdec}
\end{align}
where the subscript `$0$' means the value at present and we have substituted $g^s_0 = 3.91$~\cite{Kolb:206230} and $T_0 = 2.72\,\text{K} = 2.35\times 10^{-4}\,\text{eV}$~\cite{Aghanim:2018eyx}.

\section{Spectrum of curvature perturbations and Scalar-Induced GWs}
\label{sec:results}

\begin{figure}
        \centering \includegraphics[width=0.7\columnwidth]{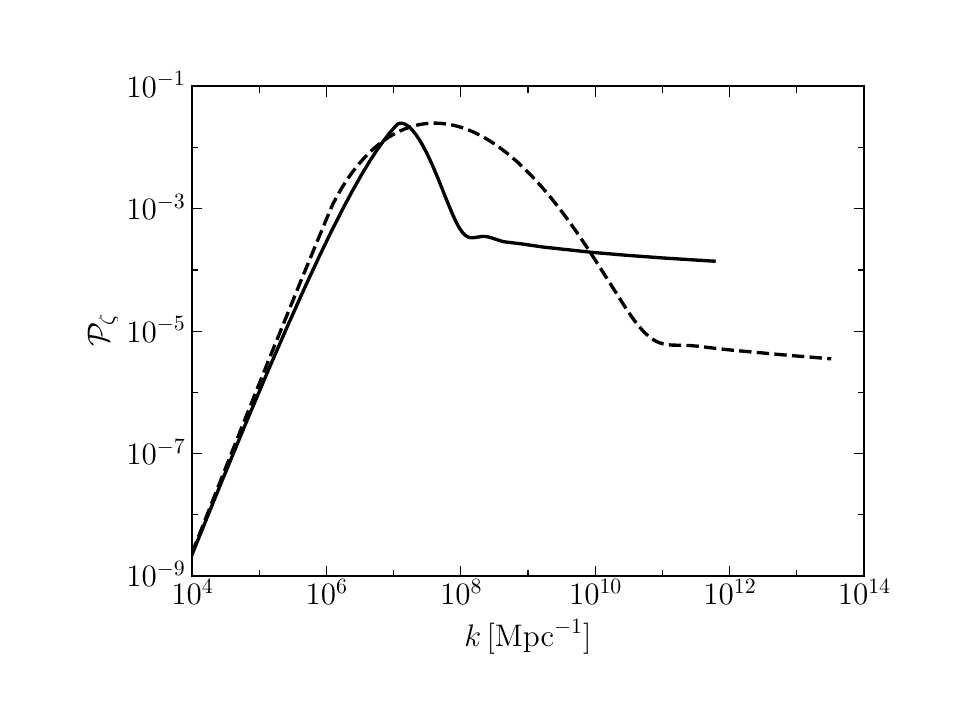}
        \caption{ The power spectrum of curvature perturbations in the parameter sets given by Eqs.~(\ref{eq:sh_para}) (solid) and (\ref{eq:br_para}) (dashed).
        We only show the power spectrum on the scales that enter the horizon after the curvaton decay. See the text between Eqs.~(\ref{eq:cal_p_zeta}) and (\ref{eq:small_scale_pzeta}) for details.
                }
        \label{fig:pzeta}
\end{figure}

Let us see the power spectrum of the curvature perturbations and the SIGWs in the axion curvaton model.
To be concrete, we consider two fiducial parameter sets for a sharp and a broad peak:
\begin{align}
  &v/M_\Pl = 2.501\times 10^{-2},\  \epsilon= 1.2\times 10^{-9}, \ g = 5.61\times 10^{-10}, \ \theta_i = 0.00265 \ \ (\text{for sharp peak}),
  \label{eq:sh_para}\\
  &v/M_\Pl = 8.733 \times 10^{-3},\ \epsilon= 1\times 10^{-9}, \ g = 7.03\times 10^{-11}, \ \theta_i = 0.0565 \ \ (\text{for broad peak}).
  \label{eq:br_para}  
\end{align}
We take $\lambda = 8.4 \times 10^{-5}$, $r = 3$, and $T_\Rf = 10^{14}\,\GeV$ for both the cases, which lead to $f_\NL = -0.438$.
Figure~\ref{fig:pzeta} shows the power spectrum of curvature perturbations with these fiducial parameter sets.
We note that $r$ is $\mathcal O(1)$ in our fiducial setup and therefore the decay of curvaton does not significantly dilute baryon number and dark matter even if they are produced before then.
However, in that case, baryon and/or CDM isocurvature perturbations are induced~\cite{Lyth:2002my}.
Their power spectrum is given by $\mathcal{P}_{S_{b}}$ or $\mathcal{P}_{S_\mathrm{CDM}} \simeq  9\mathcal{P}_\zeta$.
The induced isocurvature perturbations are only significant on the small scales ($k \gtrsim 10^{4}$~Mpc$^{-1}$) and free from the CMB constraint.
However, baryon isocurvature perturbations with $k \lesssim 10^8~\mathrm{Mpc}^{-1}$ affect the big bang nucleosynthesis and the power spectrum should satisfy  $\mathcal{P}_{S_b} (\simeq 9 \mathcal P_\zeta) \lesssim 0.016$~\cite{Inomata:2018htm}, which is violated in our fiducial cases (see Fig.~\ref{fig:pzeta}). 
This means that the baryon number must be produced after the curvaton decay in our fiducial cases.

We here briefly summarize the parameter dependence of the power spectrum.
A change of $\theta_i$ or $T_\Rf$ only affects the overall amplitude or the overall scale normalization, respectively (see Eqs.~(\ref{eq:cal_p_zeta}) and (\ref{eq:k_o_keq})).
The peak height is sensitive to $\epsilon$, where a larger $\epsilon$ decreases the peak height because a larger bias term leads to a larger value of $\varphi$ even before the roll down of $\varphi$.
The peak scale depends on the combination of $v^2\lambda/g$ (see Eq.~(\ref{eq:i_roll_down})).
The decrease $v^2 \lambda$ and $g$ with $v^2\lambda/g$ fixed leads to the broader power spectrum with the same peak scale.
This is because that change makes the potential flatter, in which $\varphi$ rolls down more slowly~\cite{Kawasaki:2021ycf}.
Also, as we can see in Eq.~(\ref{eq:small_scale_pzeta}), the amplitude of the almost scale-invariant power spectrum on the very small scales depends on the parameter combination of $v\theta_i$.
We note that the amplitudes on the peak scale and the very small scales can be determined independently since the change of $\epsilon$ only affects the amplitude on the peak scale.
We take the fiducial parameters giving different amplitudes on the very small scales in order to show that the induced GWs on the very small scales could be observed by different GW projects, as we will see below.

With the power spectrum in Fig.~\ref{fig:pzeta}, we calculate the SIGWs. 
We here briefly summarize the formulas for the SIGWs.
In the subhorizon limit during a RD era, the energy density parameter of the SIGWs per log bin ($\dd \log k$) is given by~\citep{Espinosa:2018eve,Kohri:2018awv}
\begin{align}
\label{eq:omega_gw}
\tilde \Omega_{\GW}\left(k\right) = \int_0^\infty \textrm{d}\bar v \int_{\left|1-\bar v\right|}^{1+\bar v} \textrm{d}\bar u\: \mathcal{K}\left(\bar u,\bar v\right) \mathcal{P}_\mathcal{\zeta}\left(\bar u k\right) \mathcal{P}_\mathcal{\zeta}\left(\bar v k\right),
\end{align} 
where the kernel $\mathcal{K}$ is 
\begin{align}
\mathcal{K}\left(\bar u,\bar v\right) = \frac{3\left(4\bar v^2-(1+\bar v^2-\bar u^2)^2\right)^2\left(\bar u^2+\bar v^2-3\right)^4}{1024\,\bar u^8 \bar v^8}\left[\left(\ln\left|\frac{3-(\bar u+\bar v)^2}{3-(\bar u-\bar v)^2}\right| - \frac{4\bar u\bar v}{\bar u^2+\bar v^2-3}\right)^2 + \pi^2 \Theta(\bar u+\bar v-\sqrt{3}) \right].
\end{align}
During a RD era, the energy density parameter of the SIGWs approaches a constant value, $\tilde \Omega_\GW$.
We relate the scale and the frequency of the GWs through $f = k/(2\pi)$.
Taking into account the change of the degrees of freedom and the following MD and dark-energy-dominated eras, the current energy density of the SIGWs can be expressed as
\begin{align}
    \Omega_\GW h^2 = 
    0.39 \left( \frac{g^\rho_\hh}{106.75}\right) \left( \frac{g^s_\hh}{106.75}\right)^{-4/3} \Omega_{r,0} h^2 \tilde \Omega_\GW,
    \label{eq:current_gw}
\end{align}
where the subscript `h' denotes the value at the horizon entry of the induced GW.
$\Omega_{r,0}$ is the current radiation density parameter and $h (\equiv H_0/(100 \,\text{km/s/Mpc}))$ is the normalized Hubble parameter  ($\Omega_{r,0} h^2 \simeq 4.2 \times 10^{-5}$).
Note that, in our analysis, we do not take into account the modification of the GW spectrum due to the non-Gaussianity of the curvature perturbations because the modification is small in $|f_\NL| < 1$.
See Refs.~\cite{Nakama:2016gzw, Garcia-Bellido:2017aan,  Cai:2018dig, Unal:2018yaa, Yuan:2020iwf, Atal:2021jyo, Adshead:2021hnm,Garcia-Saenz:2022tzu, Li:2023qua, Yuan:2023ofl,Li:2023xtl} for details on the modification due to the primordial non-Gaussianity. 
Furthermore, the deviation from a pure RD era during the QCD phase transition also affects the GW spectrum~\cite{Abe:2020sqb,Abe:2023yrw}, which is neglected in our analysis. 
If we take into account that effect, $\Omega_\GW$ could change by a factor of $\mathcal O(1)$~\cite{Abe:2020sqb,Abe:2023yrw}.

\begin{figure}
  \begin{minipage}[b]{0.49\linewidth}
    \centering
    \includegraphics[keepaspectratio, scale=0.55]{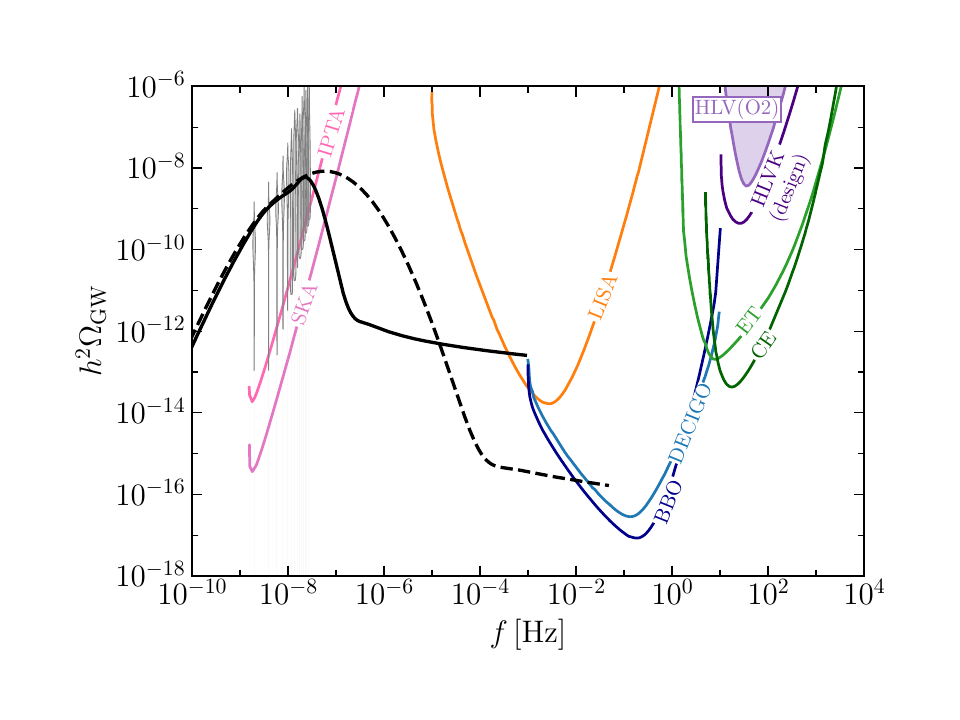}
  \end{minipage}
  \begin{minipage}[b]{0.49\linewidth}
    \centering
    \includegraphics[keepaspectratio, scale=0.55]{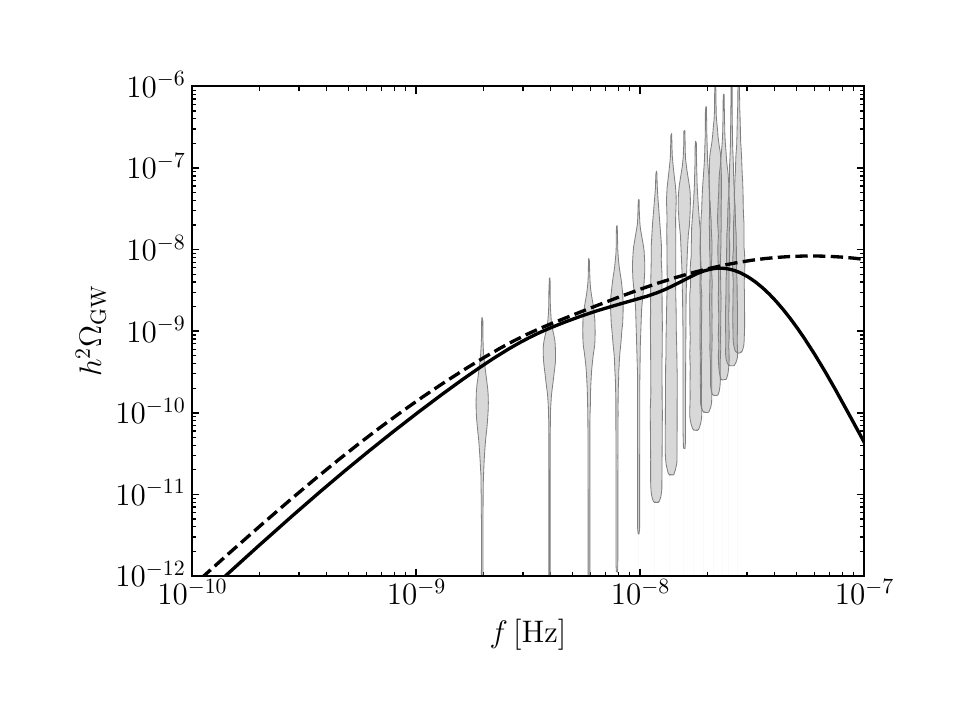}
  \end{minipage}
        \caption{ 
        [Left]: The power spectrum of the SIGWs in the cases of the sharp peak (black solid) and the broad peak (black dashed). 
        The parameters are the same as in Fig.~\ref{fig:pzeta}.
        The sensitivity curves are from Ref.~\cite{Schmitz:2020syl} and the shaded region with HLV(O2) is already excluded.
        The gray violins are from the NANOGrav result~\cite{NANOGrav:2023hvm}.
        We only show the GW spectrum on the scales that enter the horizon after the curvaton decay. See the text below Eq.~(\ref{eq:current_gw}) for details.
        [Right]: The zoom-in picture of the left panel only with the GW spectrum and the NANOGrav result~\cite{NANOGrav:2023hvm}.
        }  
        \label{fig:gw}  
\end{figure}

Figure~\ref{fig:gw} shows the spectrum of the SIGWs with the fiducial power spectra shown in Fig.~\ref{fig:pzeta}.
We can see that the SIGWs are consistent with the NANOGrav results and those on the very small scales are above the future sensitivities of LISA~\cite{LISACosmologyWorkingGroup:2022jok}, DECIGO~\cite{Kawamura:2020pcg}, and BBO~\cite{Crowder:2005nr,Corbin:2005ny}.
The integrated energy density of the SIGWs is $\int \dd \log f\, \Omega_\GW(f) h^2 = 8.9\times 10^{-9}$ for the sharp peak case and $= 2.7\times 10^{-8}$ for the broad peak case, which are consistent with the dark radiation constraints $ < 1.8 \times 10^{-6}$~\cite{Kohri:2018awv, Clarke:2020bil}.
The small-scale (high-frequency) cutoff of the GW spectrum corresponds to the horizon scale at $\eta_\Rs$ (see Eq.~(\ref{eq:horizon_tdec}) for the relation between the horizon scale and $T_\dec$). 
On the scales smaller than that, the calculation of the SIGWs should be modified because the perturbations on those scales enter the horizon during a MD era. 
If the curvaton decays to radiation perturbatively, the SIGWs on those scales ($f \gg 1/\eta_\Rs$) are expected to be smaller than those on $f < 1/\eta_\Rs$~\cite{Inomata:2019zqy}.
On the other hand, if the curvaton decay occurs much more suddenly than the perturbative decay does, the SIGWs on those scales can be much larger than those on $f < 1/\eta_\Rs$, which is known as the Poltergeist mechanism~\cite{Inomata:2019ivs,Inomata:2020lmk}. 
We leave the precise analysis of the SIGWs on those scales for future work.
From Eqs.~(\ref{eq:small_scale_pzeta}), (\ref{eq:r_exp}), and (\ref{eq:horizon_tdec}), it is evident that the cutoff scale is $k_\text{cut} \propto v^2 \theta_i^2$ and the curvature power spectrum and the GW spectrum at the cutoff scale are $\mathcal P_\zeta \propto 1/(v^2 \theta_i^2)$ and $\Omega_\GW \propto 1/(v^4 \theta_i^4)$. 

\begin{figure}
        \centering \includegraphics[width=0.7\columnwidth]{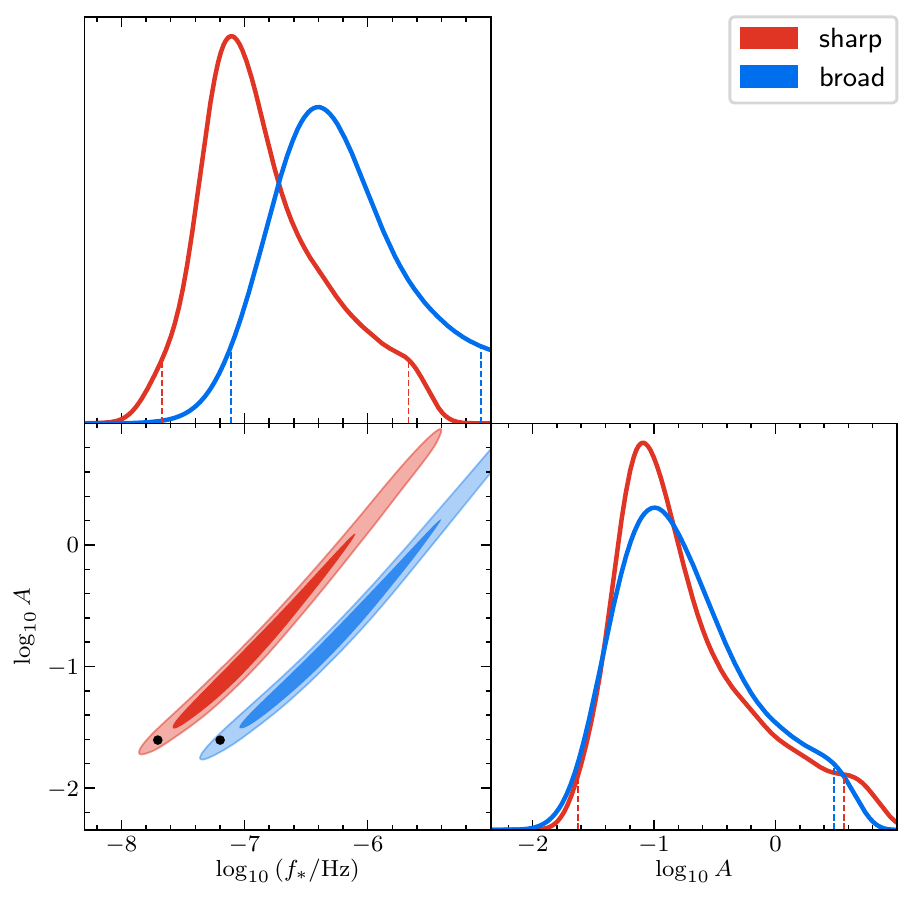}
        \caption{ The posteriors for the sharp (red) and the broad profiles (blue) based on the parameter sets Eqs.~(\ref{eq:sh_para}) and (\ref{eq:br_para}), respectively. See the text for the details on our parameterization.
        We take log-uniform distribution $[-11,-5]$ as a prior for $\log_{10}(f_*/\text{Hz})$ and the log-uniform distribution $[-3,1]$ for $\log_{10} A$.
         We use the first 14 frequency components ($f = i/T_\text{obs}, i \in \mathbb{Z}, 1 \leq i \leq 14$, $T_\text{obs} = 16.03\, \text{yr}$) to obtain the posteriors in the same way as in the NANOGrav collaborations~\cite{NANOGrav:2023gor}.
         The dark and light color regions in the 2D posteriors denote the $68\%$ and the $95\%$ credible regions.
         The black dots in the 2D posteriors correspond to the power spectra in Fig.~\ref{fig:pzeta}.
         The regions between the vertical lines in the 1D posteriors correspond to the $95\%$ credible regions.
        }
        \label{fig:posteri}
\end{figure}

To see how the changes of the peak amplitude and the peak scale affect the quality of the fitting, we obtain the posteriors on them by using $\mathtt{PTArcade}$~\cite{andrea_mitridate_2023,Mitridate:2023oar} with $\mathtt{Ceffyl}$~\cite{lamb2023need}.
To this end, we make the normalization of the amplitude and the scale of $\mathcal P_\zeta$ the free parameters with its profile (or the frequency dependence) fixed.
Specifically, we introduce the two parameters $A$ and $k_* (=2\pi f_*)$ as the amplitude and the scale (or the frequency) at the peak, which leads to $\mathcal P_\zeta(k_*) = A$ and $\mathcal P_\zeta(k) < A$ in $k \neq k_*$.
As fiducial profiles, we use the ones in Fig.~\ref{fig:pzeta}, which are obtained with the parameter sets, Eqs.~(\ref{eq:sh_para}) and (\ref{eq:br_para}).
Figure~\ref{fig:posteri} shows the posteriors on $f_*$ and $A$ for the NANOGrav 15 year results~\cite{NANOGrav:2023gor,NANOGrav:2023hvm}.
Note that the peak amplitude and scale for the power spectra in Fig.~\ref{fig:pzeta} correspond to the points around the lower-left corner in the $95\%$ credible regions of the 2D posteriors (black dots in the figure). 
Specifically, we find $\log_{10}A=-1.60$ and $\log_{10}(f_*/\text{Hz}) = -7.70$ for the sharp peak with Eq.~(\ref{eq:sh_para}) and $\log_{10}A=-1.60$ and $\log_{10}(f_*/\text{Hz}) = -7.20$ for the broad peak with Eq.~(\ref{eq:br_para}).

\section{PBH mass function}
\label{sec:pbh}

We now discuss the PBH mass function in the axion curvaton model. 
In general, once PBHs are produced during a RD era, the energy fraction of PBHs in the total energy density grows proportionally to the scale factor until the equality time $\eta_\eq$ because of their redshift difference.
This imposes the upper bound of the PBH production rate as $\beta \lesssim \mathcal O(10^{-8})$ for the PBH mass with $M_\pbh \lesssim \mathcal O(M_\odot)$, which is related to the PTA results (see Eqs.~(\ref{eq:m_pbh}) and  (\ref{eq:f_pbh}) below).
If this upper bound is violated, the energy density of PBHs dominates over the current DM energy density.
This means that the PBH production must be rare events and the threshold of its production must be located around the tail of the probability distribution function of the density perturbations. 
This results in an exponential suppression of the production rate. 
This exponential suppression is sensitive to the exponent and therefore to the change of the tail due to the non-Gaussianity of the curvature perturbations~\cite{Byrnes:2012yx,Young:2013oia}.
In the following, we will concretely see that the small non-Gaussianity with $|f_\NL| < 1$ in our fiducial setups indeed causes a large decrease of the PBH abundance and prevents the PBH overproduction.
We basically adopt the same formulas as in Ref.~\cite{Kawasaki:2019mbl}, which we summarize below.

The mass of PBH is related to the scale of the perturbations that produce the PBH ($k_\pbh$)~\cite{Inomata:2020xad}\footnote{Strictly speaking, the mass of PBH depends on not only the scale of perturbations but also the amplitude of the perturbations, called the critical phenomena~\cite{Niemeyer:1997mt,Niemeyer:1998ac,Shibata:1999zs}.
However, this effect does not much change the relation between the PBH mass and the peak scale of the perturbations~\cite{Yokoyama:1998xd} and therefore we do not take it into account for simplicity in this work.}:
\begin{align}
  M_\pbh &\simeq M_\odot \left( \frac{\gamma}{0.2}\right) \left( \frac{g^\rho_\pbh}{10.75} \right)^{1/2} \left( \frac{g^s_\pbh}{10.75} \right)^{-2/3} \left( \frac{k_\pbh}{2.0 \times 10^6 \, \Mpc^{-1}}\right)^{-2} \nonumber \\
  &\simeq M_\odot \left( \frac{\gamma}{0.2}\right) \left( \frac{g^\rho_\pbh}{10.75} \right)^{1/2} \left( \frac{g^s_\pbh}{10.75} \right)^{-2/3} \left( \frac{f_\pbh}{3.1 \times 10^{-9} \, \text{Hz}}\right)^{-2},
  \label{eq:m_pbh}
\end{align}
where $f_\pbh = k_\pbh/(2\pi)$ and $g^{\rho/s}_\pbh$ is the effective degrees of freedom at the PBH formation.
$\gamma$ is the ratio between the PBH mass and the horizon mass at the PBH formation, which we take $\gamma = 0.2$~\cite{Carr:1975qj} in this work.
We can see that the PTA frequencies $\gtrsim \mathcal O(10^{-9})\, \text{Hz}$ correspond to $M_\pbh \lesssim \mathcal O(M_\odot)$.
The current fraction of PBHs in DM (per log bin, $\dd \log M$) is given by~\cite{Kawasaki:2021ycf}\footnote{For completeness, we have modified the expression in Ref.~\cite{Kawasaki:2021ycf} so that it can be applicable even when $g^s_\pbh \neq g^\rho_\pbh$.}
\begin{align}
  f_\pbh(M_\pbh) = \left(\frac{\beta(M_\pbh)}{1.7\times 10^{-8}}\right) \left(\frac{\gamma}{0.2} \right)^{3/2} \left(\frac{g^\rho_\pbh}{10.75} \right)^{3/4} \left(\frac{g^s_\pbh}{10.75} \right)^{-1} \left(\frac{\Omega_\dm h^2}{0.12} \right)^{-1} \left( \frac{M_\pbh}{M_\odot}\right)^{-1/2},
  \label{eq:f_pbh}
\end{align}
where $\Omega_\dm$ is the energy density parameter for DM and $\beta$ is the PBH production rate per log bin.
The density perturbations that have a local-type non-Gaussianity can be expressed with Gaussian density perturbation $\delta_g$ as 
\begin{align}
  \delta(r_m) = \delta_g(r_m) + \frac{\mu_3(r_m)}{6 \sigma(r_m)} (\delta_g^2(r_m) - \sigma^2(r_m)), 
  \label{eq:delta_rm}
\end{align}
where $r_m$ is the smoothing scale, $\sigma$ is the variance of density perturbations, and $\mu_3$ is their skewness.
Specifically, the variance with the smoothing scale $r_m$, $\sigma(r_m)$, is given by~\cite{Kawasaki:2019mbl}
\begin{align} \label{eq:sigma_rm}
  \sigma^2(r_m) = \left(\frac{4}{9} \right)^2 \int \dd(\ln p r_m) |\tilde W(p r_m)|^2 (p r_m)^4 |T(p r_m)|^2 \mathcal P_\zeta(p),
\end{align}
where $\tilde W$ is a window function and $T$ is the transfer function of density perturbations given by
\begin{align}
  T(p r_m) = 3 \frac{\sin(p r_m/\sqrt{3}) - (p r_m/\sqrt{3}) \cos(p r_m/\sqrt{3})}{(p r_m/\sqrt{3})^3}.
\end{align}
As a fiducial choice of the window function\footnote{
  The uncertainty in selecting the window function results in an associated uncertainty in the PBH mass spectrum~\cite{Ando:2018qdb,Young:2019osy}.
  For example, if we take the Gaussian window function ($\tilde W(p r_m) = \ee^{-(pr_m)^2/2}$) in our formalism instead, the PBH overproduction does not necessarily occur even for the Gaussian density perturbations~\cite{NANOGrav:2023hvm,Inomata:2023zup}. 
  The real-space top hat window function is used in Ref.~\cite{Franciolini:2023pbf}, which points out the PBH overproduction problem.
}, we take the real-space top hat window function, whose expression in Fourier space is given by
\begin{align}
  \tilde W(p r_m) = 3 \left( \frac{\sin p r_m - p r_m \cos p r_m}{(p r_m)^3} \right).
\end{align}
We can express the skewness, $\mu(r_m)$, as~\cite{Kawasaki:2019mbl}
\begin{align}
  \frac{\mu_3(r_m)}{\sigma(r_m)} = Q(r_m) f_\NL - \frac{9}{4},
  \label{eq:mu_3}
\end{align}
where $Q(r_m)$ is given by
\begin{align}
  Q(r_m) =&\, \sigma^{-4}(r_m) \frac{18}{5} \left( \frac{4}{9}\right)^3 \int \frac{\dd p}{p}\tilde W(p r_m) |T(p r_m)|^2 \mathcal P_\zeta(p)(p r_m)^2 \int \frac{\dd q}{q}\tilde W(q r_m)|T(q r_m)|^2 \mathcal P_\zeta(q)(q r_m)^2  \nonumber \\
  &\,\times \int^1_{-1} \frac{\dd \cos\theta}{2} \tilde W(|\bm p + \bm q| r_m) (|\bm p + \bm q| r_m)^2. \label{eq:p_rm}
\end{align}
Note that $-\frac{9}{4}$ in Eq.~(\ref{eq:mu_3}) originates from the nonlinear relation between density perturbations and curvature perturbations~\cite{Kawasaki:2019mbl} (see also Refs.~\cite{DeLuca:2019qsy,Young:2019yug} for the effects of this nonlinear relation).

Here we summarize semi-analytic formulas to compute $\sigma (r_m)$ and $Q(r_m)$.
In Eq.~\eqref{eq:p_rm}, we may perform the integration of the angle between $\bm{p}$ and $\bm{q}$ analytically as follows
\begin{align}
  \int \frac{\dd \cos \theta}{2} \tilde W(|\bm p + \bm q| r_m) (|\bm p + \bm q| r_m)^2
  = 3 \left[ 2 \frac{\sin (p r_m)}{p r_m} \frac{\sin (q r_m)}{q r_m} - \cos (p r_m) \frac{\sin (q r_m)}{q r_m} - \cos (q r_m) \frac{\sin (p r_m)}{p r_m}  \right].
\end{align}
As the obtained functions are factorized into the $p$ and $q$ dependent parts, we can perform the $p$ and $q$ integrations separately.
We can readily see that the fundamental building blocks for $Q (r_m)$ are merely the following two one-dimensional integrals
\begin{align}
  \mathcal{I}_s (r_m) \equiv \int \frac{\dd x}{x} \tilde W (x) \left| T(x) \right|^2 \mathcal{P}_\zeta (x r_m^{-1})\,  x \sin x ,  \qquad
  \mathcal{I}_c (r_m) \equiv \int \frac{\dd x}{x} \tilde W (x) \left| T(x) \right|^2 \mathcal{P}_\zeta (x r_m^{-1})\,  x^2 \cos x .
\end{align}
By using these two integrals, we find
\begin{align}
  Q(r_m) = \sigma^{-4} (r_m) \frac{18}{5} \left( \frac{4}{9} \right)^3 \, 6 \left[ \mathcal{I}_s^2 (r_m) - \mathcal{I}_s (r_m) \mathcal{I}_c (r_m) \right].
\end{align}
Furthermore, by noting that the Fourier transform of the real-space top hat window function is $\tilde{W}(x) = 3 ( \sin x - x \cos x )/x^3$, the variance $\sigma (r_m)$ given in Eq.~\eqref{eq:sigma_rm} can also be expressed as
\begin{align}
  \sigma^2 (r_m) = \left( \frac{4}{9} \right)^2 3 \left[ \mathcal{I}_s (r_m) - \mathcal{I}_c (r_m) \right].
\end{align}
Therefore, all we need to do is to perform two one-dimensional integrals of $\mathcal{I}_s (r_m)$ and $\mathcal{I}_c (r_m)$ numerically as a function of $r_m$.

We adopt the Press-Schechter formalism and calculate the PBH production rate with the local-type non-Gaussianity as 
\begin{align}
  \beta(r_m^{-1}) = \begin{cases}
  \displaystyle\int^\infty_{\delta_{g;+}(\delta_\thre)} P_g(\delta_g) \dd \delta_g + 
  \displaystyle\int^{\delta_{g;-}(\delta_\thre)}_{-\infty} P_g(\delta_g) \dd \delta_g \ \ &\text{for } \mu_3 > 0 \vspace{10pt} \\
  \displaystyle\int^{\delta_{g;-}(\delta_\thre)}_{\delta_{g;+}(\delta_\thre)} P_g(\delta_g) \dd \delta_g \ \ &\text{for } \mu_3 < 0
  \end{cases},
  \label{eq:beta_ng}
\end{align}
where the distribution function for the Gaussian density perturbation $\delta_g$ is given by 
\begin{align}
  P_g(\delta_g) = \frac{1}{\sqrt{2\pi} \sigma(r_m)} \exp\left[ - \frac{\delta_g^2}{2\sigma^2(r_m)}\right].
\end{align}
Since we focus on a negative $f_\NL$ in this work, we only use the expression for $\mu_3 < 0$ in Eq.~(\ref{eq:beta_ng}). 
Equating the left hand side in Eq.~(\ref{eq:delta_rm}) with the PBH formation threshold $\delta_\text{th}$, we obtain the threshold values in $\delta_g$, denoted by $\delta_{g;\pm}$:
\begin{align}
  \delta_{g;\pm}(\delta_\thre) = \frac{3\sigma}{\mu_3} \left( -1 \pm \sqrt{ 1 + \frac{2\mu_3}{3} \left( \frac{\mu_3}{6} + \frac{\delta_\thre}{\sigma} \right)} \right),
\end{align}
where we have omitted the argument $r_m$ for $\sigma$ and $\mu_3$.
We take $\delta_\thre = 0.53$ as a fiducial value for a pure RD era~\cite{Harada:2015yda,Kawasaki:2019mbl}.
On top of that, we take into account the decrease of the threshold during the QCD phase transition by using the results in Ref.~\cite{Byrnes:2018clq}.\footnote{Specifically, we use the result of "Time-averaged" in Fig.~2 of Ref.~\cite{Byrnes:2018clq} with the renormalization of the base line to $\delta_\text{th} = 0.53$ in a pure RD era~\cite{Harada:2015yda,Kawasaki:2019mbl}.
See also Ref.~\cite{Musco:2023dak} for the effects of the QCD phase transition with the critical phenomena.
}
Also, in this paper, we simply relate $r_m = 1/k_\pbh$. 

\begin{figure}
        \centering \includegraphics[width=0.7\columnwidth]{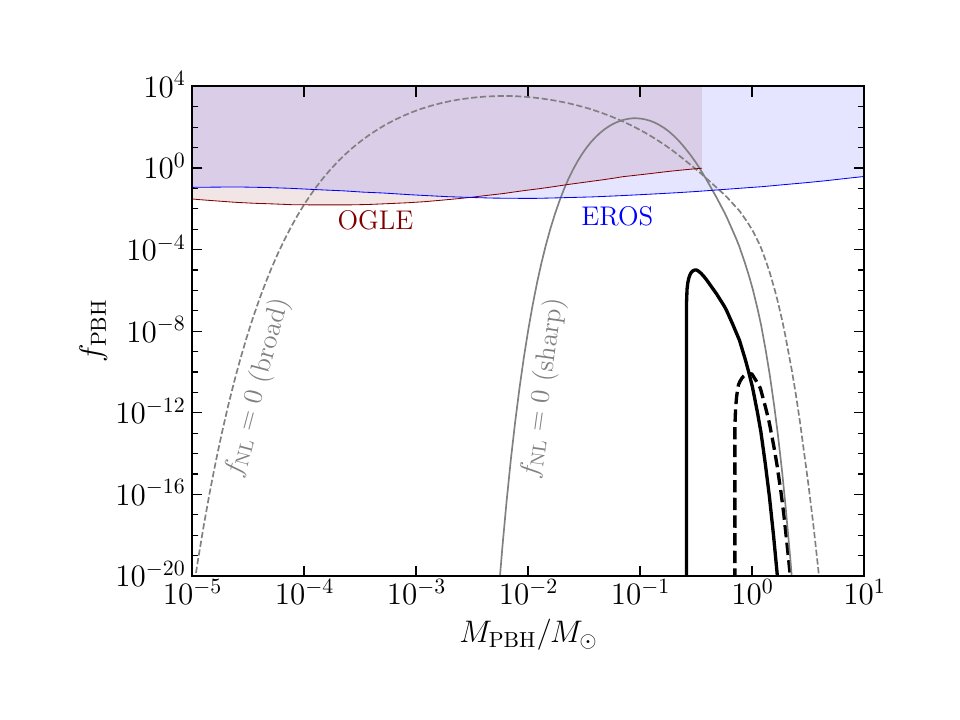}
        \caption{
        The PBH mass spectrum in the cases of the sharp peak (black solid) and the broad peak (black dashed). 
        The parameters are the same as in Fig.~\ref{fig:pzeta}.
        The shaded regions are excluded by OGLE~\cite{2017Natur.548..183M,Niikura:2019kqi} and EROS~\cite{Tisserand:2006zx}.
        The gray lines show the results in the cases without the effects of the primordial non-Gaussianity, in which we set $f_\NL = 0$ but use the same curvature power spectrum in Fig.~\ref{fig:pzeta}.
        }
        \label{fig:fpbh}
\end{figure}

Figure~\ref{fig:fpbh} shows the PBH mass spectrum for the fiducial cases, whose parameters are given in Eqs.~(\ref{eq:sh_para}) and (\ref{eq:br_para}).
From this figure, we can see that, if the effect of the primordial non-Gaussianity is neglected ($f_\NL = 0$) in our fiducial parameter sets, PBHs are overproduced.
This is qualitatively consistent with Refs.~\cite{Franciolini:2023pbf,Liu:2023ymk}.
Once we take into account the primordial non-Gaussianity ($f_\NL = -0.438$ with $r=3$), the PBH abundance is suppressed to be consistent with the current observations.
It is also worth noting that the peak of the PBH mass spectrum is around $M_\pbh \sim \mathcal O(0.1-1)M_\odot$, while the cases without the primordial non-Gaussianity ($f_\NL =0$) predict the peak at a smaller mass.
Besides, while we only discuss the fiducial two cases to show the key points for preventing the PBH overproduction in the axion curvaton model, the larger-amplitude power spectrum can, in principle, avoid the PBH overproduction by increasing $r$. Note that the largest negative value of $f_\NL$ is $f_\NL= -5/4$ in $r \gg 1$.

Finally, we mention the PBH production during the curvaton-dominated era, which can be regarded as a MD era. 
In our fiducial setups, the universe is dominated by the curvaton before its decay to radiation. 
During a MD era, PBHs can be produced even if the perturbations are smaller than the threshold value during the RD era because there is no pressure that prevents the growth of the density perturbations inside the horizon.
Then, unlike during a RD era, the PBH production rate during a MD era is governed by the deviation from spherical symmetry of the over-dense region~\cite{Khlopov:1980mg,Khlopov:1982sov}.
This results in the PBH production rate free from the exponential suppression during a MD era: $\beta \simeq 0.05556\, \sigma^5$ for $0.005 < \sigma < 0.2$~\cite{Harada:2016mhb,Harada:2017fjm}.
From this expression and that for $\sigma < 0.005$ in Ref.~\cite{Harada:2017fjm}, the power spectrum of curvature perturbations is constrained as $\mathcal P_\zeta \lesssim \mathcal O(10^{-5})$ to avoid the PBH overproduction~\cite{Kohri:2018qtx}.
However, this constraint is not applicable to our fiducial scenarios with $r=3$, where the MD era lasts only for $\mathcal O(1)$ Hubble time.
In this case, the MD era ends before the small-amplitude perturbations sufficiently grow to produce PBHs. 
Once the curvaton decays and the RD era begins, the perturbations below the Jeans scale ($\sim \mathcal O(H^{-1})$) cannot collapse to PBHs. 
The precise calculation of the PBH production rate during the short MD era is beyond the scope of this paper, but we can roughly expect the following things.
Given the fact that PBHs cannot be produced below the Jeans scale, we can expect that the short MD era only decreases the threshold by $\mathcal O(1)$, compared to the one in a RD era.
Although the PBH production rate would be enhanced accordingly by this decrease of the threshold, the exponential suppression for the PBH production rate would still remain.
In our fiducial cases, we find $\mathcal P_\zeta \lesssim \mathcal O(10^{-4})$ on the scales that enter the horizon during the MD era, which is much smaller than the required power spectrum for the sufficient PBH production during a RD era, $\mathcal P_\zeta \simeq \mathcal O(10^{-2})$~\cite{Sasaki:2018dmp}.
Given this, we can expect that the amount of PBHs produced during the short MD era would be negligibly small in our fiducial cases.

\section{Conclusion and Discussion}
\label{sec:conclusion}

The SGWB recently detected by the PTA experiments might originate from cosmological sources. 
In this work, we have focused on SIGWs as a candidate of the detected SGWB, which are induced through the nonlinear interaction between curvature and tensor perturbations. 
Following the recent release of the PTA results, it has been claimed that the large curvature power spectrum, required for the SIGWs explaining the SGWB, inevitably overproduces PBHs if the perturbations follow Gaussian distribution and they are produced during a RD era~\cite{Franciolini:2023pbf,Liu:2023ymk}.
Given this, to reconcile SIGWs with the PTA results without triggering PBH overproduction, we might need to consider the era other than the RD era or the non-Gaussianity that suppresses the PBH production. 
In this work, we have focused on the latter option.
The importance of the non-Gaussianity has already been mentioned in the previous works~\cite{Franciolini:2023pbf,Liu:2023ymk}.
However, concrete UV models that predict the desired non-Gaussianity have not been presented.

Motivated by this, we have focused on the axion curvaton model, where the axion-like field behaves as the curvaton after the inflation. 
The curvaton models generally predict the non-Gaussianity and we can use it to avoid the PBH overproduction.
Then, we have shown that the axion curvaton model can successfully explain the SGWB detected by the PTA experiments and avoid PBH overproduction at the same time.
Interestingly, in addition to the peak-like GW spectrum around the PTA frequencies, the almost scale-invariant GW spectrum is also predicted on the smaller scales with a smaller amplitude. 
This kind of SIGWs could be detected by LISA, DECIGO, and BBO.
We have also found that the primordial non-Gaussianity in the axion curvaton model can lighten the peak mass of the PBH spectrum.
Specifically, we have found that the peak mass can be around $\mathcal O(0.1-1)M_\odot$ even in the model that explains the PTA results.
Besides, we also note that it is difficult to relate the PTA results to PBHs with $\mathcal O(10) M_\odot$ for the BHs detected by the LIGO-Virgo-KAGRA collaborations~\cite{LIGOScientific:2021djp}.
This is because the PTA results favor the blue-tilted power spectrum on the PTA scales~\cite{NANOGrav:2023gor,Antoniadis:2023rey,Figueroa:2023zhu} and therefore the peak scale of the curvature power spectrum is likely to be smaller than the PTA scales. 
At the same time, we mention the possibility that future PTA data inconsistently favor a red-tilted power spectrum. If that turns out to be the case, we may explain the BHs detected by LIGO-Virgo-KAGRA collaborations~\cite{Inomata:2020xad,Kawasaki:2021ycf}.

Throughout this work, we have focused on the type II axion curvaton model as a fiducial model. 
We finally mention the advantage of type II over type I by highlighting the difference from the previous works~\cite{Ando:2018nge,Kawasaki:2021ycf}.
The previous works focused on the situation where the axion curvaton decays to radiation before it dominates the universe ($r < 1$).
However, to realize a negative $f_\NL$, the curvaton must dominate the universe before its decay ($r > 1$).
Since the universe behaves as a MD era during the curvaton domination, the PBH production rate can be enhanced due to the absence of pressure during the era. 
However, as mentioned in Sec.~\ref{sec:pbh}, the effects of the curvaton-dominated era on the PBH production rate requires careful treatment because the curvaton-dominated (equally, MD) era does not last long, which is different from the situation discussed in the literature on the PBH production during a MD era.
We leave a detailed analysis of the PBH production in this situation for future work.
In this work, we have focused on the type II model, where the power spectrum on the small-scale side of the peak scale is suppressed by some orders of magnitude. 
This suppression is realized by the increase of the field value of the saxion during the inflation.
Then, we can expect that this suppression easily enables the type II model to avoid the PBH overproduction during the MD era even if the PBH abundance is significantly enhanced during that era.
On the other hand, in the type I model, the PBH overproduction problem during the MD era could be more severe.
In the type I model, the field value of the saxion decreases during the inflation and it cannot be used to realize the suppression of the power spectrum.
Consequently, the suppression of the power spectrum on the small-scale side of the peak can only be through the decrease of the Hubble parameter during the inflation. 
Because of this, the tilt of the curvature power spectrum on the small-scale side of the peak is proportional to the slow-roll parameter~\cite{Inomata:2020xad}.
Given this, we might need to consider a large slow-roll parameter to avoid the overproduction of PBHs during the MD era in the type I model.
This requirement could put strong constraints on the inflation model.

\acknowledgments
\noindent
We thank Antonio Junior Iovino for pointing out the possible effects of the higher-order terms in the non-Gaussianity.
K.\,I.\, was supported by JSPS Postdoctoral Fellowships for Research Abroad and the Kavli Institute
for Cosmological Physics at the University of Chicago through an endowment from the Kavli Foundation and its founder Fred Kavli.
M.\,K.\, was supported by JSPS KAKENHI Grant Nos.\ 20H05851 and 21K03567,  and World Premier International Research Center Initiative (WPI Initiative), MEXT, Japan.
K.\,M.\, was supported by MEXT Leading Initiative for Excellent Young Researchers Grant No.\ JPMXS0320200430, 
and by JSPS KAKENHI Grant No.\ 	JP22K14044.
T.\,T.\,Y.\, was supported by the China Grant for Talent Scientific Start-Up Project and by Natural Science Foundation of China (NSFC) under grant No.\ 12175134, JSPS Grant-in-Aid for Scientific Research Grants No.\ 19H05810, and World Premier International Research Center Initiative (WPI Initiative), MEXT, Japan.

\appendix

\section{Relation between the scales and the temperature}
\label{app:scales}

In this appendix, we derive Eq.~(\ref{eq:k_o_keq}), which relates the perturbation scale, the inflation energy scale, and the reheating temperature in our scenario. 
We use the horizon scale at $\eta_\eq$ as a pivot scale:
\begin{align}
  \frac{k}{k_\eq} = \frac{k}{\mathcal H_\en} \frac{\mathcal H_\en}{\mathcal H_\Rf} \frac{\mathcal H_\Rf}{\mathcal H_\Rs} \frac{\mathcal H_\Rs}{\mathcal H_\eq},
  \label{eq:scales}
\end{align}
where we have used $k_\eq = \mathcal H_\eq$.
Note again that the subscript `$\bullet$' denotes the value at $\eta_\bullet$ with $\bullet$ being arbitrary characters.
We can express $\mathcal H_\en/\mathcal H_\Rf$ in Eq.~(\ref{eq:scales}) as 
\begin{align}
  \frac{\mathcal H_\en}{\mathcal H_\Rf} &= \left( \frac{\rho^\tot_\en}{\rho^\tot_\Rf}\right)^{1/6}= \left( \frac{\rho^\tot_\en}{\frac{\pi^2}{30} g^\rho_{\Rf} T_\Rf^4 }\right)^{1/6},
  \label{eq:hen_hrf}
\end{align}
where we have used the fact that the universe behaves as a MD era from $\eta_\en$ to $\eta_\Rf$. 
$\mathcal H_\Rf/\mathcal H_\dec$ can be expressed as 
\begin{align}
  \frac{\mathcal H_\Rf }{\mathcal H_\dec }  &= \left(\frac{g^s_\Rf}{g^s_\decm}\right)^{-1/3} \frac{T_\decm}{T_\Rf} \sqrt{\frac{\rho^\tot_\Rf}{\rho^\tot_\decm}} \nonumber \\
  &=  \left( \frac{g^\rho_\Rf}{g^\rho_\decm}\right)^{1/4}\left(\frac{g^s_\Rf}{g^s_\decm}\right)^{-1/3}  \left( \frac{\rho^\rr_\decm}{\rho^\rr_\Rf} \right)^{1/4} \sqrt{\frac{\rho^\tot_\Rf}{\rho^\tot_\decm}} \nonumber \\
  &\simeq (1+r)^{-1/4}\left( \frac{g^\rho_\Rf}{g^\rho_\decm}\right)^{1/4}\left(\frac{g^s_\Rf}{g^s_\decm}\right)^{-1/3} \left(\frac{\rho^\tot_\Rf}{\rho^\tot_\decm} \right)^{1/4},
\end{align}
where we have used $\rho^\rr_\Rf \simeq \rho^\tot_\Rf$.
$\mathcal H_\Rs/\mathcal H_\eq$ can be expressed as
\begin{align}
  \frac{\mathcal H_\Rs}{\mathcal H_\eq} &= 2^{-1/4}\left( \frac{\rho^\tot_\Rs}{\rho^\tot_\eq}\right)^{1/4} \left( \frac{g^\rho_{\decp}}{g^\rho_{\eq}} \right)^{1/4} \left( \frac{g^s_{\decp}}{g^s_{\eq}} \right)^{-1/3}.
  \label{eq:hdec_o_heq}
\end{align}
Substituting Eqs.~(\ref{eq:hen_hrf})-(\ref{eq:hdec_o_heq}) into Eq.~(\ref{eq:scales}), we obtain
\begin{align}
  \frac{k}{k_\eq} =\, & (2(1+r))^{-1/4}\left( \frac{\rho^\tot_\en}{\frac{\pi^2}{30} g^\rho_\Rf T_\Rf^4 }\right)^{1/6} \left( \frac{g^\rho_{\Rf}}{g^\rho_{\eq}} \right)^{1/4} \left( \frac{g^s_{\Rf}}{g^s_{\eq}} \right)^{-1/3}  \left( \frac{g^\rho_{\decp}}{g^\rho_{\decm}} \right)^{1/4} \left( \frac{g^s_{\decp}}{g^s_{\decm}} \right)^{-1/3}\left( \frac{\frac{\pi^2}{30} g^\rho_\Rf T_\Rf^4 }{\rho^\tot_\eq}\right)^{1/4} \frac{k}{\mathcal H_\en} \nonumber \\
  = \, &1.3\times 10^{24}(1+r)^{-1/4} \left( \frac{\rho^\tot_\en}{3 \times 10^{-11}\, M_\Pl^4}\right)^{1/6} \left(\frac{T_\Rf}{10^{14}\,\GeV} \right)^{1/3}  \left(\frac{g^\rho_{\Rf}}{g^s_{\Rf}} \right)^{1/3} \left( \frac{g^\rho_{\decp}}{g^\rho_{\decm}} \right)^{1/4} \left( \frac{g^s_{\decp}}{g^s_{\decm}} \right)^{-1/3} \frac{k}{\mathcal H_\en},
\end{align}
where we have used $g^\rho_\eq = 3.38$, $g^s_\eq = 3.91$, and $T_\eq = 8.0 \times 10^{-10}\,\GeV$~\cite{Inomata:2018epa}.

\small
\bibliographystyle{apsrev4-1}
\bibliography{axion_curvaton.bib}

\begin{thebibliography}{144}%
\makeatletter
\providecommand \@ifxundefined [1]{%
 \@ifx{#1\undefined}
}%
\providecommand \@ifnum [1]{%
 \ifnum #1\expandafter \@firstoftwo
 \else \expandafter \@secondoftwo
 \fi
}%
\providecommand \@ifx [1]{%
 \ifx #1\expandafter \@firstoftwo
 \else \expandafter \@secondoftwo
 \fi
}%
\providecommand \natexlab [1]{#1}%
\providecommand \enquote  [1]{``#1''}%
\providecommand \bibnamefont  [1]{#1}%
\providecommand \bibfnamefont [1]{#1}%
\providecommand \citenamefont [1]{#1}%
\providecommand \href@noop [0]{\@secondoftwo}%
\providecommand \href [0]{\begingroup \@sanitize@url \@href}%
\providecommand \@href[1]{\@@startlink{#1}\@@href}%
\providecommand \@@href[1]{\endgroup#1\@@endlink}%
\providecommand \@sanitize@url [0]{\catcode `\\12\catcode `\$12\catcode
  `\&12\catcode `\#12\catcode `\^12\catcode `\_12\catcode `\%12\relax}%
\providecommand \@@startlink[1]{}%
\providecommand \@@endlink[0]{}%
\providecommand \url  [0]{\begingroup\@sanitize@url \@url }%
\providecommand \@url [1]{\endgroup\@href {#1}{\urlprefix }}%
\providecommand \urlprefix  [0]{URL }%
\providecommand \Eprint [0]{\href }%
\providecommand \doibase [0]{http://dx.doi.org/}%
\providecommand \selectlanguage [0]{\@gobble}%
\providecommand \bibinfo  [0]{\@secondoftwo}%
\providecommand \bibfield  [0]{\@secondoftwo}%
\providecommand \translation [1]{[#1]}%
\providecommand \BibitemOpen [0]{}%
\providecommand \bibitemStop [0]{}%
\providecommand \bibitemNoStop [0]{.\EOS\space}%
\providecommand \EOS [0]{\spacefactor3000\relax}%
\providecommand \BibitemShut  [1]{\csname bibitem#1\endcsname}%
\let\auto@bib@innerbib\@empty
\bibitem [{\citenamefont {Agazie}\ \emph
  {et~al.}(2023{\natexlab{a}})\citenamefont {Agazie} \emph
  {et~al.}}]{NANOGrav:2023gor}%
  \BibitemOpen
  \bibfield  {author} {\bibinfo {author} {\bibfnamefont {G.}~\bibnamefont
  {Agazie}} \emph {et~al.} (\bibinfo {collaboration} {NANOGrav}),\ }\href
  {\doibase 10.3847/2041-8213/acdac6} {\bibfield  {journal} {\bibinfo
  {journal} {Astrophys. J. Lett.}\ }\textbf {\bibinfo {volume} {951}},\
  \bibinfo {pages} {L8} (\bibinfo {year} {2023}{\natexlab{a}})},\ \Eprint
  {http://arxiv.org/abs/2306.16213} {arXiv:2306.16213 [astro-ph.HE]}
  \BibitemShut {NoStop}%
\bibitem [{\citenamefont {Agazie}\ \emph
  {et~al.}(2023{\natexlab{b}})\citenamefont {Agazie} \emph
  {et~al.}}]{NANOGrav:2023hde}%
  \BibitemOpen
  \bibfield  {author} {\bibinfo {author} {\bibfnamefont {G.}~\bibnamefont
  {Agazie}} \emph {et~al.} (\bibinfo {collaboration} {NANOGrav}),\ }\href
  {\doibase 10.3847/2041-8213/acda9a} {\bibfield  {journal} {\bibinfo
  {journal} {Astrophys. J. Lett.}\ }\textbf {\bibinfo {volume} {951}},\
  \bibinfo {pages} {L9} (\bibinfo {year} {2023}{\natexlab{b}})},\ \Eprint
  {http://arxiv.org/abs/2306.16217} {arXiv:2306.16217 [astro-ph.HE]}
  \BibitemShut {NoStop}%
\bibitem [{\citenamefont {Afzal}\ \emph {et~al.}(2023)\citenamefont {Afzal}
  \emph {et~al.}}]{NANOGrav:2023hvm}%
  \BibitemOpen
  \bibfield  {author} {\bibinfo {author} {\bibfnamefont {A.}~\bibnamefont
  {Afzal}} \emph {et~al.} (\bibinfo {collaboration} {NANOGrav}),\ }\href
  {\doibase 10.3847/2041-8213/acdc91} {\bibfield  {journal} {\bibinfo
  {journal} {Astrophys. J. Lett.}\ }\textbf {\bibinfo {volume} {951}},\
  \bibinfo {pages} {L11} (\bibinfo {year} {2023})},\ \Eprint
  {http://arxiv.org/abs/2306.16219} {arXiv:2306.16219 [astro-ph.HE]}
  \BibitemShut {NoStop}%
\bibitem [{\citenamefont {Antoniadis}\ \emph
  {et~al.}(2023{\natexlab{a}})\citenamefont {Antoniadis} \emph
  {et~al.}}]{Antoniadis:2023rey}%
  \BibitemOpen
  \bibfield  {author} {\bibinfo {author} {\bibfnamefont {J.}~\bibnamefont
  {Antoniadis}} \emph {et~al.},\ }\href@noop {} {\  (\bibinfo {year}
  {2023}{\natexlab{a}})},\ \Eprint {http://arxiv.org/abs/2306.16214}
  {arXiv:2306.16214 [astro-ph.HE]} \BibitemShut {NoStop}%
\bibitem [{\citenamefont {Antoniadis}\ \emph
  {et~al.}(2023{\natexlab{b}})\citenamefont {Antoniadis} \emph
  {et~al.}}]{Antoniadis:2023utw}%
  \BibitemOpen
  \bibfield  {author} {\bibinfo {author} {\bibfnamefont {J.}~\bibnamefont
  {Antoniadis}} \emph {et~al.},\ }\href {\doibase 10.1051/0004-6361/202346841}
  {\  (\bibinfo {year} {2023}{\natexlab{b}}),\ 10.1051/0004-6361/202346841},\
  \Eprint {http://arxiv.org/abs/2306.16224} {arXiv:2306.16224 [astro-ph.HE]}
  \BibitemShut {NoStop}%
\bibitem [{\citenamefont {Antoniadis}\ \emph
  {et~al.}(2023{\natexlab{c}})\citenamefont {Antoniadis} \emph
  {et~al.}}]{Antoniadis:2023zhi}%
  \BibitemOpen
  \bibfield  {author} {\bibinfo {author} {\bibfnamefont {J.}~\bibnamefont
  {Antoniadis}} \emph {et~al.},\ }\href@noop {} {\  (\bibinfo {year}
  {2023}{\natexlab{c}})},\ \Eprint {http://arxiv.org/abs/2306.16227}
  {arXiv:2306.16227 [astro-ph.CO]} \BibitemShut {NoStop}%
\bibitem [{\citenamefont {Reardon}\ \emph
  {et~al.}(2023{\natexlab{a}})\citenamefont {Reardon} \emph
  {et~al.}}]{Reardon:2023gzh}%
  \BibitemOpen
  \bibfield  {author} {\bibinfo {author} {\bibfnamefont {D.~J.}\ \bibnamefont
  {Reardon}} \emph {et~al.},\ }\href {\doibase 10.3847/2041-8213/acdd02}
  {\bibfield  {journal} {\bibinfo  {journal} {Astrophys. J. Lett.}\ }\textbf
  {\bibinfo {volume} {951}},\ \bibinfo {pages} {L6} (\bibinfo {year}
  {2023}{\natexlab{a}})},\ \Eprint {http://arxiv.org/abs/2306.16215}
  {arXiv:2306.16215 [astro-ph.HE]} \BibitemShut {NoStop}%
\bibitem [{\citenamefont {Zic}\ \emph {et~al.}(2023)\citenamefont {Zic} \emph
  {et~al.}}]{Zic:2023gta}%
  \BibitemOpen
  \bibfield  {author} {\bibinfo {author} {\bibfnamefont {A.}~\bibnamefont
  {Zic}} \emph {et~al.},\ }\href@noop {} {\  (\bibinfo {year} {2023})},\
  \Eprint {http://arxiv.org/abs/2306.16230} {arXiv:2306.16230 [astro-ph.HE]}
  \BibitemShut {NoStop}%
\bibitem [{\citenamefont {Reardon}\ \emph
  {et~al.}(2023{\natexlab{b}})\citenamefont {Reardon} \emph
  {et~al.}}]{Reardon:2023zen}%
  \BibitemOpen
  \bibfield  {author} {\bibinfo {author} {\bibfnamefont {D.~J.}\ \bibnamefont
  {Reardon}} \emph {et~al.},\ }\href {\doibase 10.3847/2041-8213/acdd03}
  {\bibfield  {journal} {\bibinfo  {journal} {Astrophys. J. Lett.}\ }\textbf
  {\bibinfo {volume} {951}},\ \bibinfo {pages} {L7} (\bibinfo {year}
  {2023}{\natexlab{b}})},\ \Eprint {http://arxiv.org/abs/2306.16229}
  {arXiv:2306.16229 [astro-ph.HE]} \BibitemShut {NoStop}%
\bibitem [{\citenamefont {Xu}\ \emph {et~al.}(2023)\citenamefont {Xu} \emph
  {et~al.}}]{Xu:2023wog}%
  \BibitemOpen
  \bibfield  {author} {\bibinfo {author} {\bibfnamefont {H.}~\bibnamefont {Xu}}
  \emph {et~al.},\ }\href {\doibase 10.1088/1674-4527/acdfa5} {\bibfield
  {journal} {\bibinfo  {journal} {Res. Astron. Astrophys.}\ }\textbf {\bibinfo
  {volume} {23}},\ \bibinfo {pages} {075024} (\bibinfo {year} {2023})},\
  \Eprint {http://arxiv.org/abs/2306.16216} {arXiv:2306.16216 [astro-ph.HE]}
  \BibitemShut {NoStop}%
\bibitem [{\citenamefont {Ellis}\ \emph
  {et~al.}(2023{\natexlab{a}})\citenamefont {Ellis}, \citenamefont {Fairbairn},
  \citenamefont {H\"utsi}, \citenamefont {Raidal}, \citenamefont {Urrutia},
  \citenamefont {Vaskonen},\ and\ \citenamefont {Veerm\"ae}}]{Ellis:2023dgf}%
  \BibitemOpen
  \bibfield  {author} {\bibinfo {author} {\bibfnamefont {J.}~\bibnamefont
  {Ellis}}, \bibinfo {author} {\bibfnamefont {M.}~\bibnamefont {Fairbairn}},
  \bibinfo {author} {\bibfnamefont {G.}~\bibnamefont {H\"utsi}}, \bibinfo
  {author} {\bibfnamefont {J.}~\bibnamefont {Raidal}}, \bibinfo {author}
  {\bibfnamefont {J.}~\bibnamefont {Urrutia}}, \bibinfo {author} {\bibfnamefont
  {V.}~\bibnamefont {Vaskonen}}, \ and\ \bibinfo {author} {\bibfnamefont
  {H.}~\bibnamefont {Veerm\"ae}},\ }\href@noop {} {\  (\bibinfo {year}
  {2023}{\natexlab{a}})},\ \Eprint {http://arxiv.org/abs/2306.17021}
  {arXiv:2306.17021 [astro-ph.CO]} \BibitemShut {NoStop}%
\bibitem [{\citenamefont {Figueroa}\ \emph {et~al.}(2023)\citenamefont
  {Figueroa}, \citenamefont {Pieroni}, \citenamefont {Ricciardone},\ and\
  \citenamefont {Simakachorn}}]{Figueroa:2023zhu}%
  \BibitemOpen
  \bibfield  {author} {\bibinfo {author} {\bibfnamefont {D.~G.}\ \bibnamefont
  {Figueroa}}, \bibinfo {author} {\bibfnamefont {M.}~\bibnamefont {Pieroni}},
  \bibinfo {author} {\bibfnamefont {A.}~\bibnamefont {Ricciardone}}, \ and\
  \bibinfo {author} {\bibfnamefont {P.}~\bibnamefont {Simakachorn}},\
  }\href@noop {} {\  (\bibinfo {year} {2023})},\ \Eprint
  {http://arxiv.org/abs/2307.02399} {arXiv:2307.02399 [astro-ph.CO]}
  \BibitemShut {NoStop}%
\bibitem [{\citenamefont {Ellis}\ \emph
  {et~al.}(2023{\natexlab{b}})\citenamefont {Ellis}, \citenamefont {Fairbairn},
  \citenamefont {Franciolini}, \citenamefont {H\"utsi}, \citenamefont {Iovino},
  \citenamefont {Lewicki}, \citenamefont {Raidal}, \citenamefont {Urrutia},
  \citenamefont {Vaskonen},\ and\ \citenamefont {Veerm\"ae}}]{Ellis:2023oxs}%
  \BibitemOpen
  \bibfield  {author} {\bibinfo {author} {\bibfnamefont {J.}~\bibnamefont
  {Ellis}}, \bibinfo {author} {\bibfnamefont {M.}~\bibnamefont {Fairbairn}},
  \bibinfo {author} {\bibfnamefont {G.}~\bibnamefont {Franciolini}}, \bibinfo
  {author} {\bibfnamefont {G.}~\bibnamefont {H\"utsi}}, \bibinfo {author}
  {\bibfnamefont {A.}~\bibnamefont {Iovino}}, \bibinfo {author} {\bibfnamefont
  {M.}~\bibnamefont {Lewicki}}, \bibinfo {author} {\bibfnamefont
  {M.}~\bibnamefont {Raidal}}, \bibinfo {author} {\bibfnamefont
  {J.}~\bibnamefont {Urrutia}}, \bibinfo {author} {\bibfnamefont
  {V.}~\bibnamefont {Vaskonen}}, \ and\ \bibinfo {author} {\bibfnamefont
  {H.}~\bibnamefont {Veerm\"ae}},\ }\href@noop {} {\  (\bibinfo {year}
  {2023}{\natexlab{b}})},\ \Eprint {http://arxiv.org/abs/2308.08546}
  {arXiv:2308.08546 [astro-ph.CO]} \BibitemShut {NoStop}%
\bibitem [{\citenamefont {Ananda}\ \emph {et~al.}(2007)\citenamefont {Ananda},
  \citenamefont {Clarkson},\ and\ \citenamefont {Wands}}]{Ananda:2006af}%
  \BibitemOpen
  \bibfield  {author} {\bibinfo {author} {\bibfnamefont {K.~N.}\ \bibnamefont
  {Ananda}}, \bibinfo {author} {\bibfnamefont {C.}~\bibnamefont {Clarkson}}, \
  and\ \bibinfo {author} {\bibfnamefont {D.}~\bibnamefont {Wands}},\ }\href
  {\doibase 10.1103/PhysRevD.75.123518} {\bibfield  {journal} {\bibinfo
  {journal} {Phys. Rev.}\ }\textbf {\bibinfo {volume} {D75}},\ \bibinfo {pages}
  {123518} (\bibinfo {year} {2007})},\ \Eprint
  {http://arxiv.org/abs/gr-qc/0612013} {arXiv:gr-qc/0612013 [gr-qc]}
  \BibitemShut {NoStop}%
\bibitem [{\citenamefont {Baumann}\ \emph {et~al.}(2007)\citenamefont
  {Baumann}, \citenamefont {Steinhardt}, \citenamefont {Takahashi},\ and\
  \citenamefont {Ichiki}}]{Baumann:2007zm}%
  \BibitemOpen
  \bibfield  {author} {\bibinfo {author} {\bibfnamefont {D.}~\bibnamefont
  {Baumann}}, \bibinfo {author} {\bibfnamefont {P.~J.}\ \bibnamefont
  {Steinhardt}}, \bibinfo {author} {\bibfnamefont {K.}~\bibnamefont
  {Takahashi}}, \ and\ \bibinfo {author} {\bibfnamefont {K.}~\bibnamefont
  {Ichiki}},\ }\href {\doibase 10.1103/PhysRevD.76.084019} {\bibfield
  {journal} {\bibinfo  {journal} {Phys. Rev.}\ }\textbf {\bibinfo {volume}
  {D76}},\ \bibinfo {pages} {084019} (\bibinfo {year} {2007})},\ \Eprint
  {http://arxiv.org/abs/hep-th/0703290} {arXiv:hep-th/0703290 [hep-th]}
  \BibitemShut {NoStop}%
\bibitem [{\citenamefont {Madge}\ \emph {et~al.}(2023)\citenamefont {Madge},
  \citenamefont {Morgante}, \citenamefont {Puchades-Ib\'a\~nez}, \citenamefont
  {Ramberg}, \citenamefont {Ratzinger}, \citenamefont {Schenk},\ and\
  \citenamefont {Schwaller}}]{Madge:2023cak}%
  \BibitemOpen
  \bibfield  {author} {\bibinfo {author} {\bibfnamefont {E.}~\bibnamefont
  {Madge}}, \bibinfo {author} {\bibfnamefont {E.}~\bibnamefont {Morgante}},
  \bibinfo {author} {\bibfnamefont {C.}~\bibnamefont {Puchades-Ib\'a\~nez}},
  \bibinfo {author} {\bibfnamefont {N.}~\bibnamefont {Ramberg}}, \bibinfo
  {author} {\bibfnamefont {W.}~\bibnamefont {Ratzinger}}, \bibinfo {author}
  {\bibfnamefont {S.}~\bibnamefont {Schenk}}, \ and\ \bibinfo {author}
  {\bibfnamefont {P.}~\bibnamefont {Schwaller}},\ }\href@noop {} {\  (\bibinfo
  {year} {2023})},\ \Eprint {http://arxiv.org/abs/2306.14856} {arXiv:2306.14856
  [hep-ph]} \BibitemShut {NoStop}%
\bibitem [{\citenamefont {Franciolini}\ \emph {et~al.}(2023)\citenamefont
  {Franciolini}, \citenamefont {Iovino}, \citenamefont {Vaskonen},\ and\
  \citenamefont {Veermae}}]{Franciolini:2023pbf}%
  \BibitemOpen
  \bibfield  {author} {\bibinfo {author} {\bibfnamefont {G.}~\bibnamefont
  {Franciolini}}, \bibinfo {author} {\bibfnamefont {A.}~\bibnamefont {Iovino},
  \bibfnamefont {Junior.}}, \bibinfo {author} {\bibfnamefont {V.}~\bibnamefont
  {Vaskonen}}, \ and\ \bibinfo {author} {\bibfnamefont {H.}~\bibnamefont
  {Veermae}},\ }\href@noop {} {\  (\bibinfo {year} {2023})},\ \Eprint
  {http://arxiv.org/abs/2306.17149} {arXiv:2306.17149 [astro-ph.CO]}
  \BibitemShut {NoStop}%
\bibitem [{\citenamefont {Inomata}\ \emph {et~al.}(2023)\citenamefont
  {Inomata}, \citenamefont {Kohri},\ and\ \citenamefont
  {Terada}}]{Inomata:2023zup}%
  \BibitemOpen
  \bibfield  {author} {\bibinfo {author} {\bibfnamefont {K.}~\bibnamefont
  {Inomata}}, \bibinfo {author} {\bibfnamefont {K.}~\bibnamefont {Kohri}}, \
  and\ \bibinfo {author} {\bibfnamefont {T.}~\bibnamefont {Terada}},\
  }\href@noop {} {\  (\bibinfo {year} {2023})},\ \Eprint
  {http://arxiv.org/abs/2306.17834} {arXiv:2306.17834 [astro-ph.CO]}
  \BibitemShut {NoStop}%
\bibitem [{\citenamefont {Cai}\ \emph {et~al.}(2023)\citenamefont {Cai},
  \citenamefont {He}, \citenamefont {Ma}, \citenamefont {Yan},\ and\
  \citenamefont {Yuan}}]{Cai:2023dls}%
  \BibitemOpen
  \bibfield  {author} {\bibinfo {author} {\bibfnamefont {Y.-F.}\ \bibnamefont
  {Cai}}, \bibinfo {author} {\bibfnamefont {X.-C.}\ \bibnamefont {He}},
  \bibinfo {author} {\bibfnamefont {X.}~\bibnamefont {Ma}}, \bibinfo {author}
  {\bibfnamefont {S.-F.}\ \bibnamefont {Yan}}, \ and\ \bibinfo {author}
  {\bibfnamefont {G.-W.}\ \bibnamefont {Yuan}},\ }\href@noop {} {\  (\bibinfo
  {year} {2023})},\ \Eprint {http://arxiv.org/abs/2306.17822} {arXiv:2306.17822
  [gr-qc]} \BibitemShut {NoStop}%
\bibitem [{\citenamefont {Zhu}\ \emph {et~al.}(2023)\citenamefont {Zhu},
  \citenamefont {Zhao},\ and\ \citenamefont {Wang}}]{Zhu:2023faa}%
  \BibitemOpen
  \bibfield  {author} {\bibinfo {author} {\bibfnamefont {Q.-H.}\ \bibnamefont
  {Zhu}}, \bibinfo {author} {\bibfnamefont {Z.-C.}\ \bibnamefont {Zhao}}, \
  and\ \bibinfo {author} {\bibfnamefont {S.}~\bibnamefont {Wang}},\ }\href@noop
  {} {\  (\bibinfo {year} {2023})},\ \Eprint {http://arxiv.org/abs/2307.03095}
  {arXiv:2307.03095 [astro-ph.CO]} \BibitemShut {NoStop}%
\bibitem [{\citenamefont {Wang}\ \emph {et~al.}(2023)\citenamefont {Wang},
  \citenamefont {Zhao}, \citenamefont {Li},\ and\ \citenamefont
  {Zhu}}]{Wang:2023ost}%
  \BibitemOpen
  \bibfield  {author} {\bibinfo {author} {\bibfnamefont {S.}~\bibnamefont
  {Wang}}, \bibinfo {author} {\bibfnamefont {Z.-C.}\ \bibnamefont {Zhao}},
  \bibinfo {author} {\bibfnamefont {J.-P.}\ \bibnamefont {Li}}, \ and\ \bibinfo
  {author} {\bibfnamefont {Q.-H.}\ \bibnamefont {Zhu}},\ }\href@noop {} {\
  (\bibinfo {year} {2023})},\ \Eprint {http://arxiv.org/abs/2307.00572}
  {arXiv:2307.00572 [astro-ph.CO]} \BibitemShut {NoStop}%
\bibitem [{\citenamefont {Liu}\ \emph {et~al.}(2023{\natexlab{a}})\citenamefont
  {Liu}, \citenamefont {Chen},\ and\ \citenamefont {Huang}}]{Liu:2023ymk}%
  \BibitemOpen
  \bibfield  {author} {\bibinfo {author} {\bibfnamefont {L.}~\bibnamefont
  {Liu}}, \bibinfo {author} {\bibfnamefont {Z.-C.}\ \bibnamefont {Chen}}, \
  and\ \bibinfo {author} {\bibfnamefont {Q.-G.}\ \bibnamefont {Huang}},\
  }\href@noop {} {\  (\bibinfo {year} {2023}{\natexlab{a}})},\ \Eprint
  {http://arxiv.org/abs/2307.01102} {arXiv:2307.01102 [astro-ph.CO]}
  \BibitemShut {NoStop}%
\bibitem [{\citenamefont {Ebadi}\ \emph {et~al.}(2023)\citenamefont {Ebadi},
  \citenamefont {Kumar}, \citenamefont {McCune}, \citenamefont {Tai},\ and\
  \citenamefont {Wang}}]{Ebadi:2023xhq}%
  \BibitemOpen
  \bibfield  {author} {\bibinfo {author} {\bibfnamefont {R.}~\bibnamefont
  {Ebadi}}, \bibinfo {author} {\bibfnamefont {S.}~\bibnamefont {Kumar}},
  \bibinfo {author} {\bibfnamefont {A.}~\bibnamefont {McCune}}, \bibinfo
  {author} {\bibfnamefont {H.}~\bibnamefont {Tai}}, \ and\ \bibinfo {author}
  {\bibfnamefont {L.-T.}\ \bibnamefont {Wang}},\ }\href@noop {} {\  (\bibinfo
  {year} {2023})},\ \Eprint {http://arxiv.org/abs/2307.01248} {arXiv:2307.01248
  [astro-ph.CO]} \BibitemShut {NoStop}%
\bibitem [{\citenamefont {Abe}\ and\ \citenamefont {Tada}(2023)}]{Abe:2023yrw}%
  \BibitemOpen
  \bibfield  {author} {\bibinfo {author} {\bibfnamefont {K.~T.}\ \bibnamefont
  {Abe}}\ and\ \bibinfo {author} {\bibfnamefont {Y.}~\bibnamefont {Tada}},\
  }\href@noop {} {\  (\bibinfo {year} {2023})},\ \Eprint
  {http://arxiv.org/abs/2307.01653} {arXiv:2307.01653 [astro-ph.CO]}
  \BibitemShut {NoStop}%
\bibitem [{\citenamefont {Firouzjahi}\ and\ \citenamefont
  {Talebian}(2023)}]{Firouzjahi:2023lzg}%
  \BibitemOpen
  \bibfield  {author} {\bibinfo {author} {\bibfnamefont {H.}~\bibnamefont
  {Firouzjahi}}\ and\ \bibinfo {author} {\bibfnamefont {A.}~\bibnamefont
  {Talebian}},\ }\href@noop {} {\  (\bibinfo {year} {2023})},\ \Eprint
  {http://arxiv.org/abs/2307.03164} {arXiv:2307.03164 [gr-qc]} \BibitemShut
  {NoStop}%
\bibitem [{\citenamefont {You}\ \emph {et~al.}(2023)\citenamefont {You},
  \citenamefont {Yi},\ and\ \citenamefont {Wu}}]{You:2023rmn}%
  \BibitemOpen
  \bibfield  {author} {\bibinfo {author} {\bibfnamefont {Z.-Q.}\ \bibnamefont
  {You}}, \bibinfo {author} {\bibfnamefont {Z.}~\bibnamefont {Yi}}, \ and\
  \bibinfo {author} {\bibfnamefont {Y.}~\bibnamefont {Wu}},\ }\href@noop {} {\
  (\bibinfo {year} {2023})},\ \Eprint {http://arxiv.org/abs/2307.04419}
  {arXiv:2307.04419 [gr-qc]} \BibitemShut {NoStop}%
\bibitem [{\citenamefont {Bari}\ \emph {et~al.}(2023)\citenamefont {Bari},
  \citenamefont {Bartolo}, \citenamefont {Dom\`enech},\ and\ \citenamefont
  {Matarrese}}]{Bari:2023rcw}%
  \BibitemOpen
  \bibfield  {author} {\bibinfo {author} {\bibfnamefont {P.}~\bibnamefont
  {Bari}}, \bibinfo {author} {\bibfnamefont {N.}~\bibnamefont {Bartolo}},
  \bibinfo {author} {\bibfnamefont {G.}~\bibnamefont {Dom\`enech}}, \ and\
  \bibinfo {author} {\bibfnamefont {S.}~\bibnamefont {Matarrese}},\ }\href@noop
  {} {\  (\bibinfo {year} {2023})},\ \Eprint {http://arxiv.org/abs/2307.05404}
  {arXiv:2307.05404 [astro-ph.CO]} \BibitemShut {NoStop}%
\bibitem [{\citenamefont {Hosseini~Mansoori}\ \emph {et~al.}(2023)\citenamefont
  {Hosseini~Mansoori}, \citenamefont {Felegray}, \citenamefont {Talebian},\
  and\ \citenamefont {Sami}}]{HosseiniMansoori:2023mqh}%
  \BibitemOpen
  \bibfield  {author} {\bibinfo {author} {\bibfnamefont {S.~A.}\ \bibnamefont
  {Hosseini~Mansoori}}, \bibinfo {author} {\bibfnamefont {F.}~\bibnamefont
  {Felegray}}, \bibinfo {author} {\bibfnamefont {A.}~\bibnamefont {Talebian}},
  \ and\ \bibinfo {author} {\bibfnamefont {M.}~\bibnamefont {Sami}},\
  }\href@noop {} {\  (\bibinfo {year} {2023})},\ \Eprint
  {http://arxiv.org/abs/2307.06757} {arXiv:2307.06757 [astro-ph.CO]}
  \BibitemShut {NoStop}%
\bibitem [{\citenamefont {Balaji}\ \emph {et~al.}(2023)\citenamefont {Balaji},
  \citenamefont {Dom\`enech},\ and\ \citenamefont
  {Franciolini}}]{Balaji:2023ehk}%
  \BibitemOpen
  \bibfield  {author} {\bibinfo {author} {\bibfnamefont {S.}~\bibnamefont
  {Balaji}}, \bibinfo {author} {\bibfnamefont {G.}~\bibnamefont {Dom\`enech}},
  \ and\ \bibinfo {author} {\bibfnamefont {G.}~\bibnamefont {Franciolini}},\
  }\href@noop {} {\  (\bibinfo {year} {2023})},\ \Eprint
  {http://arxiv.org/abs/2307.08552} {arXiv:2307.08552 [gr-qc]} \BibitemShut
  {NoStop}%
\bibitem [{\citenamefont {Basilakos}\ \emph {et~al.}(2023)\citenamefont
  {Basilakos}, \citenamefont {Nanopoulos}, \citenamefont {Papanikolaou},
  \citenamefont {Saridakis},\ and\ \citenamefont
  {Tzerefos}}]{Basilakos:2023xof}%
  \BibitemOpen
  \bibfield  {author} {\bibinfo {author} {\bibfnamefont {S.}~\bibnamefont
  {Basilakos}}, \bibinfo {author} {\bibfnamefont {D.~V.}\ \bibnamefont
  {Nanopoulos}}, \bibinfo {author} {\bibfnamefont {T.}~\bibnamefont
  {Papanikolaou}}, \bibinfo {author} {\bibfnamefont {E.~N.}\ \bibnamefont
  {Saridakis}}, \ and\ \bibinfo {author} {\bibfnamefont {C.}~\bibnamefont
  {Tzerefos}},\ }\href@noop {} {\  (\bibinfo {year} {2023})},\ \Eprint
  {http://arxiv.org/abs/2307.08601} {arXiv:2307.08601 [hep-th]} \BibitemShut
  {NoStop}%
\bibitem [{\citenamefont {Jin}\ \emph {et~al.}(2023)\citenamefont {Jin},
  \citenamefont {Chen}, \citenamefont {Yi}, \citenamefont {You}, \citenamefont
  {Liu},\ and\ \citenamefont {Wu}}]{Jin:2023wri}%
  \BibitemOpen
  \bibfield  {author} {\bibinfo {author} {\bibfnamefont {J.-H.}\ \bibnamefont
  {Jin}}, \bibinfo {author} {\bibfnamefont {Z.-C.}\ \bibnamefont {Chen}},
  \bibinfo {author} {\bibfnamefont {Z.}~\bibnamefont {Yi}}, \bibinfo {author}
  {\bibfnamefont {Z.-Q.}\ \bibnamefont {You}}, \bibinfo {author} {\bibfnamefont
  {L.}~\bibnamefont {Liu}}, \ and\ \bibinfo {author} {\bibfnamefont
  {Y.}~\bibnamefont {Wu}},\ }\href@noop {} {\  (\bibinfo {year} {2023})},\
  \Eprint {http://arxiv.org/abs/2307.08687} {arXiv:2307.08687 [astro-ph.CO]}
  \BibitemShut {NoStop}%
\bibitem [{\citenamefont {Das}\ \emph {et~al.}(2023)\citenamefont {Das},
  \citenamefont {Jaman},\ and\ \citenamefont {Sami}}]{Das:2023nmm}%
  \BibitemOpen
  \bibfield  {author} {\bibinfo {author} {\bibfnamefont {B.}~\bibnamefont
  {Das}}, \bibinfo {author} {\bibfnamefont {N.}~\bibnamefont {Jaman}}, \ and\
  \bibinfo {author} {\bibfnamefont {M.}~\bibnamefont {Sami}},\ }\href@noop {}
  {\  (\bibinfo {year} {2023})},\ \Eprint {http://arxiv.org/abs/2307.12913}
  {arXiv:2307.12913 [gr-qc]} \BibitemShut {NoStop}%
\bibitem [{\citenamefont {Zhao}\ \emph {et~al.}(2023)\citenamefont {Zhao},
  \citenamefont {Zhu}, \citenamefont {Wang},\ and\ \citenamefont
  {Zhang}}]{Zhao:2023joc}%
  \BibitemOpen
  \bibfield  {author} {\bibinfo {author} {\bibfnamefont {Z.-C.}\ \bibnamefont
  {Zhao}}, \bibinfo {author} {\bibfnamefont {Q.-H.}\ \bibnamefont {Zhu}},
  \bibinfo {author} {\bibfnamefont {S.}~\bibnamefont {Wang}}, \ and\ \bibinfo
  {author} {\bibfnamefont {X.}~\bibnamefont {Zhang}},\ }\href@noop {} {\
  (\bibinfo {year} {2023})},\ \Eprint {http://arxiv.org/abs/2307.13574}
  {arXiv:2307.13574 [astro-ph.CO]} \BibitemShut {NoStop}%
\bibitem [{\citenamefont {Liu}\ \emph {et~al.}(2023{\natexlab{b}})\citenamefont
  {Liu}, \citenamefont {Chen},\ and\ \citenamefont {Huang}}]{Liu:2023pau}%
  \BibitemOpen
  \bibfield  {author} {\bibinfo {author} {\bibfnamefont {L.}~\bibnamefont
  {Liu}}, \bibinfo {author} {\bibfnamefont {Z.-C.}\ \bibnamefont {Chen}}, \
  and\ \bibinfo {author} {\bibfnamefont {Q.-G.}\ \bibnamefont {Huang}},\
  }\href@noop {} {\  (\bibinfo {year} {2023}{\natexlab{b}})},\ \Eprint
  {http://arxiv.org/abs/2307.14911} {arXiv:2307.14911 [astro-ph.CO]}
  \BibitemShut {NoStop}%
\bibitem [{\citenamefont {Yi}\ \emph {et~al.}(2023{\natexlab{a}})\citenamefont
  {Yi}, \citenamefont {You},\ and\ \citenamefont {Wu}}]{Yi:2023tdk}%
  \BibitemOpen
  \bibfield  {author} {\bibinfo {author} {\bibfnamefont {Z.}~\bibnamefont
  {Yi}}, \bibinfo {author} {\bibfnamefont {Z.-Q.}\ \bibnamefont {You}}, \ and\
  \bibinfo {author} {\bibfnamefont {Y.}~\bibnamefont {Wu}},\ }\href@noop {} {\
  (\bibinfo {year} {2023}{\natexlab{a}})},\ \Eprint
  {http://arxiv.org/abs/2308.05632} {arXiv:2308.05632 [astro-ph.CO]}
  \BibitemShut {NoStop}%
\bibitem [{\citenamefont {Frosina}\ and\ \citenamefont
  {Urbano}(2023)}]{Frosina:2023nxu}%
  \BibitemOpen
  \bibfield  {author} {\bibinfo {author} {\bibfnamefont {L.}~\bibnamefont
  {Frosina}}\ and\ \bibinfo {author} {\bibfnamefont {A.}~\bibnamefont
  {Urbano}},\ }\href@noop {} {\  (\bibinfo {year} {2023})},\ \Eprint
  {http://arxiv.org/abs/2308.06915} {arXiv:2308.06915 [astro-ph.CO]}
  \BibitemShut {NoStop}%
\bibitem [{\citenamefont {Yuan}\ \emph {et~al.}(2023)\citenamefont {Yuan},
  \citenamefont {Meng},\ and\ \citenamefont {Huang}}]{Yuan:2023ofl}%
  \BibitemOpen
  \bibfield  {author} {\bibinfo {author} {\bibfnamefont {C.}~\bibnamefont
  {Yuan}}, \bibinfo {author} {\bibfnamefont {D.-S.}\ \bibnamefont {Meng}}, \
  and\ \bibinfo {author} {\bibfnamefont {Q.-G.}\ \bibnamefont {Huang}},\
  }\href@noop {} {\  (\bibinfo {year} {2023})},\ \Eprint
  {http://arxiv.org/abs/2308.07155} {arXiv:2308.07155 [astro-ph.CO]}
  \BibitemShut {NoStop}%
\bibitem [{\citenamefont {Bhaumik}\ \emph {et~al.}(2023)\citenamefont
  {Bhaumik}, \citenamefont {Jain},\ and\ \citenamefont
  {Lewicki}}]{Bhaumik:2023wmw}%
  \BibitemOpen
  \bibfield  {author} {\bibinfo {author} {\bibfnamefont {N.}~\bibnamefont
  {Bhaumik}}, \bibinfo {author} {\bibfnamefont {R.~K.}\ \bibnamefont {Jain}}, \
  and\ \bibinfo {author} {\bibfnamefont {M.}~\bibnamefont {Lewicki}},\
  }\href@noop {} {\  (\bibinfo {year} {2023})},\ \Eprint
  {http://arxiv.org/abs/2308.07912} {arXiv:2308.07912 [astro-ph.CO]}
  \BibitemShut {NoStop}%
\bibitem [{\citenamefont {Choudhury}\ \emph {et~al.}(2023)\citenamefont
  {Choudhury}, \citenamefont {Karde}, \citenamefont {Panda},\ and\
  \citenamefont {Sami}}]{Choudhury:2023wrm}%
  \BibitemOpen
  \bibfield  {author} {\bibinfo {author} {\bibfnamefont {S.}~\bibnamefont
  {Choudhury}}, \bibinfo {author} {\bibfnamefont {A.}~\bibnamefont {Karde}},
  \bibinfo {author} {\bibfnamefont {S.}~\bibnamefont {Panda}}, \ and\ \bibinfo
  {author} {\bibfnamefont {M.}~\bibnamefont {Sami}},\ }\href@noop {} {\
  (\bibinfo {year} {2023})},\ \Eprint {http://arxiv.org/abs/2308.09273}
  {arXiv:2308.09273 [astro-ph.CO]} \BibitemShut {NoStop}%
\bibitem [{\citenamefont {Kawasaki}\ and\ \citenamefont
  {Murai}(2023)}]{Kawasaki:2023rfx}%
  \BibitemOpen
  \bibfield  {author} {\bibinfo {author} {\bibfnamefont {M.}~\bibnamefont
  {Kawasaki}}\ and\ \bibinfo {author} {\bibfnamefont {K.}~\bibnamefont
  {Murai}},\ }\href@noop {} {\  (\bibinfo {year} {2023})},\ \Eprint
  {http://arxiv.org/abs/2308.13134} {arXiv:2308.13134 [astro-ph.CO]}
  \BibitemShut {NoStop}%
\bibitem [{\citenamefont {Yi}\ \emph {et~al.}(2023{\natexlab{b}})\citenamefont
  {Yi}, \citenamefont {You}, \citenamefont {Wu}, \citenamefont {Chen},\ and\
  \citenamefont {Liu}}]{Yi:2023npi}%
  \BibitemOpen
  \bibfield  {author} {\bibinfo {author} {\bibfnamefont {Z.}~\bibnamefont
  {Yi}}, \bibinfo {author} {\bibfnamefont {Z.-Q.}\ \bibnamefont {You}},
  \bibinfo {author} {\bibfnamefont {Y.}~\bibnamefont {Wu}}, \bibinfo {author}
  {\bibfnamefont {Z.-C.}\ \bibnamefont {Chen}}, \ and\ \bibinfo {author}
  {\bibfnamefont {L.}~\bibnamefont {Liu}},\ }\href@noop {} {\  (\bibinfo {year}
  {2023}{\natexlab{b}})},\ \Eprint {http://arxiv.org/abs/2308.14688}
  {arXiv:2308.14688 [astro-ph.CO]} \BibitemShut {NoStop}%
\bibitem [{\citenamefont {Bhattacharya}\ \emph {et~al.}(2023)\citenamefont
  {Bhattacharya}, \citenamefont {Choudhury}, \citenamefont {Dey}, \citenamefont
  {Ghosh}, \citenamefont {Karde},\ and\ \citenamefont
  {Mishra}}]{Bhattacharya:2023ysp}%
  \BibitemOpen
  \bibfield  {author} {\bibinfo {author} {\bibfnamefont {G.}~\bibnamefont
  {Bhattacharya}}, \bibinfo {author} {\bibfnamefont {S.}~\bibnamefont
  {Choudhury}}, \bibinfo {author} {\bibfnamefont {K.}~\bibnamefont {Dey}},
  \bibinfo {author} {\bibfnamefont {S.}~\bibnamefont {Ghosh}}, \bibinfo
  {author} {\bibfnamefont {A.}~\bibnamefont {Karde}}, \ and\ \bibinfo {author}
  {\bibfnamefont {N.~S.}\ \bibnamefont {Mishra}},\ }\href@noop {} {\  (\bibinfo
  {year} {2023})},\ \Eprint {http://arxiv.org/abs/2309.00973} {arXiv:2309.00973
  [astro-ph.CO]} \BibitemShut {NoStop}%
\bibitem [{\citenamefont {Gangopadhyay}\ \emph {et~al.}(2023)\citenamefont
  {Gangopadhyay}, \citenamefont {Godithi}, \citenamefont {Ichiki},
  \citenamefont {Inui}, \citenamefont {Kajino}, \citenamefont {Manusankar},
  \citenamefont {Mathews},\ and\ \citenamefont
  {Yogesh}}]{Gangopadhyay:2023qjr}%
  \BibitemOpen
  \bibfield  {author} {\bibinfo {author} {\bibfnamefont {M.~R.}\ \bibnamefont
  {Gangopadhyay}}, \bibinfo {author} {\bibfnamefont {V.~V.}\ \bibnamefont
  {Godithi}}, \bibinfo {author} {\bibfnamefont {K.}~\bibnamefont {Ichiki}},
  \bibinfo {author} {\bibfnamefont {R.}~\bibnamefont {Inui}}, \bibinfo {author}
  {\bibfnamefont {T.}~\bibnamefont {Kajino}}, \bibinfo {author} {\bibfnamefont
  {A.}~\bibnamefont {Manusankar}}, \bibinfo {author} {\bibfnamefont {G.~J.}\
  \bibnamefont {Mathews}}, \ and\ \bibinfo {author} {\bibnamefont {Yogesh}},\
  }\href@noop {} {\  (\bibinfo {year} {2023})},\ \Eprint
  {http://arxiv.org/abs/2309.03101} {arXiv:2309.03101 [astro-ph.CO]}
  \BibitemShut {NoStop}%
\bibitem [{\citenamefont {Chang}\ \emph {et~al.}(2023)\citenamefont {Chang},
  \citenamefont {Kuang}, \citenamefont {Wu},\ and\ \citenamefont
  {Zhou}}]{Chang:2023ist}%
  \BibitemOpen
  \bibfield  {author} {\bibinfo {author} {\bibfnamefont {Z.}~\bibnamefont
  {Chang}}, \bibinfo {author} {\bibfnamefont {Y.-T.}\ \bibnamefont {Kuang}},
  \bibinfo {author} {\bibfnamefont {D.}~\bibnamefont {Wu}}, \ and\ \bibinfo
  {author} {\bibfnamefont {J.-Z.}\ \bibnamefont {Zhou}},\ }\href@noop {} {\
  (\bibinfo {year} {2023})},\ \Eprint {http://arxiv.org/abs/2309.06676}
  {arXiv:2309.06676 [astro-ph.CO]} \BibitemShut {NoStop}%
\bibitem [{\citenamefont {Vaskonen}\ and\ \citenamefont
  {Veerm\"ae}(2021)}]{Vaskonen:2020lbd}%
  \BibitemOpen
  \bibfield  {author} {\bibinfo {author} {\bibfnamefont {V.}~\bibnamefont
  {Vaskonen}}\ and\ \bibinfo {author} {\bibfnamefont {H.}~\bibnamefont
  {Veerm\"ae}},\ }\href {\doibase 10.1103/PhysRevLett.126.051303} {\bibfield
  {journal} {\bibinfo  {journal} {Phys. Rev. Lett.}\ }\textbf {\bibinfo
  {volume} {126}},\ \bibinfo {pages} {051303} (\bibinfo {year} {2021})},\
  \Eprint {http://arxiv.org/abs/2009.07832} {arXiv:2009.07832 [astro-ph.CO]}
  \BibitemShut {NoStop}%
\bibitem [{\citenamefont {De~Luca}\ \emph {et~al.}(2021)\citenamefont
  {De~Luca}, \citenamefont {Franciolini},\ and\ \citenamefont
  {Riotto}}]{DeLuca:2020agl}%
  \BibitemOpen
  \bibfield  {author} {\bibinfo {author} {\bibfnamefont {V.}~\bibnamefont
  {De~Luca}}, \bibinfo {author} {\bibfnamefont {G.}~\bibnamefont
  {Franciolini}}, \ and\ \bibinfo {author} {\bibfnamefont {A.}~\bibnamefont
  {Riotto}},\ }\href {\doibase 10.1103/PhysRevLett.126.041303} {\bibfield
  {journal} {\bibinfo  {journal} {Phys. Rev. Lett.}\ }\textbf {\bibinfo
  {volume} {126}},\ \bibinfo {pages} {041303} (\bibinfo {year} {2021})},\
  \Eprint {http://arxiv.org/abs/2009.08268} {arXiv:2009.08268 [astro-ph.CO]}
  \BibitemShut {NoStop}%
\bibitem [{\citenamefont {Kohri}\ and\ \citenamefont
  {Terada}(2021)}]{Kohri:2020qqd}%
  \BibitemOpen
  \bibfield  {author} {\bibinfo {author} {\bibfnamefont {K.}~\bibnamefont
  {Kohri}}\ and\ \bibinfo {author} {\bibfnamefont {T.}~\bibnamefont {Terada}},\
  }\href {\doibase 10.1016/j.physletb.2020.136040} {\bibfield  {journal}
  {\bibinfo  {journal} {Phys. Lett. B}\ }\textbf {\bibinfo {volume} {813}},\
  \bibinfo {pages} {136040} (\bibinfo {year} {2021})},\ \Eprint
  {http://arxiv.org/abs/2009.11853} {arXiv:2009.11853 [astro-ph.CO]}
  \BibitemShut {NoStop}%
\bibitem [{\citenamefont {Dom\`enech}\ and\ \citenamefont
  {Pi}(2022)}]{Domenech:2020ers}%
  \BibitemOpen
  \bibfield  {author} {\bibinfo {author} {\bibfnamefont {G.}~\bibnamefont
  {Dom\`enech}}\ and\ \bibinfo {author} {\bibfnamefont {S.}~\bibnamefont
  {Pi}},\ }\href {\doibase 10.1007/s11433-021-1839-6} {\bibfield  {journal}
  {\bibinfo  {journal} {Sci. China Phys. Mech. Astron.}\ }\textbf {\bibinfo
  {volume} {65}},\ \bibinfo {pages} {230411} (\bibinfo {year} {2022})},\
  \Eprint {http://arxiv.org/abs/2010.03976} {arXiv:2010.03976 [astro-ph.CO]}
  \BibitemShut {NoStop}%
\bibitem [{\citenamefont {Papanikolaou}\ \emph {et~al.}(2021)\citenamefont
  {Papanikolaou}, \citenamefont {Vennin},\ and\ \citenamefont
  {Langlois}}]{Papanikolaou:2020qtd}%
  \BibitemOpen
  \bibfield  {author} {\bibinfo {author} {\bibfnamefont {T.}~\bibnamefont
  {Papanikolaou}}, \bibinfo {author} {\bibfnamefont {V.}~\bibnamefont
  {Vennin}}, \ and\ \bibinfo {author} {\bibfnamefont {D.}~\bibnamefont
  {Langlois}},\ }\href {\doibase 10.1088/1475-7516/2021/03/053} {\bibfield
  {journal} {\bibinfo  {journal} {JCAP}\ }\textbf {\bibinfo {volume} {03}},\
  \bibinfo {pages} {053} (\bibinfo {year} {2021})},\ \Eprint
  {http://arxiv.org/abs/2010.11573} {arXiv:2010.11573 [astro-ph.CO]}
  \BibitemShut {NoStop}%
\bibitem [{\citenamefont {Inomata}\ \emph {et~al.}(2021)\citenamefont
  {Inomata}, \citenamefont {Kawasaki}, \citenamefont {Mukaida},\ and\
  \citenamefont {Yanagida}}]{Inomata:2020xad}%
  \BibitemOpen
  \bibfield  {author} {\bibinfo {author} {\bibfnamefont {K.}~\bibnamefont
  {Inomata}}, \bibinfo {author} {\bibfnamefont {M.}~\bibnamefont {Kawasaki}},
  \bibinfo {author} {\bibfnamefont {K.}~\bibnamefont {Mukaida}}, \ and\
  \bibinfo {author} {\bibfnamefont {T.~T.}\ \bibnamefont {Yanagida}},\ }\href
  {\doibase 10.1103/PhysRevLett.126.131301} {\bibfield  {journal} {\bibinfo
  {journal} {Phys. Rev. Lett.}\ }\textbf {\bibinfo {volume} {126}},\ \bibinfo
  {pages} {131301} (\bibinfo {year} {2021})},\ \Eprint
  {http://arxiv.org/abs/2011.01270} {arXiv:2011.01270 [astro-ph.CO]}
  \BibitemShut {NoStop}%
\bibitem [{\citenamefont {Kawasaki}\ and\ \citenamefont
  {Nakatsuka}(2021)}]{Kawasaki:2021ycf}%
  \BibitemOpen
  \bibfield  {author} {\bibinfo {author} {\bibfnamefont {M.}~\bibnamefont
  {Kawasaki}}\ and\ \bibinfo {author} {\bibfnamefont {H.}~\bibnamefont
  {Nakatsuka}},\ }\href {\doibase 10.1088/1475-7516/2021/05/023} {\bibfield
  {journal} {\bibinfo  {journal} {JCAP}\ }\textbf {\bibinfo {volume} {05}},\
  \bibinfo {pages} {023} (\bibinfo {year} {2021})},\ \Eprint
  {http://arxiv.org/abs/2101.11244} {arXiv:2101.11244 [astro-ph.CO]}
  \BibitemShut {NoStop}%
\bibitem [{\citenamefont {Dandoy}\ \emph {et~al.}(2023)\citenamefont {Dandoy},
  \citenamefont {Domcke},\ and\ \citenamefont {Rompineve}}]{Dandoy:2023jot}%
  \BibitemOpen
  \bibfield  {author} {\bibinfo {author} {\bibfnamefont {V.}~\bibnamefont
  {Dandoy}}, \bibinfo {author} {\bibfnamefont {V.}~\bibnamefont {Domcke}}, \
  and\ \bibinfo {author} {\bibfnamefont {F.}~\bibnamefont {Rompineve}},\
  }\href@noop {} {\  (\bibinfo {year} {2023})},\ \Eprint
  {http://arxiv.org/abs/2302.07901} {arXiv:2302.07901 [astro-ph.CO]}
  \BibitemShut {NoStop}%
\bibitem [{\citenamefont {Arzoumanian}\ \emph {et~al.}(2020)\citenamefont
  {Arzoumanian} \emph {et~al.}}]{NANOGrav:2020bcs}%
  \BibitemOpen
  \bibfield  {author} {\bibinfo {author} {\bibfnamefont {Z.}~\bibnamefont
  {Arzoumanian}} \emph {et~al.} (\bibinfo {collaboration} {NANOGrav}),\ }\href
  {\doibase 10.3847/2041-8213/abd401} {\bibfield  {journal} {\bibinfo
  {journal} {Astrophys. J. Lett.}\ }\textbf {\bibinfo {volume} {905}},\
  \bibinfo {pages} {L34} (\bibinfo {year} {2020})},\ \Eprint
  {http://arxiv.org/abs/2009.04496} {arXiv:2009.04496 [astro-ph.HE]}
  \BibitemShut {NoStop}%
\bibitem [{\citenamefont {Antoniadis}\ \emph {et~al.}(2022)\citenamefont
  {Antoniadis} \emph {et~al.}}]{Antoniadis:2022pcn}%
  \BibitemOpen
  \bibfield  {author} {\bibinfo {author} {\bibfnamefont {J.}~\bibnamefont
  {Antoniadis}} \emph {et~al.},\ }\href {\doibase 10.1093/mnras/stab3418}
  {\bibfield  {journal} {\bibinfo  {journal} {Mon. Not. Roy. Astron. Soc.}\
  }\textbf {\bibinfo {volume} {510}},\ \bibinfo {pages} {4873} (\bibinfo {year}
  {2022})},\ \Eprint {http://arxiv.org/abs/2201.03980} {arXiv:2201.03980
  [astro-ph.HE]} \BibitemShut {NoStop}%
\bibitem [{\citenamefont {Saito}\ and\ \citenamefont
  {Yokoyama}(2009)}]{Saito:2008jc}%
  \BibitemOpen
  \bibfield  {author} {\bibinfo {author} {\bibfnamefont {R.}~\bibnamefont
  {Saito}}\ and\ \bibinfo {author} {\bibfnamefont {J.}~\bibnamefont
  {Yokoyama}},\ }\href {\doibase 10.1103/PhysRevLett.102.161101,
  10.1103/PhysRevLett.107.069901} {\bibfield  {journal} {\bibinfo  {journal}
  {Phys. Rev. Lett.}\ }\textbf {\bibinfo {volume} {102}},\ \bibinfo {pages}
  {161101} (\bibinfo {year} {2009})},\ \bibinfo {note} {[Erratum: Phys. Rev.
  Lett.107,069901(2011)]},\ \Eprint {http://arxiv.org/abs/0812.4339}
  {arXiv:0812.4339 [astro-ph]} \BibitemShut {NoStop}%
\bibitem [{\citenamefont {Saito}\ and\ \citenamefont
  {Yokoyama}(2010)}]{Saito:2009jt}%
  \BibitemOpen
  \bibfield  {author} {\bibinfo {author} {\bibfnamefont {R.}~\bibnamefont
  {Saito}}\ and\ \bibinfo {author} {\bibfnamefont {J.}~\bibnamefont
  {Yokoyama}},\ }\href {\doibase 10.1143/PTP.126.351, 10.1143/PTP.123.867}
  {\bibfield  {journal} {\bibinfo  {journal} {Prog. Theor. Phys.}\ }\textbf
  {\bibinfo {volume} {123}},\ \bibinfo {pages} {867} (\bibinfo {year}
  {2010})},\ \bibinfo {note} {[Erratum: Prog. Theor. Phys.126,351(2011)]},\
  \Eprint {http://arxiv.org/abs/0912.5317} {arXiv:0912.5317 [astro-ph.CO]}
  \BibitemShut {NoStop}%
\bibitem [{\citenamefont {Inomata}\ \emph {et~al.}(2017)\citenamefont
  {Inomata}, \citenamefont {Kawasaki}, \citenamefont {Mukaida}, \citenamefont
  {Tada},\ and\ \citenamefont {Yanagida}}]{Inomata:2016rbd}%
  \BibitemOpen
  \bibfield  {author} {\bibinfo {author} {\bibfnamefont {K.}~\bibnamefont
  {Inomata}}, \bibinfo {author} {\bibfnamefont {M.}~\bibnamefont {Kawasaki}},
  \bibinfo {author} {\bibfnamefont {K.}~\bibnamefont {Mukaida}}, \bibinfo
  {author} {\bibfnamefont {Y.}~\bibnamefont {Tada}}, \ and\ \bibinfo {author}
  {\bibfnamefont {T.~T.}\ \bibnamefont {Yanagida}},\ }\href {\doibase
  10.1103/PhysRevD.95.123510} {\bibfield  {journal} {\bibinfo  {journal} {Phys.
  Rev.}\ }\textbf {\bibinfo {volume} {D95}},\ \bibinfo {pages} {123510}
  (\bibinfo {year} {2017})},\ \Eprint {http://arxiv.org/abs/1611.06130}
  {arXiv:1611.06130 [astro-ph.CO]} \BibitemShut {NoStop}%
\bibitem [{\citenamefont {Orlofsky}\ \emph {et~al.}(2017)\citenamefont
  {Orlofsky}, \citenamefont {Pierce},\ and\ \citenamefont
  {Wells}}]{Orlofsky:2016vbd}%
  \BibitemOpen
  \bibfield  {author} {\bibinfo {author} {\bibfnamefont {N.}~\bibnamefont
  {Orlofsky}}, \bibinfo {author} {\bibfnamefont {A.}~\bibnamefont {Pierce}}, \
  and\ \bibinfo {author} {\bibfnamefont {J.~D.}\ \bibnamefont {Wells}},\ }\href
  {\doibase 10.1103/PhysRevD.95.063518} {\bibfield  {journal} {\bibinfo
  {journal} {Phys. Rev.}\ }\textbf {\bibinfo {volume} {D95}},\ \bibinfo {pages}
  {063518} (\bibinfo {year} {2017})},\ \Eprint
  {http://arxiv.org/abs/1612.05279} {arXiv:1612.05279 [astro-ph.CO]}
  \BibitemShut {NoStop}%
\bibitem [{\citenamefont {Nakama}\ \emph {et~al.}(2017)\citenamefont {Nakama},
  \citenamefont {Silk},\ and\ \citenamefont {Kamionkowski}}]{Nakama:2016gzw}%
  \BibitemOpen
  \bibfield  {author} {\bibinfo {author} {\bibfnamefont {T.}~\bibnamefont
  {Nakama}}, \bibinfo {author} {\bibfnamefont {J.}~\bibnamefont {Silk}}, \ and\
  \bibinfo {author} {\bibfnamefont {M.}~\bibnamefont {Kamionkowski}},\ }\href
  {\doibase 10.1103/PhysRevD.95.043511} {\bibfield  {journal} {\bibinfo
  {journal} {Phys. Rev.}\ }\textbf {\bibinfo {volume} {D95}},\ \bibinfo {pages}
  {043511} (\bibinfo {year} {2017})},\ \Eprint
  {http://arxiv.org/abs/1612.06264} {arXiv:1612.06264 [astro-ph.CO]}
  \BibitemShut {NoStop}%
\bibitem [{\citenamefont {Garcia-Bellido}\ \emph {et~al.}(2017)\citenamefont
  {Garcia-Bellido}, \citenamefont {Peloso},\ and\ \citenamefont
  {Unal}}]{Garcia-Bellido:2017aan}%
  \BibitemOpen
  \bibfield  {author} {\bibinfo {author} {\bibfnamefont {J.}~\bibnamefont
  {Garcia-Bellido}}, \bibinfo {author} {\bibfnamefont {M.}~\bibnamefont
  {Peloso}}, \ and\ \bibinfo {author} {\bibfnamefont {C.}~\bibnamefont
  {Unal}},\ }\href {\doibase 10.1088/1475-7516/2017/09/013} {\bibfield
  {journal} {\bibinfo  {journal} {JCAP}\ }\textbf {\bibinfo {volume} {1709}},\
  \bibinfo {pages} {013} (\bibinfo {year} {2017})},\ \Eprint
  {http://arxiv.org/abs/1707.02441} {arXiv:1707.02441 [astro-ph.CO]}
  \BibitemShut {NoStop}%
\bibitem [{\citenamefont {Di}\ and\ \citenamefont {Gong}(2018)}]{Di:2017ndc}%
  \BibitemOpen
  \bibfield  {author} {\bibinfo {author} {\bibfnamefont {H.}~\bibnamefont
  {Di}}\ and\ \bibinfo {author} {\bibfnamefont {Y.}~\bibnamefont {Gong}},\
  }\href {\doibase 10.1088/1475-7516/2018/07/007} {\bibfield  {journal}
  {\bibinfo  {journal} {JCAP}\ }\textbf {\bibinfo {volume} {07}},\ \bibinfo
  {pages} {007} (\bibinfo {year} {2018})},\ \Eprint
  {http://arxiv.org/abs/1707.09578} {arXiv:1707.09578 [astro-ph.CO]}
  \BibitemShut {NoStop}%
\bibitem [{\citenamefont {Ando}\ \emph
  {et~al.}(2018{\natexlab{a}})\citenamefont {Ando}, \citenamefont {Inomata},
  \citenamefont {Kawasaki}, \citenamefont {Mukaida},\ and\ \citenamefont
  {Yanagida}}]{Ando:2017veq}%
  \BibitemOpen
  \bibfield  {author} {\bibinfo {author} {\bibfnamefont {K.}~\bibnamefont
  {Ando}}, \bibinfo {author} {\bibfnamefont {K.}~\bibnamefont {Inomata}},
  \bibinfo {author} {\bibfnamefont {M.}~\bibnamefont {Kawasaki}}, \bibinfo
  {author} {\bibfnamefont {K.}~\bibnamefont {Mukaida}}, \ and\ \bibinfo
  {author} {\bibfnamefont {T.~T.}\ \bibnamefont {Yanagida}},\ }\href {\doibase
  10.1103/PhysRevD.97.123512} {\bibfield  {journal} {\bibinfo  {journal} {Phys.
  Rev.}\ }\textbf {\bibinfo {volume} {D97}},\ \bibinfo {pages} {123512}
  (\bibinfo {year} {2018}{\natexlab{a}})},\ \Eprint
  {http://arxiv.org/abs/1711.08956} {arXiv:1711.08956 [astro-ph.CO]}
  \BibitemShut {NoStop}%
\bibitem [{\citenamefont {Cheng}\ \emph {et~al.}(2018)\citenamefont {Cheng},
  \citenamefont {Lee},\ and\ \citenamefont {Ng}}]{Cheng:2018yyr}%
  \BibitemOpen
  \bibfield  {author} {\bibinfo {author} {\bibfnamefont {S.-L.}\ \bibnamefont
  {Cheng}}, \bibinfo {author} {\bibfnamefont {W.}~\bibnamefont {Lee}}, \ and\
  \bibinfo {author} {\bibfnamefont {K.-W.}\ \bibnamefont {Ng}},\ }\href
  {\doibase 10.1088/1475-7516/2018/07/001} {\bibfield  {journal} {\bibinfo
  {journal} {JCAP}\ }\textbf {\bibinfo {volume} {07}},\ \bibinfo {pages} {001}
  (\bibinfo {year} {2018})},\ \Eprint {http://arxiv.org/abs/1801.09050}
  {arXiv:1801.09050 [astro-ph.CO]} \BibitemShut {NoStop}%
\bibitem [{\citenamefont {Kohri}\ and\ \citenamefont
  {Terada}(2018{\natexlab{a}})}]{Kohri:2018qtx}%
  \BibitemOpen
  \bibfield  {author} {\bibinfo {author} {\bibfnamefont {K.}~\bibnamefont
  {Kohri}}\ and\ \bibinfo {author} {\bibfnamefont {T.}~\bibnamefont {Terada}},\
  }\href {\doibase 10.1088/1361-6382/aaea18} {\bibfield  {journal} {\bibinfo
  {journal} {Class. Quant. Grav.}\ }\textbf {\bibinfo {volume} {35}},\ \bibinfo
  {pages} {235017} (\bibinfo {year} {2018}{\natexlab{a}})},\ \Eprint
  {http://arxiv.org/abs/1802.06785} {arXiv:1802.06785 [astro-ph.CO]}
  \BibitemShut {NoStop}%
\bibitem [{\citenamefont {Chen}\ \emph {et~al.}(2020)\citenamefont {Chen},
  \citenamefont {Yuan},\ and\ \citenamefont {Huang}}]{Chen:2019xse}%
  \BibitemOpen
  \bibfield  {author} {\bibinfo {author} {\bibfnamefont {Z.-C.}\ \bibnamefont
  {Chen}}, \bibinfo {author} {\bibfnamefont {C.}~\bibnamefont {Yuan}}, \ and\
  \bibinfo {author} {\bibfnamefont {Q.-G.}\ \bibnamefont {Huang}},\ }\href
  {\doibase 10.1103/PhysRevLett.124.251101} {\bibfield  {journal} {\bibinfo
  {journal} {Phys. Rev. Lett.}\ }\textbf {\bibinfo {volume} {124}},\ \bibinfo
  {pages} {251101} (\bibinfo {year} {2020})},\ \Eprint
  {http://arxiv.org/abs/1910.12239} {arXiv:1910.12239 [astro-ph.CO]}
  \BibitemShut {NoStop}%
\bibitem [{\citenamefont {Espinosa}\ \emph {et~al.}(2018)\citenamefont
  {Espinosa}, \citenamefont {Racco},\ and\ \citenamefont
  {Riotto}}]{Espinosa:2018eve}%
  \BibitemOpen
  \bibfield  {author} {\bibinfo {author} {\bibfnamefont {J.~R.}\ \bibnamefont
  {Espinosa}}, \bibinfo {author} {\bibfnamefont {D.}~\bibnamefont {Racco}}, \
  and\ \bibinfo {author} {\bibfnamefont {A.}~\bibnamefont {Riotto}},\ }\href
  {\doibase 10.1088/1475-7516/2018/09/012} {\bibfield  {journal} {\bibinfo
  {journal} {JCAP}\ }\textbf {\bibinfo {volume} {1809}},\ \bibinfo {pages}
  {012} (\bibinfo {year} {2018})},\ \Eprint {http://arxiv.org/abs/1804.07732}
  {arXiv:1804.07732 [hep-ph]} \BibitemShut {NoStop}%
\bibitem [{\citenamefont {Kohri}\ and\ \citenamefont
  {Terada}(2018{\natexlab{b}})}]{Kohri:2018awv}%
  \BibitemOpen
  \bibfield  {author} {\bibinfo {author} {\bibfnamefont {K.}~\bibnamefont
  {Kohri}}\ and\ \bibinfo {author} {\bibfnamefont {T.}~\bibnamefont {Terada}},\
  }\href {\doibase 10.1103/PhysRevD.97.123532} {\bibfield  {journal} {\bibinfo
  {journal} {Phys. Rev.}\ }\textbf {\bibinfo {volume} {D97}},\ \bibinfo {pages}
  {123532} (\bibinfo {year} {2018}{\natexlab{b}})},\ \Eprint
  {http://arxiv.org/abs/1804.08577} {arXiv:1804.08577 [gr-qc]} \BibitemShut
  {NoStop}%
\bibitem [{\citenamefont {Cai}\ \emph {et~al.}(2018)\citenamefont {Cai},
  \citenamefont {Pi},\ and\ \citenamefont {Sasaki}}]{Cai:2018dig}%
  \BibitemOpen
  \bibfield  {author} {\bibinfo {author} {\bibfnamefont {R.-g.}\ \bibnamefont
  {Cai}}, \bibinfo {author} {\bibfnamefont {S.}~\bibnamefont {Pi}}, \ and\
  \bibinfo {author} {\bibfnamefont {M.}~\bibnamefont {Sasaki}},\ }\href@noop {}
  {\  (\bibinfo {year} {2018})},\ \Eprint {http://arxiv.org/abs/1810.11000}
  {arXiv:1810.11000 [astro-ph.CO]} \BibitemShut {NoStop}%
\bibitem [{\citenamefont {Bartolo}\ \emph
  {et~al.}(2018{\natexlab{a}})\citenamefont {Bartolo}, \citenamefont {De~Luca},
  \citenamefont {Franciolini}, \citenamefont {Peloso},\ and\ \citenamefont
  {Riotto}}]{Bartolo:2018evs}%
  \BibitemOpen
  \bibfield  {author} {\bibinfo {author} {\bibfnamefont {N.}~\bibnamefont
  {Bartolo}}, \bibinfo {author} {\bibfnamefont {V.}~\bibnamefont {De~Luca}},
  \bibinfo {author} {\bibfnamefont {G.}~\bibnamefont {Franciolini}}, \bibinfo
  {author} {\bibfnamefont {M.}~\bibnamefont {Peloso}}, \ and\ \bibinfo {author}
  {\bibfnamefont {A.}~\bibnamefont {Riotto}},\ }\href@noop {} {\  (\bibinfo
  {year} {2018}{\natexlab{a}})},\ \Eprint {http://arxiv.org/abs/1810.12218}
  {arXiv:1810.12218 [astro-ph.CO]} \BibitemShut {NoStop}%
\bibitem [{\citenamefont {Bartolo}\ \emph
  {et~al.}(2018{\natexlab{b}})\citenamefont {Bartolo}, \citenamefont {De~Luca},
  \citenamefont {Franciolini}, \citenamefont {Peloso}, \citenamefont {Racco},\
  and\ \citenamefont {Riotto}}]{Bartolo:2018rku}%
  \BibitemOpen
  \bibfield  {author} {\bibinfo {author} {\bibfnamefont {N.}~\bibnamefont
  {Bartolo}}, \bibinfo {author} {\bibfnamefont {V.}~\bibnamefont {De~Luca}},
  \bibinfo {author} {\bibfnamefont {G.}~\bibnamefont {Franciolini}}, \bibinfo
  {author} {\bibfnamefont {M.}~\bibnamefont {Peloso}}, \bibinfo {author}
  {\bibfnamefont {D.}~\bibnamefont {Racco}}, \ and\ \bibinfo {author}
  {\bibfnamefont {A.}~\bibnamefont {Riotto}},\ }\href@noop {} {\  (\bibinfo
  {year} {2018}{\natexlab{b}})},\ \Eprint {http://arxiv.org/abs/1810.12224}
  {arXiv:1810.12224 [astro-ph.CO]} \BibitemShut {NoStop}%
\bibitem [{\citenamefont {Unal}(2018)}]{Unal:2018yaa}%
  \BibitemOpen
  \bibfield  {author} {\bibinfo {author} {\bibfnamefont {C.}~\bibnamefont
  {Unal}},\ }\href@noop {} {\  (\bibinfo {year} {2018})},\ \Eprint
  {http://arxiv.org/abs/1811.09151} {arXiv:1811.09151 [astro-ph.CO]}
  \BibitemShut {NoStop}%
\bibitem [{\citenamefont {Byrnes}\ \emph
  {et~al.}(2018{\natexlab{a}})\citenamefont {Byrnes}, \citenamefont {Cole},\
  and\ \citenamefont {Patil}}]{Byrnes:2018txb}%
  \BibitemOpen
  \bibfield  {author} {\bibinfo {author} {\bibfnamefont {C.~T.}\ \bibnamefont
  {Byrnes}}, \bibinfo {author} {\bibfnamefont {P.~S.}\ \bibnamefont {Cole}}, \
  and\ \bibinfo {author} {\bibfnamefont {S.~P.}\ \bibnamefont {Patil}},\
  }\href@noop {} {\  (\bibinfo {year} {2018}{\natexlab{a}})},\ \Eprint
  {http://arxiv.org/abs/1811.11158} {arXiv:1811.11158 [astro-ph.CO]}
  \BibitemShut {NoStop}%
\bibitem [{\citenamefont {Inomata}\ and\ \citenamefont
  {Nakama}(2019)}]{Inomata:2018epa}%
  \BibitemOpen
  \bibfield  {author} {\bibinfo {author} {\bibfnamefont {K.}~\bibnamefont
  {Inomata}}\ and\ \bibinfo {author} {\bibfnamefont {T.}~\bibnamefont
  {Nakama}},\ }\href {\doibase 10.1103/PhysRevD.99.043511} {\bibfield
  {journal} {\bibinfo  {journal} {Phys. Rev.}\ }\textbf {\bibinfo {volume}
  {D99}},\ \bibinfo {pages} {043511} (\bibinfo {year} {2019})},\ \Eprint
  {http://arxiv.org/abs/1812.00674} {arXiv:1812.00674 [astro-ph.CO]}
  \BibitemShut {NoStop}%
\bibitem [{\citenamefont {Clesse}\ \emph {et~al.}(2018)\citenamefont {Clesse},
  \citenamefont {Garc\'\i{}a-Bellido},\ and\ \citenamefont
  {Orani}}]{Clesse:2018ogk}%
  \BibitemOpen
  \bibfield  {author} {\bibinfo {author} {\bibfnamefont {S.}~\bibnamefont
  {Clesse}}, \bibinfo {author} {\bibfnamefont {J.}~\bibnamefont
  {Garc\'\i{}a-Bellido}}, \ and\ \bibinfo {author} {\bibfnamefont
  {S.}~\bibnamefont {Orani}},\ }\href@noop {} {\  (\bibinfo {year} {2018})},\
  \Eprint {http://arxiv.org/abs/1812.11011} {arXiv:1812.11011 [astro-ph.CO]}
  \BibitemShut {NoStop}%
\bibitem [{\citenamefont {Cai}\ \emph {et~al.}(2019{\natexlab{a}})\citenamefont
  {Cai}, \citenamefont {Pi}, \citenamefont {Wang},\ and\ \citenamefont
  {Yang}}]{Cai:2019amo}%
  \BibitemOpen
  \bibfield  {author} {\bibinfo {author} {\bibfnamefont {R.-G.}\ \bibnamefont
  {Cai}}, \bibinfo {author} {\bibfnamefont {S.}~\bibnamefont {Pi}}, \bibinfo
  {author} {\bibfnamefont {S.-J.}\ \bibnamefont {Wang}}, \ and\ \bibinfo
  {author} {\bibfnamefont {X.-Y.}\ \bibnamefont {Yang}},\ }\href {\doibase
  10.1088/1475-7516/2019/05/013} {\bibfield  {journal} {\bibinfo  {journal}
  {JCAP}\ }\textbf {\bibinfo {volume} {05}},\ \bibinfo {pages} {013} (\bibinfo
  {year} {2019}{\natexlab{a}})},\ \Eprint {http://arxiv.org/abs/1901.10152}
  {arXiv:1901.10152 [astro-ph.CO]} \BibitemShut {NoStop}%
\bibitem [{\citenamefont {Cai}\ \emph {et~al.}(2019{\natexlab{b}})\citenamefont
  {Cai}, \citenamefont {Chen}, \citenamefont {Tong}, \citenamefont {Wang},\
  and\ \citenamefont {Yan}}]{Cai:2019jah}%
  \BibitemOpen
  \bibfield  {author} {\bibinfo {author} {\bibfnamefont {Y.-F.}\ \bibnamefont
  {Cai}}, \bibinfo {author} {\bibfnamefont {C.}~\bibnamefont {Chen}}, \bibinfo
  {author} {\bibfnamefont {X.}~\bibnamefont {Tong}}, \bibinfo {author}
  {\bibfnamefont {D.-G.}\ \bibnamefont {Wang}}, \ and\ \bibinfo {author}
  {\bibfnamefont {S.-F.}\ \bibnamefont {Yan}},\ }\href {\doibase
  10.1103/PhysRevD.100.043518} {\bibfield  {journal} {\bibinfo  {journal}
  {Phys. Rev. D}\ }\textbf {\bibinfo {volume} {100}},\ \bibinfo {pages}
  {043518} (\bibinfo {year} {2019}{\natexlab{b}})},\ \Eprint
  {http://arxiv.org/abs/1902.08187} {arXiv:1902.08187 [astro-ph.CO]}
  \BibitemShut {NoStop}%
\bibitem [{\citenamefont {Dom\`enech}(2021)}]{Domenech:2021ztg}%
  \BibitemOpen
  \bibfield  {author} {\bibinfo {author} {\bibfnamefont {G.}~\bibnamefont
  {Dom\`enech}},\ }\href {\doibase 10.3390/universe7110398} {\bibfield
  {journal} {\bibinfo  {journal} {Universe}\ }\textbf {\bibinfo {volume} {7}},\
  \bibinfo {pages} {398} (\bibinfo {year} {2021})},\ \Eprint
  {http://arxiv.org/abs/2109.01398} {arXiv:2109.01398 [gr-qc]} \BibitemShut
  {NoStop}%
\bibitem [{\citenamefont {Yoo}\ \emph {et~al.}(2021)\citenamefont {Yoo},
  \citenamefont {Harada}, \citenamefont {Hirano},\ and\ \citenamefont
  {Kohri}}]{Yoo:2020dkz}%
  \BibitemOpen
  \bibfield  {author} {\bibinfo {author} {\bibfnamefont {C.-M.}\ \bibnamefont
  {Yoo}}, \bibinfo {author} {\bibfnamefont {T.}~\bibnamefont {Harada}},
  \bibinfo {author} {\bibfnamefont {S.}~\bibnamefont {Hirano}}, \ and\ \bibinfo
  {author} {\bibfnamefont {K.}~\bibnamefont {Kohri}},\ }\href {\doibase
  10.1093/ptep/ptaa155} {\bibfield  {journal} {\bibinfo  {journal} {PTEP}\
  }\textbf {\bibinfo {volume} {2021}},\ \bibinfo {pages} {013E02} (\bibinfo
  {year} {2021})},\ \Eprint {http://arxiv.org/abs/2008.02425} {arXiv:2008.02425
  [astro-ph.CO]} \BibitemShut {NoStop}%
\bibitem [{\citenamefont {De~Luca}\ \emph {et~al.}(2023)\citenamefont
  {De~Luca}, \citenamefont {Kehagias},\ and\ \citenamefont
  {Riotto}}]{DeLuca:2023tun}%
  \BibitemOpen
  \bibfield  {author} {\bibinfo {author} {\bibfnamefont {V.}~\bibnamefont
  {De~Luca}}, \bibinfo {author} {\bibfnamefont {A.}~\bibnamefont {Kehagias}}, \
  and\ \bibinfo {author} {\bibfnamefont {A.}~\bibnamefont {Riotto}},\
  }\href@noop {} {\  (\bibinfo {year} {2023})},\ \Eprint
  {http://arxiv.org/abs/2307.13633} {arXiv:2307.13633 [astro-ph.CO]}
  \BibitemShut {NoStop}%
\bibitem [{\citenamefont {Harigaya}\ \emph {et~al.}(2023)\citenamefont
  {Harigaya}, \citenamefont {Inomata},\ and\ \citenamefont
  {Terada}}]{Harigaya:2023pmw}%
  \BibitemOpen
  \bibfield  {author} {\bibinfo {author} {\bibfnamefont {K.}~\bibnamefont
  {Harigaya}}, \bibinfo {author} {\bibfnamefont {K.}~\bibnamefont {Inomata}}, \
  and\ \bibinfo {author} {\bibfnamefont {T.}~\bibnamefont {Terada}},\
  }\href@noop {} {\  (\bibinfo {year} {2023})},\ \Eprint
  {http://arxiv.org/abs/2309.00228} {arXiv:2309.00228 [astro-ph.CO]}
  \BibitemShut {NoStop}%
\bibitem [{\citenamefont {Enqvist}\ and\ \citenamefont
  {Sloth}(2002)}]{Enqvist:2001zp}%
  \BibitemOpen
  \bibfield  {author} {\bibinfo {author} {\bibfnamefont {K.}~\bibnamefont
  {Enqvist}}\ and\ \bibinfo {author} {\bibfnamefont {M.~S.}\ \bibnamefont
  {Sloth}},\ }\href {\doibase 10.1016/S0550-3213(02)00043-3} {\bibfield
  {journal} {\bibinfo  {journal} {Nucl. Phys. B}\ }\textbf {\bibinfo {volume}
  {626}},\ \bibinfo {pages} {395} (\bibinfo {year} {2002})},\ \Eprint
  {http://arxiv.org/abs/hep-ph/0109214} {arXiv:hep-ph/0109214} \BibitemShut
  {NoStop}%
\bibitem [{\citenamefont {Lyth}\ and\ \citenamefont
  {Wands}(2002)}]{Lyth:2001nq}%
  \BibitemOpen
  \bibfield  {author} {\bibinfo {author} {\bibfnamefont {D.~H.}\ \bibnamefont
  {Lyth}}\ and\ \bibinfo {author} {\bibfnamefont {D.}~\bibnamefont {Wands}},\
  }\href {\doibase 10.1016/S0370-2693(01)01366-1} {\bibfield  {journal}
  {\bibinfo  {journal} {Phys. Lett. B}\ }\textbf {\bibinfo {volume} {524}},\
  \bibinfo {pages} {5} (\bibinfo {year} {2002})},\ \Eprint
  {http://arxiv.org/abs/hep-ph/0110002} {arXiv:hep-ph/0110002} \BibitemShut
  {NoStop}%
\bibitem [{\citenamefont {Moroi}\ and\ \citenamefont
  {Takahashi}(2001)}]{Moroi:2001ct}%
  \BibitemOpen
  \bibfield  {author} {\bibinfo {author} {\bibfnamefont {T.}~\bibnamefont
  {Moroi}}\ and\ \bibinfo {author} {\bibfnamefont {T.}~\bibnamefont
  {Takahashi}},\ }\href {\doibase 10.1016/S0370-2693(01)01295-3} {\bibfield
  {journal} {\bibinfo  {journal} {Phys. Lett. B}\ }\textbf {\bibinfo {volume}
  {522}},\ \bibinfo {pages} {215} (\bibinfo {year} {2001})},\ \bibinfo {note}
  {[Erratum: Phys.Lett.B 539, 303--303 (2002)]},\ \Eprint
  {http://arxiv.org/abs/hep-ph/0110096} {arXiv:hep-ph/0110096} \BibitemShut
  {NoStop}%
\bibitem [{\citenamefont {Pi}\ and\ \citenamefont {Sasaki}(2021)}]{Pi:2021dft}%
  \BibitemOpen
  \bibfield  {author} {\bibinfo {author} {\bibfnamefont {S.}~\bibnamefont
  {Pi}}\ and\ \bibinfo {author} {\bibfnamefont {M.}~\bibnamefont {Sasaki}},\
  }\href@noop {} {\  (\bibinfo {year} {2021})},\ \Eprint
  {http://arxiv.org/abs/2112.12680} {arXiv:2112.12680 [astro-ph.CO]}
  \BibitemShut {NoStop}%
\bibitem [{\citenamefont {Kasuya}\ and\ \citenamefont
  {Kawasaki}(2009)}]{Kasuya:2009up}%
  \BibitemOpen
  \bibfield  {author} {\bibinfo {author} {\bibfnamefont {S.}~\bibnamefont
  {Kasuya}}\ and\ \bibinfo {author} {\bibfnamefont {M.}~\bibnamefont
  {Kawasaki}},\ }\href {\doibase 10.1103/PhysRevD.80.023516} {\bibfield
  {journal} {\bibinfo  {journal} {Phys. Rev.}\ }\textbf {\bibinfo {volume}
  {D80}},\ \bibinfo {pages} {023516} (\bibinfo {year} {2009})},\ \Eprint
  {http://arxiv.org/abs/0904.3800} {arXiv:0904.3800 [astro-ph.CO]} \BibitemShut
  {NoStop}%
\bibitem [{\citenamefont {Kawasaki}\ \emph
  {et~al.}(2013{\natexlab{a}})\citenamefont {Kawasaki}, \citenamefont
  {Kitajima},\ and\ \citenamefont {Yanagida}}]{Kawasaki:2012wr}%
  \BibitemOpen
  \bibfield  {author} {\bibinfo {author} {\bibfnamefont {M.}~\bibnamefont
  {Kawasaki}}, \bibinfo {author} {\bibfnamefont {N.}~\bibnamefont {Kitajima}},
  \ and\ \bibinfo {author} {\bibfnamefont {T.~T.}\ \bibnamefont {Yanagida}},\
  }\href {\doibase 10.1103/PhysRevD.87.063519} {\bibfield  {journal} {\bibinfo
  {journal} {Phys. Rev.}\ }\textbf {\bibinfo {volume} {D87}},\ \bibinfo {pages}
  {063519} (\bibinfo {year} {2013}{\natexlab{a}})},\ \Eprint
  {http://arxiv.org/abs/1207.2550} {arXiv:1207.2550 [hep-ph]} \BibitemShut
  {NoStop}%
\bibitem [{\citenamefont {Kawasaki}\ \emph
  {et~al.}(2013{\natexlab{b}})\citenamefont {Kawasaki}, \citenamefont
  {Kitajima},\ and\ \citenamefont {Yokoyama}}]{Kawasaki:2013xsa}%
  \BibitemOpen
  \bibfield  {author} {\bibinfo {author} {\bibfnamefont {M.}~\bibnamefont
  {Kawasaki}}, \bibinfo {author} {\bibfnamefont {N.}~\bibnamefont {Kitajima}},
  \ and\ \bibinfo {author} {\bibfnamefont {S.}~\bibnamefont {Yokoyama}},\
  }\href {\doibase 10.1088/1475-7516/2013/08/042} {\bibfield  {journal}
  {\bibinfo  {journal} {JCAP}\ }\textbf {\bibinfo {volume} {08}},\ \bibinfo
  {pages} {042} (\bibinfo {year} {2013}{\natexlab{b}})},\ \Eprint
  {http://arxiv.org/abs/1305.4464} {arXiv:1305.4464 [astro-ph.CO]} \BibitemShut
  {NoStop}%
\bibitem [{\citenamefont {Ferrante}\ \emph
  {et~al.}(2023{\natexlab{a}})\citenamefont {Ferrante}, \citenamefont
  {Franciolini}, \citenamefont {Iovino},\ and\ \citenamefont
  {Urbano}}]{Ferrante:2023bgz}%
  \BibitemOpen
  \bibfield  {author} {\bibinfo {author} {\bibfnamefont {G.}~\bibnamefont
  {Ferrante}}, \bibinfo {author} {\bibfnamefont {G.}~\bibnamefont
  {Franciolini}}, \bibinfo {author} {\bibfnamefont {A.}~\bibnamefont {Iovino},
  \bibfnamefont {Junior.}}, \ and\ \bibinfo {author} {\bibfnamefont
  {A.}~\bibnamefont {Urbano}},\ }\href {\doibase 10.1088/1475-7516/2023/06/057}
  {\bibfield  {journal} {\bibinfo  {journal} {JCAP}\ }\textbf {\bibinfo
  {volume} {06}},\ \bibinfo {pages} {057} (\bibinfo {year}
  {2023}{\natexlab{a}})},\ \Eprint {http://arxiv.org/abs/2305.13382}
  {arXiv:2305.13382 [astro-ph.CO]} \BibitemShut {NoStop}%
\bibitem [{\citenamefont {Ando}\ \emph
  {et~al.}(2018{\natexlab{b}})\citenamefont {Ando}, \citenamefont {Kawasaki},\
  and\ \citenamefont {Nakatsuka}}]{Ando:2018nge}%
  \BibitemOpen
  \bibfield  {author} {\bibinfo {author} {\bibfnamefont {K.}~\bibnamefont
  {Ando}}, \bibinfo {author} {\bibfnamefont {M.}~\bibnamefont {Kawasaki}}, \
  and\ \bibinfo {author} {\bibfnamefont {H.}~\bibnamefont {Nakatsuka}},\ }\href
  {\doibase 10.1103/PhysRevD.98.083508} {\bibfield  {journal} {\bibinfo
  {journal} {Phys. Rev.}\ }\textbf {\bibinfo {volume} {D98}},\ \bibinfo {pages}
  {083508} (\bibinfo {year} {2018}{\natexlab{b}})},\ \Eprint
  {http://arxiv.org/abs/1805.07757} {arXiv:1805.07757 [astro-ph.CO]}
  \BibitemShut {NoStop}%
\bibitem [{\citenamefont {Sasaki}\ \emph {et~al.}(2006)\citenamefont {Sasaki},
  \citenamefont {Valiviita},\ and\ \citenamefont {Wands}}]{Sasaki:2006kq}%
  \BibitemOpen
  \bibfield  {author} {\bibinfo {author} {\bibfnamefont {M.}~\bibnamefont
  {Sasaki}}, \bibinfo {author} {\bibfnamefont {J.}~\bibnamefont {Valiviita}}, \
  and\ \bibinfo {author} {\bibfnamefont {D.}~\bibnamefont {Wands}},\ }\href
  {\doibase 10.1103/PhysRevD.74.103003} {\bibfield  {journal} {\bibinfo
  {journal} {Phys. Rev. D}\ }\textbf {\bibinfo {volume} {74}},\ \bibinfo
  {pages} {103003} (\bibinfo {year} {2006})},\ \Eprint
  {http://arxiv.org/abs/astro-ph/0607627} {arXiv:astro-ph/0607627} \BibitemShut
  {NoStop}%
\bibitem [{\citenamefont {Kawasaki}\ \emph {et~al.}(2011)\citenamefont
  {Kawasaki}, \citenamefont {Kobayashi},\ and\ \citenamefont
  {Takahashi}}]{Kawasaki:2011pd}%
  \BibitemOpen
  \bibfield  {author} {\bibinfo {author} {\bibfnamefont {M.}~\bibnamefont
  {Kawasaki}}, \bibinfo {author} {\bibfnamefont {T.}~\bibnamefont {Kobayashi}},
  \ and\ \bibinfo {author} {\bibfnamefont {F.}~\bibnamefont {Takahashi}},\
  }\href {\doibase 10.1103/PhysRevD.84.123506} {\bibfield  {journal} {\bibinfo
  {journal} {Phys. Rev. D}\ }\textbf {\bibinfo {volume} {84}},\ \bibinfo
  {pages} {123506} (\bibinfo {year} {2011})},\ \Eprint
  {http://arxiv.org/abs/1107.6011} {arXiv:1107.6011 [astro-ph.CO]} \BibitemShut
  {NoStop}%
\bibitem [{\citenamefont {Ferrante}\ \emph
  {et~al.}(2023{\natexlab{b}})\citenamefont {Ferrante}, \citenamefont
  {Franciolini}, \citenamefont {Iovino},\ and\ \citenamefont
  {Urbano}}]{Ferrante:2022mui}%
  \BibitemOpen
  \bibfield  {author} {\bibinfo {author} {\bibfnamefont {G.}~\bibnamefont
  {Ferrante}}, \bibinfo {author} {\bibfnamefont {G.}~\bibnamefont
  {Franciolini}}, \bibinfo {author} {\bibfnamefont {A.}~\bibnamefont {Iovino},
  \bibfnamefont {Junior.}}, \ and\ \bibinfo {author} {\bibfnamefont
  {A.}~\bibnamefont {Urbano}},\ }\href {\doibase 10.1103/PhysRevD.107.043520}
  {\bibfield  {journal} {\bibinfo  {journal} {Phys. Rev. D}\ }\textbf {\bibinfo
  {volume} {107}},\ \bibinfo {pages} {043520} (\bibinfo {year}
  {2023}{\natexlab{b}})},\ \Eprint {http://arxiv.org/abs/2211.01728}
  {arXiv:2211.01728 [astro-ph.CO]} \BibitemShut {NoStop}%
\bibitem [{\citenamefont {Byrnes}\ \emph {et~al.}(2012)\citenamefont {Byrnes},
  \citenamefont {Copeland},\ and\ \citenamefont {Green}}]{Byrnes:2012yx}%
  \BibitemOpen
  \bibfield  {author} {\bibinfo {author} {\bibfnamefont {C.~T.}\ \bibnamefont
  {Byrnes}}, \bibinfo {author} {\bibfnamefont {E.~J.}\ \bibnamefont
  {Copeland}}, \ and\ \bibinfo {author} {\bibfnamefont {A.~M.}\ \bibnamefont
  {Green}},\ }\href {\doibase 10.1103/PhysRevD.86.043512} {\bibfield  {journal}
  {\bibinfo  {journal} {Phys. Rev.}\ }\textbf {\bibinfo {volume} {D86}},\
  \bibinfo {pages} {043512} (\bibinfo {year} {2012})},\ \Eprint
  {http://arxiv.org/abs/1206.4188} {arXiv:1206.4188 [astro-ph.CO]} \BibitemShut
  {NoStop}%
\bibitem [{\citenamefont {Young}\ and\ \citenamefont
  {Byrnes}(2013)}]{Young:2013oia}%
  \BibitemOpen
  \bibfield  {author} {\bibinfo {author} {\bibfnamefont {S.}~\bibnamefont
  {Young}}\ and\ \bibinfo {author} {\bibfnamefont {C.~T.}\ \bibnamefont
  {Byrnes}},\ }\href {\doibase 10.1088/1475-7516/2013/08/052} {\bibfield
  {journal} {\bibinfo  {journal} {JCAP}\ }\textbf {\bibinfo {volume} {1308}},\
  \bibinfo {pages} {052} (\bibinfo {year} {2013})},\ \Eprint
  {http://arxiv.org/abs/1307.4995} {arXiv:1307.4995 [astro-ph.CO]} \BibitemShut
  {NoStop}%
\bibitem [{\citenamefont {Firouzjahi}\ and\ \citenamefont
  {Riotto}(2023)}]{Firouzjahi:2023xke}%
  \BibitemOpen
  \bibfield  {author} {\bibinfo {author} {\bibfnamefont {H.}~\bibnamefont
  {Firouzjahi}}\ and\ \bibinfo {author} {\bibfnamefont {A.}~\bibnamefont
  {Riotto}},\ }\href@noop {} {\  (\bibinfo {year} {2023})},\ \Eprint
  {http://arxiv.org/abs/2309.10536} {arXiv:2309.10536 [astro-ph.CO]}
  \BibitemShut {NoStop}%
\bibitem [{\citenamefont {Starobinsky}(1980)}]{Starobinsky:1980te}%
  \BibitemOpen
  \bibfield  {author} {\bibinfo {author} {\bibfnamefont {A.~A.}\ \bibnamefont
  {Starobinsky}},\ }\href {\doibase 10.1016/0370-2693(80)90670-X} {\bibfield
  {journal} {\bibinfo  {journal} {Phys. Lett. B}\ }\textbf {\bibinfo {volume}
  {91}},\ \bibinfo {pages} {99} (\bibinfo {year} {1980})}\BibitemShut {NoStop}%
\bibitem [{\citenamefont {Akrami}\ \emph {et~al.}(2020)\citenamefont {Akrami}
  \emph {et~al.}}]{Planck:2018jri}%
  \BibitemOpen
  \bibfield  {author} {\bibinfo {author} {\bibfnamefont {Y.}~\bibnamefont
  {Akrami}} \emph {et~al.} (\bibinfo {collaboration} {Planck}),\ }\href
  {\doibase 10.1051/0004-6361/201833887} {\bibfield  {journal} {\bibinfo
  {journal} {Astron. Astrophys.}\ }\textbf {\bibinfo {volume} {641}},\ \bibinfo
  {pages} {A10} (\bibinfo {year} {2020})},\ \Eprint
  {http://arxiv.org/abs/1807.06211} {arXiv:1807.06211 [astro-ph.CO]}
  \BibitemShut {NoStop}%
\bibitem [{\citenamefont {Aghanim}\ \emph {et~al.}(2018)\citenamefont {Aghanim}
  \emph {et~al.}}]{Aghanim:2018eyx}%
  \BibitemOpen
  \bibfield  {author} {\bibinfo {author} {\bibfnamefont {N.}~\bibnamefont
  {Aghanim}} \emph {et~al.} (\bibinfo {collaboration} {Planck}),\ }\href@noop
  {} {\  (\bibinfo {year} {2018})},\ \Eprint {http://arxiv.org/abs/1807.06209}
  {arXiv:1807.06209 [astro-ph.CO]} \BibitemShut {NoStop}%
\bibitem [{\citenamefont {Dom\`enech}\ \emph {et~al.}(2022)\citenamefont
  {Dom\`enech}, \citenamefont {Passaglia},\ and\ \citenamefont
  {Renaux-Petel}}]{Domenech:2021and}%
  \BibitemOpen
  \bibfield  {author} {\bibinfo {author} {\bibfnamefont {G.}~\bibnamefont
  {Dom\`enech}}, \bibinfo {author} {\bibfnamefont {S.}~\bibnamefont
  {Passaglia}}, \ and\ \bibinfo {author} {\bibfnamefont {S.}~\bibnamefont
  {Renaux-Petel}},\ }\href {\doibase 10.1088/1475-7516/2022/03/023} {\bibfield
  {journal} {\bibinfo  {journal} {JCAP}\ }\textbf {\bibinfo {volume} {03}},\
  \bibinfo {pages} {023} (\bibinfo {year} {2022})},\ \Eprint
  {http://arxiv.org/abs/2112.10163} {arXiv:2112.10163 [astro-ph.CO]}
  \BibitemShut {NoStop}%
\bibitem [{\citenamefont {Inomata}\ \emph
  {et~al.}(2019{\natexlab{a}})\citenamefont {Inomata}, \citenamefont {Kohri},
  \citenamefont {Nakama},\ and\ \citenamefont {Terada}}]{Inomata:2019zqy}%
  \BibitemOpen
  \bibfield  {author} {\bibinfo {author} {\bibfnamefont {K.}~\bibnamefont
  {Inomata}}, \bibinfo {author} {\bibfnamefont {K.}~\bibnamefont {Kohri}},
  \bibinfo {author} {\bibfnamefont {T.}~\bibnamefont {Nakama}}, \ and\ \bibinfo
  {author} {\bibfnamefont {T.}~\bibnamefont {Terada}},\ }\href {\doibase
  10.1088/1475-7516/2019/10/071} {\bibfield  {journal} {\bibinfo  {journal}
  {JCAP}\ }\textbf {\bibinfo {volume} {10}},\ \bibinfo {pages} {071} (\bibinfo
  {year} {2019}{\natexlab{a}})},\ \Eprint {http://arxiv.org/abs/1904.12878}
  {arXiv:1904.12878 [astro-ph.CO]} \BibitemShut {NoStop}%
\bibitem [{\citenamefont {Inomata}\ \emph
  {et~al.}(2019{\natexlab{b}})\citenamefont {Inomata}, \citenamefont {Kohri},
  \citenamefont {Nakama},\ and\ \citenamefont {Terada}}]{Inomata:2019ivs}%
  \BibitemOpen
  \bibfield  {author} {\bibinfo {author} {\bibfnamefont {K.}~\bibnamefont
  {Inomata}}, \bibinfo {author} {\bibfnamefont {K.}~\bibnamefont {Kohri}},
  \bibinfo {author} {\bibfnamefont {T.}~\bibnamefont {Nakama}}, \ and\ \bibinfo
  {author} {\bibfnamefont {T.}~\bibnamefont {Terada}},\ }\href {\doibase
  10.1103/PhysRevD.100.043532} {\bibfield  {journal} {\bibinfo  {journal}
  {Phys. Rev.}\ }\textbf {\bibinfo {volume} {D100}},\ \bibinfo {pages} {043532}
  (\bibinfo {year} {2019}{\natexlab{b}})},\ \Eprint
  {http://arxiv.org/abs/1904.12879} {arXiv:1904.12879 [astro-ph.CO]}
  \BibitemShut {NoStop}%
\bibitem [{\citenamefont {Inomata}\ \emph {et~al.}(2020)\citenamefont
  {Inomata}, \citenamefont {Kawasaki}, \citenamefont {Mukaida}, \citenamefont
  {Terada},\ and\ \citenamefont {Yanagida}}]{Inomata:2020lmk}%
  \BibitemOpen
  \bibfield  {author} {\bibinfo {author} {\bibfnamefont {K.}~\bibnamefont
  {Inomata}}, \bibinfo {author} {\bibfnamefont {M.}~\bibnamefont {Kawasaki}},
  \bibinfo {author} {\bibfnamefont {K.}~\bibnamefont {Mukaida}}, \bibinfo
  {author} {\bibfnamefont {T.}~\bibnamefont {Terada}}, \ and\ \bibinfo {author}
  {\bibfnamefont {T.~T.}\ \bibnamefont {Yanagida}},\ }\href {\doibase
  10.1103/PhysRevD.101.123533} {\bibfield  {journal} {\bibinfo  {journal}
  {Phys. Rev. D}\ }\textbf {\bibinfo {volume} {101}},\ \bibinfo {pages}
  {123533} (\bibinfo {year} {2020})},\ \Eprint
  {http://arxiv.org/abs/2003.10455} {arXiv:2003.10455 [astro-ph.CO]}
  \BibitemShut {NoStop}%
\bibitem [{\citenamefont {Saikawa}\ and\ \citenamefont
  {Shirai}(2018)}]{Saikawa:2018rcs}%
  \BibitemOpen
  \bibfield  {author} {\bibinfo {author} {\bibfnamefont {K.}~\bibnamefont
  {Saikawa}}\ and\ \bibinfo {author} {\bibfnamefont {S.}~\bibnamefont
  {Shirai}},\ }\href {\doibase 10.1088/1475-7516/2018/05/035} {\bibfield
  {journal} {\bibinfo  {journal} {JCAP}\ }\textbf {\bibinfo {volume} {1805}},\
  \bibinfo {pages} {035} (\bibinfo {year} {2018})},\ \Eprint
  {http://arxiv.org/abs/1803.01038} {arXiv:1803.01038 [hep-ph]} \BibitemShut
  {NoStop}%
\bibitem [{\citenamefont {Kolb}\ and\ \citenamefont
  {Turner}(1990)}]{Kolb:206230}%
  \BibitemOpen
  \bibfield  {author} {\bibinfo {author} {\bibfnamefont {E.~W.}\ \bibnamefont
  {Kolb}}\ and\ \bibinfo {author} {\bibfnamefont {M.~S.}\ \bibnamefont
  {Turner}},\ }\href {https://cds.cern.ch/record/206230} {\emph {\bibinfo
  {title} {{The early universe}}}},\ Frontiers in Physics\ (\bibinfo
  {publisher} {Westview Press},\ \bibinfo {address} {Boulder, CO},\ \bibinfo
  {year} {1990})\BibitemShut {NoStop}%
\bibitem [{\citenamefont {Lyth}\ \emph {et~al.}(2003)\citenamefont {Lyth},
  \citenamefont {Ungarelli},\ and\ \citenamefont {Wands}}]{Lyth:2002my}%
  \BibitemOpen
  \bibfield  {author} {\bibinfo {author} {\bibfnamefont {D.~H.}\ \bibnamefont
  {Lyth}}, \bibinfo {author} {\bibfnamefont {C.}~\bibnamefont {Ungarelli}}, \
  and\ \bibinfo {author} {\bibfnamefont {D.}~\bibnamefont {Wands}},\ }\href
  {\doibase 10.1103/PhysRevD.67.023503} {\bibfield  {journal} {\bibinfo
  {journal} {Phys. Rev.}\ }\textbf {\bibinfo {volume} {D67}},\ \bibinfo {pages}
  {023503} (\bibinfo {year} {2003})},\ \Eprint
  {http://arxiv.org/abs/astro-ph/0208055} {arXiv:astro-ph/0208055 [astro-ph]}
  \BibitemShut {NoStop}%
\bibitem [{\citenamefont {Inomata}\ \emph {et~al.}(2018)\citenamefont
  {Inomata}, \citenamefont {Kawasaki}, \citenamefont {Kusenko},\ and\
  \citenamefont {Yang}}]{Inomata:2018htm}%
  \BibitemOpen
  \bibfield  {author} {\bibinfo {author} {\bibfnamefont {K.}~\bibnamefont
  {Inomata}}, \bibinfo {author} {\bibfnamefont {M.}~\bibnamefont {Kawasaki}},
  \bibinfo {author} {\bibfnamefont {A.}~\bibnamefont {Kusenko}}, \ and\
  \bibinfo {author} {\bibfnamefont {L.}~\bibnamefont {Yang}},\ }\href {\doibase
  10.1088/1475-7516/2018/12/003} {\bibfield  {journal} {\bibinfo  {journal}
  {JCAP}\ }\textbf {\bibinfo {volume} {12}},\ \bibinfo {pages} {003} (\bibinfo
  {year} {2018})},\ \Eprint {http://arxiv.org/abs/1806.00123} {arXiv:1806.00123
  [astro-ph.CO]} \BibitemShut {NoStop}%
\bibitem [{\citenamefont {Yuan}\ and\ \citenamefont
  {Huang}(2021)}]{Yuan:2020iwf}%
  \BibitemOpen
  \bibfield  {author} {\bibinfo {author} {\bibfnamefont {C.}~\bibnamefont
  {Yuan}}\ and\ \bibinfo {author} {\bibfnamefont {Q.-G.}\ \bibnamefont
  {Huang}},\ }\href {\doibase 10.1016/j.physletb.2021.136606} {\bibfield
  {journal} {\bibinfo  {journal} {Phys. Lett. B}\ }\textbf {\bibinfo {volume}
  {821}},\ \bibinfo {pages} {136606} (\bibinfo {year} {2021})},\ \Eprint
  {http://arxiv.org/abs/2007.10686} {arXiv:2007.10686 [astro-ph.CO]}
  \BibitemShut {NoStop}%
\bibitem [{\citenamefont {Atal}\ and\ \citenamefont
  {Dom\`enech}(2021)}]{Atal:2021jyo}%
  \BibitemOpen
  \bibfield  {author} {\bibinfo {author} {\bibfnamefont {V.}~\bibnamefont
  {Atal}}\ and\ \bibinfo {author} {\bibfnamefont {G.}~\bibnamefont
  {Dom\`enech}},\ }\href {\doibase 10.1088/1475-7516/2021/06/001} {\bibfield
  {journal} {\bibinfo  {journal} {JCAP}\ }\textbf {\bibinfo {volume} {06}},\
  \bibinfo {pages} {001} (\bibinfo {year} {2021})},\ \Eprint
  {http://arxiv.org/abs/2103.01056} {arXiv:2103.01056 [astro-ph.CO]}
  \BibitemShut {NoStop}%
\bibitem [{\citenamefont {Adshead}\ \emph {et~al.}(2021)\citenamefont
  {Adshead}, \citenamefont {Lozanov},\ and\ \citenamefont
  {Weiner}}]{Adshead:2021hnm}%
  \BibitemOpen
  \bibfield  {author} {\bibinfo {author} {\bibfnamefont {P.}~\bibnamefont
  {Adshead}}, \bibinfo {author} {\bibfnamefont {K.~D.}\ \bibnamefont
  {Lozanov}}, \ and\ \bibinfo {author} {\bibfnamefont {Z.~J.}\ \bibnamefont
  {Weiner}},\ }\href {\doibase 10.1088/1475-7516/2021/10/080} {\bibfield
  {journal} {\bibinfo  {journal} {JCAP}\ }\textbf {\bibinfo {volume} {10}},\
  \bibinfo {pages} {080} (\bibinfo {year} {2021})},\ \Eprint
  {http://arxiv.org/abs/2105.01659} {arXiv:2105.01659 [astro-ph.CO]}
  \BibitemShut {NoStop}%
\bibitem [{\citenamefont {Garcia-Saenz}\ \emph {et~al.}(2023)\citenamefont
  {Garcia-Saenz}, \citenamefont {Pinol}, \citenamefont {Renaux-Petel},\ and\
  \citenamefont {Werth}}]{Garcia-Saenz:2022tzu}%
  \BibitemOpen
  \bibfield  {author} {\bibinfo {author} {\bibfnamefont {S.}~\bibnamefont
  {Garcia-Saenz}}, \bibinfo {author} {\bibfnamefont {L.}~\bibnamefont {Pinol}},
  \bibinfo {author} {\bibfnamefont {S.}~\bibnamefont {Renaux-Petel}}, \ and\
  \bibinfo {author} {\bibfnamefont {D.}~\bibnamefont {Werth}},\ }\href
  {\doibase 10.1088/1475-7516/2023/03/057} {\bibfield  {journal} {\bibinfo
  {journal} {JCAP}\ }\textbf {\bibinfo {volume} {03}},\ \bibinfo {pages} {057}
  (\bibinfo {year} {2023})},\ \Eprint {http://arxiv.org/abs/2207.14267}
  {arXiv:2207.14267 [astro-ph.CO]} \BibitemShut {NoStop}%
\bibitem [{\citenamefont {Li}\ \emph {et~al.}(2023{\natexlab{a}})\citenamefont
  {Li}, \citenamefont {Wang}, \citenamefont {Zhao},\ and\ \citenamefont
  {Kohri}}]{Li:2023qua}%
  \BibitemOpen
  \bibfield  {author} {\bibinfo {author} {\bibfnamefont {J.-P.}\ \bibnamefont
  {Li}}, \bibinfo {author} {\bibfnamefont {S.}~\bibnamefont {Wang}}, \bibinfo
  {author} {\bibfnamefont {Z.-C.}\ \bibnamefont {Zhao}}, \ and\ \bibinfo
  {author} {\bibfnamefont {K.}~\bibnamefont {Kohri}},\ }\href@noop {} {\
  (\bibinfo {year} {2023}{\natexlab{a}})},\ \Eprint
  {http://arxiv.org/abs/2305.19950} {arXiv:2305.19950 [astro-ph.CO]}
  \BibitemShut {NoStop}%
\bibitem [{\citenamefont {Li}\ \emph {et~al.}(2023{\natexlab{b}})\citenamefont
  {Li}, \citenamefont {Wang}, \citenamefont {Zhao},\ and\ \citenamefont
  {Kohri}}]{Li:2023xtl}%
  \BibitemOpen
  \bibfield  {author} {\bibinfo {author} {\bibfnamefont {J.-P.}\ \bibnamefont
  {Li}}, \bibinfo {author} {\bibfnamefont {S.}~\bibnamefont {Wang}}, \bibinfo
  {author} {\bibfnamefont {Z.-C.}\ \bibnamefont {Zhao}}, \ and\ \bibinfo
  {author} {\bibfnamefont {K.}~\bibnamefont {Kohri}},\ }\href@noop {} {\
  (\bibinfo {year} {2023}{\natexlab{b}})},\ \Eprint
  {http://arxiv.org/abs/2309.07792} {arXiv:2309.07792 [astro-ph.CO]}
  \BibitemShut {NoStop}%
\bibitem [{\citenamefont {Abe}\ \emph {et~al.}(2021)\citenamefont {Abe},
  \citenamefont {Tada},\ and\ \citenamefont {Ueda}}]{Abe:2020sqb}%
  \BibitemOpen
  \bibfield  {author} {\bibinfo {author} {\bibfnamefont {K.~T.}\ \bibnamefont
  {Abe}}, \bibinfo {author} {\bibfnamefont {Y.}~\bibnamefont {Tada}}, \ and\
  \bibinfo {author} {\bibfnamefont {I.}~\bibnamefont {Ueda}},\ }\href {\doibase
  10.1088/1475-7516/2021/06/048} {\bibfield  {journal} {\bibinfo  {journal}
  {JCAP}\ }\textbf {\bibinfo {volume} {06}},\ \bibinfo {pages} {048} (\bibinfo
  {year} {2021})},\ \Eprint {http://arxiv.org/abs/2010.06193} {arXiv:2010.06193
  [astro-ph.CO]} \BibitemShut {NoStop}%
\bibitem [{\citenamefont {Auclair}\ \emph {et~al.}(2022)\citenamefont {Auclair}
  \emph {et~al.}}]{LISACosmologyWorkingGroup:2022jok}%
  \BibitemOpen
  \bibfield  {author} {\bibinfo {author} {\bibfnamefont {P.}~\bibnamefont
  {Auclair}} \emph {et~al.} (\bibinfo {collaboration} {LISA Cosmology Working
  Group}),\ }\href@noop {} {\  (\bibinfo {year} {2022})},\ \Eprint
  {http://arxiv.org/abs/2204.05434} {arXiv:2204.05434 [astro-ph.CO]}
  \BibitemShut {NoStop}%
\bibitem [{\citenamefont {Kawamura}\ \emph {et~al.}(2021)\citenamefont
  {Kawamura} \emph {et~al.}}]{Kawamura:2020pcg}%
  \BibitemOpen
  \bibfield  {author} {\bibinfo {author} {\bibfnamefont {S.}~\bibnamefont
  {Kawamura}} \emph {et~al.},\ }\href {\doibase 10.1093/ptep/ptab019}
  {\bibfield  {journal} {\bibinfo  {journal} {PTEP}\ }\textbf {\bibinfo
  {volume} {2021}},\ \bibinfo {pages} {05A105} (\bibinfo {year} {2021})},\
  \Eprint {http://arxiv.org/abs/2006.13545} {arXiv:2006.13545 [gr-qc]}
  \BibitemShut {NoStop}%
\bibitem [{\citenamefont {Crowder}\ and\ \citenamefont
  {Cornish}(2005)}]{Crowder:2005nr}%
  \BibitemOpen
  \bibfield  {author} {\bibinfo {author} {\bibfnamefont {J.}~\bibnamefont
  {Crowder}}\ and\ \bibinfo {author} {\bibfnamefont {N.~J.}\ \bibnamefont
  {Cornish}},\ }\href {\doibase 10.1103/PhysRevD.72.083005} {\bibfield
  {journal} {\bibinfo  {journal} {Phys. Rev. D}\ }\textbf {\bibinfo {volume}
  {72}},\ \bibinfo {pages} {083005} (\bibinfo {year} {2005})},\ \Eprint
  {http://arxiv.org/abs/gr-qc/0506015} {arXiv:gr-qc/0506015} \BibitemShut
  {NoStop}%
\bibitem [{\citenamefont {Corbin}\ and\ \citenamefont
  {Cornish}(2006)}]{Corbin:2005ny}%
  \BibitemOpen
  \bibfield  {author} {\bibinfo {author} {\bibfnamefont {V.}~\bibnamefont
  {Corbin}}\ and\ \bibinfo {author} {\bibfnamefont {N.~J.}\ \bibnamefont
  {Cornish}},\ }\href {\doibase 10.1088/0264-9381/23/7/014} {\bibfield
  {journal} {\bibinfo  {journal} {Class. Quant. Grav.}\ }\textbf {\bibinfo
  {volume} {23}},\ \bibinfo {pages} {2435} (\bibinfo {year} {2006})},\ \Eprint
  {http://arxiv.org/abs/gr-qc/0512039} {arXiv:gr-qc/0512039} \BibitemShut
  {NoStop}%
\bibitem [{\citenamefont {Clarke}\ \emph {et~al.}(2020)\citenamefont {Clarke},
  \citenamefont {Copeland},\ and\ \citenamefont {Moss}}]{Clarke:2020bil}%
  \BibitemOpen
  \bibfield  {author} {\bibinfo {author} {\bibfnamefont {T.~J.}\ \bibnamefont
  {Clarke}}, \bibinfo {author} {\bibfnamefont {E.~J.}\ \bibnamefont
  {Copeland}}, \ and\ \bibinfo {author} {\bibfnamefont {A.}~\bibnamefont
  {Moss}},\ }\href {\doibase 10.1088/1475-7516/2020/10/002} {\bibfield
  {journal} {\bibinfo  {journal} {JCAP}\ }\textbf {\bibinfo {volume} {10}},\
  \bibinfo {pages} {002} (\bibinfo {year} {2020})},\ \Eprint
  {http://arxiv.org/abs/2004.11396} {arXiv:2004.11396 [astro-ph.CO]}
  \BibitemShut {NoStop}%
\bibitem [{\citenamefont {Schmitz}(2021)}]{Schmitz:2020syl}%
  \BibitemOpen
  \bibfield  {author} {\bibinfo {author} {\bibfnamefont {K.}~\bibnamefont
  {Schmitz}},\ }\href {\doibase 10.1007/JHEP01(2021)097} {\bibfield  {journal}
  {\bibinfo  {journal} {JHEP}\ }\textbf {\bibinfo {volume} {01}},\ \bibinfo
  {pages} {097} (\bibinfo {year} {2021})},\ \Eprint
  {http://arxiv.org/abs/2002.04615} {arXiv:2002.04615 [hep-ph]} \BibitemShut
  {NoStop}%
\bibitem [{\citenamefont {Mitridate}(2023)}]{andrea_mitridate_2023}%
  \BibitemOpen
  \bibfield  {author} {\bibinfo {author} {\bibfnamefont {A.}~\bibnamefont
  {Mitridate}},\ }\href {\doibase 10.5281/zenodo.7876430} {\  (\bibinfo {year}
  {2023}),\ 10.5281/zenodo.7876430}\BibitemShut {NoStop}%
\bibitem [{\citenamefont {Mitridate}\ \emph {et~al.}(2023)\citenamefont
  {Mitridate}, \citenamefont {Wright}, \citenamefont {von Eckardstein},
  \citenamefont {Schr\"oder}, \citenamefont {Nay}, \citenamefont {Olum},
  \citenamefont {Schmitz},\ and\ \citenamefont {Trickle}}]{Mitridate:2023oar}%
  \BibitemOpen
  \bibfield  {author} {\bibinfo {author} {\bibfnamefont {A.}~\bibnamefont
  {Mitridate}}, \bibinfo {author} {\bibfnamefont {D.}~\bibnamefont {Wright}},
  \bibinfo {author} {\bibfnamefont {R.}~\bibnamefont {von Eckardstein}},
  \bibinfo {author} {\bibfnamefont {T.}~\bibnamefont {Schr\"oder}}, \bibinfo
  {author} {\bibfnamefont {J.}~\bibnamefont {Nay}}, \bibinfo {author}
  {\bibfnamefont {K.}~\bibnamefont {Olum}}, \bibinfo {author} {\bibfnamefont
  {K.}~\bibnamefont {Schmitz}}, \ and\ \bibinfo {author} {\bibfnamefont
  {T.}~\bibnamefont {Trickle}},\ }\href@noop {} {\  (\bibinfo {year} {2023})},\
  \Eprint {http://arxiv.org/abs/2306.16377} {arXiv:2306.16377 [hep-ph]}
  \BibitemShut {NoStop}%
\bibitem [{\citenamefont {Lamb}\ \emph {et~al.}(2023)\citenamefont {Lamb},
  \citenamefont {Taylor},\ and\ \citenamefont {van Haasteren}}]{lamb2023need}%
  \BibitemOpen
  \bibfield  {author} {\bibinfo {author} {\bibfnamefont {W.~G.}\ \bibnamefont
  {Lamb}}, \bibinfo {author} {\bibfnamefont {S.~R.}\ \bibnamefont {Taylor}}, \
  and\ \bibinfo {author} {\bibfnamefont {R.}~\bibnamefont {van Haasteren}},\
  }\href@noop {} {\enquote {\bibinfo {title} {The need for speed: Rapid
  refitting techniques for bayesian spectral characterization of the
  gravitational wave background using ptas},}\ } (\bibinfo {year} {2023}),\
  \Eprint {http://arxiv.org/abs/2303.15442} {arXiv:2303.15442 [astro-ph.HE]}
  \BibitemShut {NoStop}%
\bibitem [{\citenamefont {Kawasaki}\ and\ \citenamefont
  {Nakatsuka}(2019)}]{Kawasaki:2019mbl}%
  \BibitemOpen
  \bibfield  {author} {\bibinfo {author} {\bibfnamefont {M.}~\bibnamefont
  {Kawasaki}}\ and\ \bibinfo {author} {\bibfnamefont {H.}~\bibnamefont
  {Nakatsuka}},\ }\href {\doibase 10.1103/PhysRevD.99.123501} {\bibfield
  {journal} {\bibinfo  {journal} {Phys. Rev. D}\ }\textbf {\bibinfo {volume}
  {99}},\ \bibinfo {pages} {123501} (\bibinfo {year} {2019})},\ \Eprint
  {http://arxiv.org/abs/1903.02994} {arXiv:1903.02994 [astro-ph.CO]}
  \BibitemShut {NoStop}%
\bibitem [{\citenamefont {Niemeyer}\ and\ \citenamefont
  {Jedamzik}(1998)}]{Niemeyer:1997mt}%
  \BibitemOpen
  \bibfield  {author} {\bibinfo {author} {\bibfnamefont {J.~C.}\ \bibnamefont
  {Niemeyer}}\ and\ \bibinfo {author} {\bibfnamefont {K.}~\bibnamefont
  {Jedamzik}},\ }\href {\doibase 10.1103/PhysRevLett.80.5481} {\bibfield
  {journal} {\bibinfo  {journal} {Phys. Rev. Lett.}\ }\textbf {\bibinfo
  {volume} {80}},\ \bibinfo {pages} {5481} (\bibinfo {year} {1998})},\ \Eprint
  {http://arxiv.org/abs/astro-ph/9709072} {arXiv:astro-ph/9709072 [astro-ph]}
  \BibitemShut {NoStop}%
\bibitem [{\citenamefont {Niemeyer}(1998)}]{Niemeyer:1998ac}%
  \BibitemOpen
  \bibfield  {author} {\bibinfo {author} {\bibfnamefont {J.~C.}\ \bibnamefont
  {Niemeyer}},\ }in\ \href@noop {} {\emph {\bibinfo {booktitle} {{Sources and
  detection of dark matter in the universe. Proceedings, 3rd International
  Symposium, and Workshop on Primordial Black Holes and Hawking Radiation,
  Marina del Rey, USA, February 17-20, 1998}}}}\ (\bibinfo {year} {1998})\
  \Eprint {http://arxiv.org/abs/astro-ph/9806043} {arXiv:astro-ph/9806043
  [astro-ph]} \BibitemShut {NoStop}%
\bibitem [{\citenamefont {Shibata}\ and\ \citenamefont
  {Sasaki}(1999)}]{Shibata:1999zs}%
  \BibitemOpen
  \bibfield  {author} {\bibinfo {author} {\bibfnamefont {M.}~\bibnamefont
  {Shibata}}\ and\ \bibinfo {author} {\bibfnamefont {M.}~\bibnamefont
  {Sasaki}},\ }\href {\doibase 10.1103/PhysRevD.60.084002} {\bibfield
  {journal} {\bibinfo  {journal} {Phys. Rev.}\ }\textbf {\bibinfo {volume}
  {D60}},\ \bibinfo {pages} {084002} (\bibinfo {year} {1999})},\ \Eprint
  {http://arxiv.org/abs/gr-qc/9905064} {arXiv:gr-qc/9905064 [gr-qc]}
  \BibitemShut {NoStop}%
\bibitem [{\citenamefont {Yokoyama}(1998)}]{Yokoyama:1998xd}%
  \BibitemOpen
  \bibfield  {author} {\bibinfo {author} {\bibfnamefont {J.}~\bibnamefont
  {Yokoyama}},\ }\href {\doibase 10.1103/PhysRevD.58.107502} {\bibfield
  {journal} {\bibinfo  {journal} {Phys. Rev. D}\ }\textbf {\bibinfo {volume}
  {58}},\ \bibinfo {pages} {107502} (\bibinfo {year} {1998})},\ \Eprint
  {http://arxiv.org/abs/gr-qc/9804041} {arXiv:gr-qc/9804041} \BibitemShut
  {NoStop}%
\bibitem [{\citenamefont {Carr}(1975)}]{Carr:1975qj}%
  \BibitemOpen
  \bibfield  {author} {\bibinfo {author} {\bibfnamefont {B.~J.}\ \bibnamefont
  {Carr}},\ }\href {\doibase 10.1086/153853} {\bibfield  {journal} {\bibinfo
  {journal} {Astrophys. J.}\ }\textbf {\bibinfo {volume} {201}},\ \bibinfo
  {pages} {1} (\bibinfo {year} {1975})}\BibitemShut {NoStop}%
\bibitem [{\citenamefont {Ando}\ \emph
  {et~al.}(2018{\natexlab{c}})\citenamefont {Ando}, \citenamefont {Inomata},\
  and\ \citenamefont {Kawasaki}}]{Ando:2018qdb}%
  \BibitemOpen
  \bibfield  {author} {\bibinfo {author} {\bibfnamefont {K.}~\bibnamefont
  {Ando}}, \bibinfo {author} {\bibfnamefont {K.}~\bibnamefont {Inomata}}, \
  and\ \bibinfo {author} {\bibfnamefont {M.}~\bibnamefont {Kawasaki}},\ }\href
  {\doibase 10.1103/PhysRevD.97.103528} {\bibfield  {journal} {\bibinfo
  {journal} {Phys. Rev.}\ }\textbf {\bibinfo {volume} {D97}},\ \bibinfo {pages}
  {103528} (\bibinfo {year} {2018}{\natexlab{c}})},\ \Eprint
  {http://arxiv.org/abs/1802.06393} {arXiv:1802.06393 [astro-ph.CO]}
  \BibitemShut {NoStop}%
\bibitem [{\citenamefont {Young}(2019)}]{Young:2019osy}%
  \BibitemOpen
  \bibfield  {author} {\bibinfo {author} {\bibfnamefont {S.}~\bibnamefont
  {Young}},\ }\href {\doibase 10.1142/S0218271820300025} {\bibfield  {journal}
  {\bibinfo  {journal} {Int. J. Mod. Phys. D}\ }\textbf {\bibinfo {volume}
  {29}},\ \bibinfo {pages} {2030002} (\bibinfo {year} {2019})},\ \Eprint
  {http://arxiv.org/abs/1905.01230} {arXiv:1905.01230 [astro-ph.CO]}
  \BibitemShut {NoStop}%
\bibitem [{\citenamefont {De~Luca}\ \emph {et~al.}(2019)\citenamefont
  {De~Luca}, \citenamefont {Franciolini}, \citenamefont {Kehagias},
  \citenamefont {Peloso}, \citenamefont {Riotto},\ and\ \citenamefont
  {\"Unal}}]{DeLuca:2019qsy}%
  \BibitemOpen
  \bibfield  {author} {\bibinfo {author} {\bibfnamefont {V.}~\bibnamefont
  {De~Luca}}, \bibinfo {author} {\bibfnamefont {G.}~\bibnamefont
  {Franciolini}}, \bibinfo {author} {\bibfnamefont {A.}~\bibnamefont
  {Kehagias}}, \bibinfo {author} {\bibfnamefont {M.}~\bibnamefont {Peloso}},
  \bibinfo {author} {\bibfnamefont {A.}~\bibnamefont {Riotto}}, \ and\ \bibinfo
  {author} {\bibfnamefont {C.}~\bibnamefont {\"Unal}},\ }\href {\doibase
  10.1088/1475-7516/2019/07/048} {\bibfield  {journal} {\bibinfo  {journal}
  {JCAP}\ }\textbf {\bibinfo {volume} {07}},\ \bibinfo {pages} {048} (\bibinfo
  {year} {2019})},\ \Eprint {http://arxiv.org/abs/1904.00970} {arXiv:1904.00970
  [astro-ph.CO]} \BibitemShut {NoStop}%
\bibitem [{\citenamefont {Young}\ \emph {et~al.}(2019)\citenamefont {Young},
  \citenamefont {Musco},\ and\ \citenamefont {Byrnes}}]{Young:2019yug}%
  \BibitemOpen
  \bibfield  {author} {\bibinfo {author} {\bibfnamefont {S.}~\bibnamefont
  {Young}}, \bibinfo {author} {\bibfnamefont {I.}~\bibnamefont {Musco}}, \ and\
  \bibinfo {author} {\bibfnamefont {C.~T.}\ \bibnamefont {Byrnes}},\ }\href
  {\doibase 10.1088/1475-7516/2019/11/012} {\bibfield  {journal} {\bibinfo
  {journal} {JCAP}\ }\textbf {\bibinfo {volume} {11}},\ \bibinfo {pages} {012}
  (\bibinfo {year} {2019})},\ \Eprint {http://arxiv.org/abs/1904.00984}
  {arXiv:1904.00984 [astro-ph.CO]} \BibitemShut {NoStop}%
\bibitem [{\citenamefont {Harada}\ \emph {et~al.}(2015)\citenamefont {Harada},
  \citenamefont {Yoo}, \citenamefont {Nakama},\ and\ \citenamefont
  {Koga}}]{Harada:2015yda}%
  \BibitemOpen
  \bibfield  {author} {\bibinfo {author} {\bibfnamefont {T.}~\bibnamefont
  {Harada}}, \bibinfo {author} {\bibfnamefont {C.-M.}\ \bibnamefont {Yoo}},
  \bibinfo {author} {\bibfnamefont {T.}~\bibnamefont {Nakama}}, \ and\ \bibinfo
  {author} {\bibfnamefont {Y.}~\bibnamefont {Koga}},\ }\href {\doibase
  10.1103/PhysRevD.91.084057} {\bibfield  {journal} {\bibinfo  {journal} {Phys.
  Rev. D}\ }\textbf {\bibinfo {volume} {91}},\ \bibinfo {pages} {084057}
  (\bibinfo {year} {2015})},\ \Eprint {http://arxiv.org/abs/1503.03934}
  {arXiv:1503.03934 [gr-qc]} \BibitemShut {NoStop}%
\bibitem [{\citenamefont {Byrnes}\ \emph
  {et~al.}(2018{\natexlab{b}})\citenamefont {Byrnes}, \citenamefont
  {Hindmarsh}, \citenamefont {Young},\ and\ \citenamefont
  {Hawkins}}]{Byrnes:2018clq}%
  \BibitemOpen
  \bibfield  {author} {\bibinfo {author} {\bibfnamefont {C.~T.}\ \bibnamefont
  {Byrnes}}, \bibinfo {author} {\bibfnamefont {M.}~\bibnamefont {Hindmarsh}},
  \bibinfo {author} {\bibfnamefont {S.}~\bibnamefont {Young}}, \ and\ \bibinfo
  {author} {\bibfnamefont {M.~R.~S.}\ \bibnamefont {Hawkins}},\ }\href
  {\doibase 10.1088/1475-7516/2018/08/041} {\bibfield  {journal} {\bibinfo
  {journal} {JCAP}\ }\textbf {\bibinfo {volume} {08}},\ \bibinfo {pages} {041}
  (\bibinfo {year} {2018}{\natexlab{b}})},\ \Eprint
  {http://arxiv.org/abs/1801.06138} {arXiv:1801.06138 [astro-ph.CO]}
  \BibitemShut {NoStop}%
\bibitem [{\citenamefont {Musco}\ \emph {et~al.}(2023)\citenamefont {Musco},
  \citenamefont {Jedamzik},\ and\ \citenamefont {Young}}]{Musco:2023dak}%
  \BibitemOpen
  \bibfield  {author} {\bibinfo {author} {\bibfnamefont {I.}~\bibnamefont
  {Musco}}, \bibinfo {author} {\bibfnamefont {K.}~\bibnamefont {Jedamzik}}, \
  and\ \bibinfo {author} {\bibfnamefont {S.}~\bibnamefont {Young}},\
  }\href@noop {} {\  (\bibinfo {year} {2023})},\ \Eprint
  {http://arxiv.org/abs/2303.07980} {arXiv:2303.07980 [astro-ph.CO]}
  \BibitemShut {NoStop}%
\bibitem [{\citenamefont {{Mr{\'o}z}}\ \emph {et~al.}(2017)\citenamefont
  {{Mr{\'o}z}}, \citenamefont {{Udalski}}, \citenamefont {{Skowron}},
  \citenamefont {{Poleski}}, \citenamefont {{Koz{\l}owski}}, \citenamefont
  {{Szyma{\'n}ski}}, \citenamefont {{Soszy{\'n}ski}}, \citenamefont
  {{Wyrzykowski}}, \citenamefont {{Pietrukowicz}}, \citenamefont {{Ulaczyk}},
  \citenamefont {{Skowron}},\ and\ \citenamefont
  {{Pawlak}}}]{2017Natur.548..183M}%
  \BibitemOpen
  \bibfield  {author} {\bibinfo {author} {\bibfnamefont {P.}~\bibnamefont
  {{Mr{\'o}z}}}, \bibinfo {author} {\bibfnamefont {A.}~\bibnamefont
  {{Udalski}}}, \bibinfo {author} {\bibfnamefont {J.}~\bibnamefont
  {{Skowron}}}, \bibinfo {author} {\bibfnamefont {R.}~\bibnamefont
  {{Poleski}}}, \bibinfo {author} {\bibfnamefont {S.}~\bibnamefont
  {{Koz{\l}owski}}}, \bibinfo {author} {\bibfnamefont {M.~K.}\ \bibnamefont
  {{Szyma{\'n}ski}}}, \bibinfo {author} {\bibfnamefont {I.}~\bibnamefont
  {{Soszy{\'n}ski}}}, \bibinfo {author} {\bibfnamefont {{\L}.}~\bibnamefont
  {{Wyrzykowski}}}, \bibinfo {author} {\bibfnamefont {P.}~\bibnamefont
  {{Pietrukowicz}}}, \bibinfo {author} {\bibfnamefont {K.}~\bibnamefont
  {{Ulaczyk}}}, \bibinfo {author} {\bibfnamefont {D.}~\bibnamefont
  {{Skowron}}}, \ and\ \bibinfo {author} {\bibfnamefont {M.}~\bibnamefont
  {{Pawlak}}},\ }\href {\doibase 10.1038/nature23276} {\bibfield  {journal}
  {\bibinfo  {journal} {\nat}\ }\textbf {\bibinfo {volume} {548}},\ \bibinfo
  {pages} {183} (\bibinfo {year} {2017})},\ \Eprint
  {http://arxiv.org/abs/1707.07634} {arXiv:1707.07634 [astro-ph.EP]}
  \BibitemShut {NoStop}%
\bibitem [{\citenamefont {Niikura}\ \emph {et~al.}(2019)\citenamefont
  {Niikura}, \citenamefont {Takada}, \citenamefont {Yokoyama}, \citenamefont
  {Sumi},\ and\ \citenamefont {Masaki}}]{Niikura:2019kqi}%
  \BibitemOpen
  \bibfield  {author} {\bibinfo {author} {\bibfnamefont {H.}~\bibnamefont
  {Niikura}}, \bibinfo {author} {\bibfnamefont {M.}~\bibnamefont {Takada}},
  \bibinfo {author} {\bibfnamefont {S.}~\bibnamefont {Yokoyama}}, \bibinfo
  {author} {\bibfnamefont {T.}~\bibnamefont {Sumi}}, \ and\ \bibinfo {author}
  {\bibfnamefont {S.}~\bibnamefont {Masaki}},\ }\href {\doibase
  10.1103/PhysRevD.99.083503} {\bibfield  {journal} {\bibinfo  {journal} {Phys.
  Rev. D}\ }\textbf {\bibinfo {volume} {99}},\ \bibinfo {pages} {083503}
  (\bibinfo {year} {2019})},\ \Eprint {http://arxiv.org/abs/1901.07120}
  {arXiv:1901.07120 [astro-ph.CO]} \BibitemShut {NoStop}%
\bibitem [{\citenamefont {Tisserand}\ \emph {et~al.}(2007)\citenamefont
  {Tisserand} \emph {et~al.}}]{Tisserand:2006zx}%
  \BibitemOpen
  \bibfield  {author} {\bibinfo {author} {\bibfnamefont {P.}~\bibnamefont
  {Tisserand}} \emph {et~al.} (\bibinfo {collaboration} {EROS-2}),\ }\href
  {\doibase 10.1051/0004-6361:20066017} {\bibfield  {journal} {\bibinfo
  {journal} {Astron. Astrophys.}\ }\textbf {\bibinfo {volume} {469}},\ \bibinfo
  {pages} {387} (\bibinfo {year} {2007})},\ \Eprint
  {http://arxiv.org/abs/astro-ph/0607207} {arXiv:astro-ph/0607207 [astro-ph]}
  \BibitemShut {NoStop}%
\bibitem [{\citenamefont {Khlopov}\ and\ \citenamefont
  {Polnarev}(1980)}]{Khlopov:1980mg}%
  \BibitemOpen
  \bibfield  {author} {\bibinfo {author} {\bibfnamefont {M.~Y.}\ \bibnamefont
  {Khlopov}}\ and\ \bibinfo {author} {\bibfnamefont {A.~G.}\ \bibnamefont
  {Polnarev}},\ }\href {\doibase 10.1016/0370-2693(80)90624-3} {\bibfield
  {journal} {\bibinfo  {journal} {Phys. Lett. B}\ }\textbf {\bibinfo {volume}
  {97}},\ \bibinfo {pages} {383} (\bibinfo {year} {1980})}\BibitemShut
  {NoStop}%
\bibitem [{\citenamefont {Polnarev}\ and\ \citenamefont
  {Khlopov}(1982)}]{Khlopov:1982sov}%
  \BibitemOpen
  \bibfield  {author} {\bibinfo {author} {\bibfnamefont {A.~G.}\ \bibnamefont
  {Polnarev}}\ and\ \bibinfo {author} {\bibfnamefont {M.~{\relax Yu}.}\
  \bibnamefont {Khlopov}},\ }\href {\doibase 10.1016/0370-2693(80)90624-3}
  {\bibfield  {journal} {\bibinfo  {journal} {Sov. Astron.}\ }\textbf {\bibinfo
  {volume} {26}},\ \bibinfo {pages} {9} (\bibinfo {year} {1982})}\BibitemShut
  {NoStop}%
\bibitem [{\citenamefont {Harada}\ \emph {et~al.}(2016)\citenamefont {Harada},
  \citenamefont {Yoo}, \citenamefont {Kohri}, \citenamefont {Nakao},\ and\
  \citenamefont {Jhingan}}]{Harada:2016mhb}%
  \BibitemOpen
  \bibfield  {author} {\bibinfo {author} {\bibfnamefont {T.}~\bibnamefont
  {Harada}}, \bibinfo {author} {\bibfnamefont {C.-M.}\ \bibnamefont {Yoo}},
  \bibinfo {author} {\bibfnamefont {K.}~\bibnamefont {Kohri}}, \bibinfo
  {author} {\bibfnamefont {K.-i.}\ \bibnamefont {Nakao}}, \ and\ \bibinfo
  {author} {\bibfnamefont {S.}~\bibnamefont {Jhingan}},\ }\href {\doibase
  10.3847/1538-4357/833/1/61} {\bibfield  {journal} {\bibinfo  {journal}
  {Astrophys. J.}\ }\textbf {\bibinfo {volume} {833}},\ \bibinfo {pages} {61}
  (\bibinfo {year} {2016})},\ \Eprint {http://arxiv.org/abs/1609.01588}
  {arXiv:1609.01588 [astro-ph.CO]} \BibitemShut {NoStop}%
\bibitem [{\citenamefont {Harada}\ \emph {et~al.}(2017)\citenamefont {Harada},
  \citenamefont {Yoo}, \citenamefont {Kohri},\ and\ \citenamefont
  {Nakao}}]{Harada:2017fjm}%
  \BibitemOpen
  \bibfield  {author} {\bibinfo {author} {\bibfnamefont {T.}~\bibnamefont
  {Harada}}, \bibinfo {author} {\bibfnamefont {C.-M.}\ \bibnamefont {Yoo}},
  \bibinfo {author} {\bibfnamefont {K.}~\bibnamefont {Kohri}}, \ and\ \bibinfo
  {author} {\bibfnamefont {K.-I.}\ \bibnamefont {Nakao}},\ }\href {\doibase
  10.1103/PhysRevD.96.083517} {\bibfield  {journal} {\bibinfo  {journal} {Phys.
  Rev. D}\ }\textbf {\bibinfo {volume} {96}},\ \bibinfo {pages} {083517}
  (\bibinfo {year} {2017})},\ \bibinfo {note} {[Erratum: Phys.Rev.D 99, 069904
  (2019)]},\ \Eprint {http://arxiv.org/abs/1707.03595} {arXiv:1707.03595
  [gr-qc]} \BibitemShut {NoStop}%
\bibitem [{\citenamefont {Sasaki}\ \emph {et~al.}(2018)\citenamefont {Sasaki},
  \citenamefont {Suyama}, \citenamefont {Tanaka},\ and\ \citenamefont
  {Yokoyama}}]{Sasaki:2018dmp}%
  \BibitemOpen
  \bibfield  {author} {\bibinfo {author} {\bibfnamefont {M.}~\bibnamefont
  {Sasaki}}, \bibinfo {author} {\bibfnamefont {T.}~\bibnamefont {Suyama}},
  \bibinfo {author} {\bibfnamefont {T.}~\bibnamefont {Tanaka}}, \ and\ \bibinfo
  {author} {\bibfnamefont {S.}~\bibnamefont {Yokoyama}},\ }\href {\doibase
  10.1088/1361-6382/aaa7b4} {\bibfield  {journal} {\bibinfo  {journal} {Class.
  Quant. Grav.}\ }\textbf {\bibinfo {volume} {35}},\ \bibinfo {pages} {063001}
  (\bibinfo {year} {2018})},\ \Eprint {http://arxiv.org/abs/1801.05235}
  {arXiv:1801.05235 [astro-ph.CO]} \BibitemShut {NoStop}%
\bibitem [{\citenamefont {Abbott}\ \emph {et~al.}(2021)\citenamefont {Abbott}
  \emph {et~al.}}]{LIGOScientific:2021djp}%
  \BibitemOpen
  \bibfield  {author} {\bibinfo {author} {\bibfnamefont {R.}~\bibnamefont
  {Abbott}} \emph {et~al.} (\bibinfo {collaboration} {LIGO Scientific, VIRGO,
  KAGRA}),\ }\href@noop {} {\  (\bibinfo {year} {2021})},\ \Eprint
  {http://arxiv.org/abs/2111.03606} {arXiv:2111.03606 [gr-qc]} \BibitemShut
  {NoStop}%
\end{thebibliography}%

\end{document}